\documentclass[MSNbibl,nameyear,dvips]{arxstspdf}
\usepackage{flushend}
\usepackage{stfloats}
\usepackage{graphicx}
\volume{28}
\issue{1}
\pubyear{2013}
\firstpage{69}
\lastpage{94}
\doi{10.1214/12-STS401} 

\makeatletter

\newcommand{\cal}{\mathcal}

\newcommand{\citecs}[1]{\citeauthor{#1}, \citeyear{#1}}
\renewcommand{\citep}[1]{(\citeauthor{#1}, \citeyear{#1})}

\newtheorem{propo}{Proposition}

\newcommand{\utwi}[1]{\mathbf{#1}}
\newcommand{\bg}{{\utwi{g}}}
\newcommand{\bh}{{\utwi{h}}}
\newcommand{\bx}{{\utwi{x}}}
\newcommand{\by}{{\utwi{y}}}
\newcommand{\bz}{{\utwi{z}}}
\newcommand{\bF}{{\utwi{F}}}
\newcommand{\bI}{{\utwi{I}}}
\newcommand{\dff}{\stackrel{\triangle}{=}}

\setattribute{abstract}   {width}  {345pt}
\setattribute{keyword}    {width}  {345pt}
\makeatother

\begin{document}
\begin{frontmatter}

\title{Lookahead Strategies for Sequential Monte Carlo}
\runtitle{Lookahead Strategies for SMC}

\begin{aug}
\author[a]{\fnms{Ming} \snm{Lin}},
\author[b]{\fnms{Rong} \snm{Chen}\corref{}\ead[label=e1]{rongchen@stat.rutgers.edu}}
\and
\author[c]{\fnms{Jun S.} \snm{Liu}}
\runauthor{M. Lin, R. Chen and J. S. Liu}

\affiliation{Xiamen University, Rutgers University and Harvard University}

\address[a]{Ming Lin is Associate Professor,
Wang Yanan Institute for Studies in Economics, Xiamen University,
Xiamen, Fujian 361005, China.}
\address[b]{Rong Chen is Professor, Department of Statistics, Rutgers
University,
Piscataway, New Jersey 08854, USA \printead{e1}.}
\address[c]{Jun S. Liu is Professor, Department of Statistics, Harvard
University,
Cambridge, Massachusetts 02138, USA.}

\end{aug}

%
\begin{abstract}
Based on the principles of importance sampling and resampling,
sequential Monte Carlo (SMC) encompasses a large set of powerful
techniques dealing with complex stochastic dynamic systems. Many of
these systems possess strong memory, with which future information
can help sharpen the inference about the current state. By
providing theoretical justification of several existing algorithms
and introducing several new ones, we study systematically how to
construct efficient SMC algorithms to take advantage of the ``future''
information without creating a substantially high computational burden. The
main idea is to allow for lookahead in the Monte Carlo process so that
future information can be utilized in weighting and generating Monte
Carlo samples, or resampling from samples of the current state.
\end{abstract}

%
\begin{keyword}
\kwd{Sequential Monte Carlo}
\kwd{lookahead weighting}
\kwd{lookahead sampling}
\kwd{pilot lookahead}
\kwd{multilevel}
\kwd{adaptive lookahead}
\end{keyword}

\end{frontmatter}

\section{Introduction}\label{sec1}

Sequential Monte Carlo (SMC) methods have been widely used to deal
with stochastic dynamic systems often encountered in engineering,
bioinformatics, finance and many other fields
(Gordon, Salmond and Smith, \citeyear{Gordonetal93}; \citecs{Kong94}; \citecs{Avit95};
\citecs{HurzKuns95}; Liu and Chen,\break \citeyear{LiuChen95}; \citecs{Kita96};
Kim, Shephard and Chib, \citeyear{Kim1998}; \citecs{LiuChen98}; \citecs{PittShep99};
\citecs{Chen00}; Doucet, de~Freitas and Gordon, \citeyear{Douc01}; \citecs{Liu01}; \citecs{Fong02};
Godsill, Doucet and West, \citeyear{Godsill04}).
They utilize the sequential nature of stochastic dynamic systems
to generate sequentially weighted Monte Carlo samples of the
unobservable state variables or other latent variables, and use
these weighted samples for statistical inference of the system or
finding the stochastic optimization solution. A general framework for
SMC is provided in \citet{LiuChen98} and \citet{DelMoral04}. Many
successful applications of SMC in diverse areas of science and
engineering can be found in \citet{Douc01} and \citet{Liu01}.

Dynamic systems often possess strong memory so that future
information is often critical for sharpening the inference about the
current state. For example, in target tracking systems
(\citecs{GodsillVermaak2004}; \citecs{Ikomaetal01}), at each time
point along
the trajectory of a moving object, one observes a function of the
object's location with noise. Such observations obtained in the
future contain substantial information about the current true
location, velocity and acceleration of the object. In protein
structure prediction problems, often the objective is to find an
optimal polymer conformation that minimizes certain energy
function. By ``growing'' the polymer sequentially
\citep{RoseRose55}, the construction of
polymer conformations can be turned into a stochastic dynamic
system with long memory. In such cases, lookahead techniques have
been proven very useful \citep{ZhangLiu02}.

To utilize the strong memory effect, \citet{ClappGodsill99} studied
fixed-lag smoothing using the information from the future.
Independently, \citet{Chen00} proposed the \textit{delayed-sample}
method that generates samples of the current state by integrating
(marginalizing) out the future states, and showed that this method is
effective in solving the problem of adaptive detection and decoding in
a wireless communication problem. The computational complexity of this
method, however, can be substantial when the number of future states
being marginalized out is large. \citet{Wang02SP} developed the
\textit{delayed-pilot} sampling method which generates random pilot
streams to partially explore the space of future states, as well as the
\textit{hybrid-pilot} method that combines the delayed-sam\-ple method
and the delayed-pilot sampling method. \citet{Guoetal04} proposed a
\textit{multilevel} method to reduce complexity for the large state
space system. These low-complexity techniques have been shown to be
effective in the flat-fading channel problem treated in \citet{Chen00}.
\citet{Doucet2006}
proposed a block sampling strategy to utilize future information in
generating better samples of the current states. \citet{ZhangLiu02}
developed the \textit{pilot-exploration resampling} method, which
utilizes multiple random pilot paths for each particle of the current
state to gather future information, and showed that it is effective in
finding the minimum-energy polymer conformation.

In this paper we formalize the general principle of lookahead in
SMC. Several existing methods are then systematically summarized
and studied under this principle, with more detailed theoretical
justifications. In addition, we propose an adaptive lookahead
scheme. The rest of this paper is organized as follows. In Section
\ref{sec2} we briefly overview the general framework of SMC. Section
\ref{sec3}
introduces the general principle of lookahead. Section~\ref{sec4} discusses
several lookahead methods in detail.
In Section~\ref{sec5} we discuss adaptive lookahead. Section~\ref{sec6}
presents several applications. The proofs of all theorems are
presented in the \hyperref[app]{Appendix}.

\section{Sequential Monte Carlo (SMC)}\label{sec2} \label{smc}
Following \citet{LiuChen98}, we define a sto\-chastic dynamic system
as a sequence of evolving probability distributions $\pi_0(\bx_0),
\pi_1(\bx_1), \ldots, \pi_t(\bx_t),\break \ldots\,$, where $\bx_t$ is
called the state variable. We focus on the case when the state
variable evolves with increasing dimension, that is,
$\bx_t=(x_0,x_1,\ldots, x_t)=(\bx_{t-1}, x_t)$, where $x_t$ can be
multidimensional. For example, in the state space
model, the latent state $x_t$ evolves through state dynamic
$x_t\sim g_t(\cdot\mid\bx_{t-1})$, and ``information'' $y_t\sim
f_t(\cdot\mid\bx_t)$ is observed at each time $t$. In this case,
\begin{eqnarray*}
\pi_t(\bx_t)&=&p(\bx_t\mid\by_t)\\
&\propto& g_0(x_0)\prod_{s=1}^t
g_s(x_s\mid\bx_{s-1})f_s(y_s
\mid\bx_s).
\end{eqnarray*}
In this paper we use the notation
$\pi_t(x_t\mid\bx_{t-1})\equiv p(x_t\mid\bx_{t-1},\by_t)$
and
$\pi_{t-1}(x_t\mid\bx_{t-1})\equiv p(x_t\mid\bx_{t-1},\break\by_{t-1})$.
Usually, the goal is to make inference of
a certain function $h(\bx_t)$ given all past information
$\by_t=(y_1,\ldots,y_t)$.


With all the information up to time $t$, we see that the minimum mean
squared error (MMSE) estimator of $h(\bx_t)$, which minimizes
$E_{\pi_t} [\widehat{h}-h(\bx_t) ]^2$, is
$\widehat{h}=E_{\pi_t} (h(\bx_t) )$. When an analytic
solution of $E_{\pi_t} (h(\bx_t) )$ is not available,
an importance sampling Monte Carlo scheme can be employed
(\citecs{Mars56}; \citecs{Liu01}).
Specifically, we can draw samples $\bx_t^{(j)}$, $j=1,\ldots,m$,
from a trial distribution $r_t(\bx_t)$, given that $r_t(\bx_t)$'s
support covers $\pi_t(\bx_t)$'s support, then
$E_{\pi_t} (h(\bx_t) )$ can be estimated by
\[
\frac{1}{m}\sum_{j=1}^m
w_t^{(j)}h\bigl(\bx_t^{(j)}\bigr)
\quad\mbox{or}\quad \frac{1}{\sum_{j=1}^m w_t^{(j)}} \sum
_{j=1}^m w_t^{(j)}h\bigl(
\bx_t^{(j)}\bigr),
\]
where
$w_t^{(j)}=w_t(\bx_t^{(j)})=\pi_t(\bx_t^{(j)})/r_t(\bx_t^{(j)})$
is referred to as a \textit{proper importance weight} for
$\bx_t^{(j)}$ with respect to $\pi_t(\bx_t)$. Although the second
estimator is biased, it is often less variable and easier to use
since in this case $w_t$ only needs to be evaluated up to a
multiplicative constant. Throughout this paper, we will use
$\bx_t$ and $\bx_t^{(j)}$ to denote the true state and the Monte
Carlo sample, respectively.

The basis of all SMC methods is the so-called ``sequential
importance sampling (SIS)'' (Kong, Liu and Wong, \citeyear{Kong94}; \citecs{Liu01}), which
sequentially builds up a high-dimensional sample according to the
chain rule. More precisely, the sample $\bx_t^{(j)}$ is built up
sequentially according to a series of low-dimensional conditional
distributions:
\[
r_t(\bx_t)=q_0(x_0)q_1(x_1\mid
\bx_0)q_2(x_2\mid\bx_1)\cdots
q_t(x_t\mid\bx_{t-1}).
\]
The importance weight for the sample can
be updated sequentially as
\[
w_t\bigl(\bx_t^{(j)}\bigr)=w_{t-1}
\bigl(\bx_{t-1}^{(j)}\bigr)u_t\bigl(
\bx_t^{(j)}\bigr),
\]
where
\[
u_t\bigl(\bx_t^{(j)}\bigr)=\frac{\pi_t(\bx_t^{(j)})}{\pi_{t-1}(\bx_{t-1}^{(j)})
q_t(x_t^{(j)}\mid\bx_{t-1}^{(j)})}
\]
is called the \textit{incremental
weight}. The choice of the trial
distribution $r_t$ (or $q_t$) has a significant impact on the
accuracy and efficiency of the algorithm. As a general principle,
a good trial distribution should be close to the target
distribution. An obvious choice of $q_t$ in the dynamic system
setting is
$q_t(x_t\mid\bx_{t-1})=
\pi_{t-1}(x_t\mid\bx_{t-1})$
(\citecs{Avit95}; Gordon, Salmond and Smith, \citeyear{Gordonetal93}; \citecs{Kita96}).
\citet{Kong94} and \citet{LiuChen98}
argued that $q_t(x_t\mid\bx_{t-1})=
\pi_t(x_t\mid\bx_{t-1})$ is a better trial distribution because of
its usage of the most ``up-to-date'' information to generate
$x_t$. More choices of $q_t(x_t\mid\bx_{t-1})$ can be found in
\citet{ChenLiu00}, \citet{KoteDjur03}, \citet{Linetal05},
\citet{LiuChen98}, \citet{Merweetal02}
and \citet{PittShep99}.

As $t$ increases, the distribution of the importance weight $w_t$
often becomes increasingly skewed (Kong, Liu and Wong, \citeyear{Kong94}), resulting in many
unrepresentative samples of $\bx_t$. A resampling scheme is often
used to alleviate this problem (Gordon, Salmond and Smith, \citeyear{Gordonetal93};
\citecs{LiuChen95}; Kitagawa,\break \citeyear{Kita96};
\citecs{LiuChen98}; \citecs{PittShep99}; \citecs{Chopin04};
\citecs{DelMoral04}). The basic idea is
to imagine implementing multiple SIS procedures in parallel, that is,
to generate
$\{\bx_t^{(1)},\ldots, \bx_t^{(m)}\}$ at each step~$t$, with
corresponding weights $\{w_t^{(1)},\ldots,w_t^{(m)}\}$, and resample
from the set according to a certain ``priority score.''
More precisely, suppose we have obtained
$\{(\bx_t^{(j)},w_t^{(j)}),j=1,\ldots,m\}$ that is properly weighted
with respect to $\pi_t(\bx_t)$, then we create a new set of weighted
samples as follows:

\textit{Resampling scheme.}

\begin{itemize}
\item For each sample $\bx_t^{(j)}$, $j=1,\ldots,m$, assign a \textit{priority
score} $\alpha_t^{(j)}>0$.
\item For $j=1,\ldots,m$,
\begin{itemize}
\item Randomly draw
$\bx_t^{*(j)}$ from the set $\{\bx_t^{(j)},j=1,\ldots,m\}$ with
probabilities proportional to $\{\alpha_t^{(j)}$, $j=1,\ldots,m\}$;
\item If $\bx_t^{*(j)}=\bx_t^{(j_0)}$, then set the new weight
associated with $\bx_t^{*(j)}$ to be
$w_t^{*(j)}=w_t^{(j_0)}/\alpha_t^{(j_0)}$.
\end{itemize}
\item Return the new set of weighted samples
$\{(\bx_t^{*(j)}$, $w_t^{*(j)}),j=1,\ldots,m\}$.
\end{itemize}
This new set of weighted samples is also approximately properly
weighted\vspace*{1pt} with respect to $\pi_t(\bx_t)$. Often, $\alpha_t^{(j)}$ are
chosen to be proportional to $w_t^{(j)}$, so that the new samples are
equally weighted. Some improved resampling schemes can be found in
\citet{LiuChen98}, \citet{Carpenter99}, \citet{CrisanLyons2002}, \citet{Liang02} and
\citet{Pitt2002}.

Resampling plays an important role in SMC. Cho\-pin (\citeyear{Chopin04}) and
\citet{DelMoral04} provide asymptotic results on its effect, but
its finite sample effects have not been fully understood.
Performing resampling at every step $t$ is usually neither
necessary nor efficient since it induces excessive variations
\citep{LiuChen95}. \citet{LiuChen98} suggests to use either a
deterministic schedule, in which resampling only takes place at
time $T,2T,3T,\ldots\,$, or a dynamic schedule, in which resampling
is performed when the effective sample size \citep{Kong94}
$\mathrm{ESS}=m/(1+v_t(w))$ is less than a certain threshold, where
$v_t(w)$ is the estimated coefficient of variation, that is,
%
\begin{equation}\label{EqESS}\quad
v_t(w)=\frac{(\sum_{j=1}^m (w_t^{(j)}-\sum_{j=1}^m
w_t^{(j)}/m )^2)/m}{ (\sum_{j=1}^m w_t^{(j)}/m )^2}.
\end{equation}

In problems that the state variable $x_t$ takes values in a finite
set ${\cal A}=\{a_1,\ldots,a_{|{\mathcal A}|}\}$, duplicated
samples produced in sampling or resampling steps result in
repeated calculation and a waste of resources. Using an idea
related to the rejection control (Liu, Chen and Wong, \citeyear{LiuChenWong98}),
\citet{FearnheadClifford03} developed a more efficient scheme that
combines sampling and resampling in one step and guarantees to
generate distinctive samples.

Most of the SMC algorithms are designed for filtering and
smoothing problems. It is a challenging problem when the system
has unknown fixed parameters to be estimated and learned. Some new
development can be found in
\citet{GilksBerzuini2001}, \citet{Chopin2002},
\citet{Fearnhead2002}, \citet{Andrieu2010} and
\citet{Carvalho2010}. In this paper we assume all the parameters
are known.


\section{The Principle of Lookahead}\label{sec3}

To formalize our argument that the ``future'' information
is helpful for the inference about the current state, we assume that
the dynamic
system $\pi_t$ offers more and more ``information''\vadjust{\goodbreak} of the state
variables as $t$ increases. A simple way to quantify this concept
is to assume that the information available at time $t$ takes the
form $\by_t=(y_{1},y_{2},\ldots,y_t)$ and increments to
$(\by_t,y_{t+1})$ at time $t+1$. The dynamic system $\pi_t(\bx_t)$
simply takes the form of $\pi_t(\bx_t)=p(\bx_t\mid\by_t)$.
Although this framework is not all-inclusive, it is sufficiently
broad and our theoretical results are all under this setting. The
basic lookahead principle is to use ``future'' information for the
inference of the current state. That is, we believe that $E(h(\bx_t)\mid
\by_{t+\Delta})$ results in a better inference of the current state
$h(\bx_t)$ than $E(h(\bx_t)\mid\by_{t})$ for any $\Delta>0$.
Thus, if the added computational burden is not considered, we would like
to use a Monte Carlo estimate of $E(h(\bx_t)\mid\by_{t+\Delta})$ to
make inference on $h(\bx_t)$.


Here we\vspace*{2pt} study the benefit of the lookahead strategy rigorously.
Let $\widehat{h}_{t+\Delta}$ be a consistent Monte Carlo estimator
of $E_{\pi_{t+\Delta}} (h(\bx_t) )=E (h(\bx_t)\mid
\by_{t+\Delta} )$ and\break $\widehat{h}_{t+\Delta}$ is independent
of the true state $\bx_t$ conditional on $\by_{t+\Delta}$. The
mean squared difference between
$h(\bx_t)$ and its estimator $\widehat{h}_{t+\Delta}$, averaged
over the Monte Carlo samples, the true state and the future
observations can be decomposed as
%
\begin{eqnarray}
\label{MSE1}
&&
E_{\pi_t} \bigl[\widehat{h}_{t+\Delta}-h(
\bx_t) \bigr]^2 \nonumber\\
&&\quad= E_{\pi_t} \bigl[
\widehat{h}_{t+\Delta}-E \bigl(h(\bx_t)\mid\by_{t+\Delta}
\bigr) \bigr]^2\nonumber\\[-8pt]\\[-8pt]
&&\qquad{} +E_{\pi_t} \bigl[E \bigl(h(\bx_t)
\mid\by_{t+\Delta} \bigr)-h(\bx_t) \bigr]^2
\nonumber\\
&&\quad\dff I(\Delta) + \mathit{II}(\Delta).\nonumber
\end{eqnarray}
As the Monte Carlo sample size tends to infinity, $I(\Delta)$, which
is the variance of the consistent estimator, tends to zero. For
$\mathit{II}(\Delta)$,
we can show the following:
%
\begin{propo}\label{prop1} For any square integrable function
$h(\cdot)$, $\mathit{II}(\Delta)$ decreases as $\Delta$ increases.
\end{propo}
The proof is given in the \hyperref[app]{Appendix}.

When the Monte Carlo sample size is
sufficiently large, $I(\Delta)$ becomes negligible relative to
$\mathit{II}(\Delta)$. Hence, the above proposition implies that
a consistent Monte Carlo estimator of
$E (h(\bx_t)\mid\by_{t+\Delta} )$ 
is always more accurate with larger $\Delta$ when the Monte Carlo
sample size is sufficiently
large.

However, this gain of accuracy is not always desirable in practice
because of the additional computational costs. Most of the time
additional computational resources are needed to obtain consistent
estimators of $E (h(\bx_t)\mid\by_{t+\Delta} )$ with larger
$\Delta$. Furthermore, $I(\Delta)$ sometimes increases sharply as
$\Delta$ increases when the Monte Carlo sample size is fixed.
More detailed analysis is shown in Section~\ref{secadaptive}.\vadjust{\goodbreak}

In order to achieve the goal of estimating the
\textit{lookahead} expectation
$E_{\pi_{t+\Delta}} (h(\bx_t) )$ effectively using SMC, we
may consider defining
a new stochastic dynamic system with the probability distribution at
step $t$ being the $\Delta$-step lookahead (or delayed)
distribution, that is,
%
\begin{eqnarray}
\label{new-SDS}
\pi_t^*(\bx_t)&=&\pi_{t+\Delta}(
\bx_t)\nonumber\\[-8pt]\\[-8pt]
&=&\int\pi_{t+\Delta}(\bx_{t+\Delta})\,dx_{t+1}
\cdots dx_{t+\Delta}.\nonumber
\end{eqnarray}
With the system defined by
$\{\pi_0^*(\bx_0),\pi_1^*(\bx_1),\ldots\}$, the same SMC recursion
can be carried out.

In practice, however, it is often difficult to use this modified
system directly since the analytic evaluation of the
integration/summation in (\ref{new-SDS}) is impossible for most
systems. Even when the state variables take values from a finite
set so that $\pi_{t+\Delta}(\bx_t)$ can be calculated exactly
through summation, the number of terms in the summation grows
exponentially with $\Delta$. Nonetheless, the lookahead system
$\{\pi_0^*(\bx_0),\pi_1^*(\bx_1),\ldots\}$ suggests a potential direction
that we can work toward.

There are three possible ways to
make use of the future information: (i) for choosing a good trial
distribution $r_t(\bx_t)$ close to $\pi_t^*(\bx_t)$; (ii)
for calculating and keeping
track of the importance weight for $\bx_t$ using $\pi_t^*(\bx_t)$
as the target distribution; and (iii) for setting up an effective
resampling priority score function $\alpha_t(\bx_t)$ using information
provided by $\pi_t^*(\bx_t)$. Detailed algorithms are given in the next
section.

We note here that \textit{lookahead} (into the ``future'') strategies
are mathematically equivalent to \textit{delay}\break strategies (i.e.,
making inference after seeing more data) in \citet{Chen00}. In our
setup, we assume that the current time is $t+\Delta$ and we
observe $y_1, \ldots, y_{t+\Delta}$. In fact, some of the
algorithms we covered were initially named ``delay algorithms,''
under the notion that the system allows certain delay in
estimation. The reason that we choose to use the term ``lookahead'' instead
of ``delay'' is that we focus on sampling of $x_t$, using
information after time $t$ (i.e., its own future). It is easier to
discuss and compare the same $x_t$ when looking further into the
future (increasing $\Delta$), rather than a longer delay (with a
fixed current time and to discuss the estimation of
$x_{t-\Delta}$ with changing~$\Delta$).

The lookahead algorithms we discuss here are clos\-ely related to the
smoothing problem in state space models where one is interested in
making inference with respect to $p(x_t\mid y_1,\ldots,y_T)$ for
$t=1,\ldots,T$.\break Many algorithms, some are closely related\vadjust{\goodbreak} to our
approach, can be found in
\citet{Godsill04}, \citet{Douc2012}, \citet{Briers2010},
\citet{Carvalho2010}, \citet{Fearnhead2010} and
others. However, in this paper we emphasize on dynamically processing
of $p(x_t\mid y_1,\ldots,y_{t+\Delta})$ for $t=1,\ldots,n$. It has the
characteristic of both filtering (updating as new information comes~in)
and smoothing (inference with future information).

Another possible benefit of the proposed lookahead strategy is that it
tends to be more robust to
outliers, since the future information will correct
the misinformation from the outliers. This is particularly helpful during
resampling stages. With an outlier, the ``good samples'' that are
close to the true state will be mistakenly
given smaller weights. Resampling according to weights will then be
more likely
to remove these ``good samples.'' Lookahead that takes into account more
information will be very useful in such a situation.

A ``true'' lookahead would utilize
the expected (but unobserved) future information in generating
samples of current $x_t$. The popular and
powerful \textit{auxiliary particle filter} \citep{PittShep99}
is based on such an insight, though it only looks ahead one step.
Our experience shows that the improvement is limited with more
steps of such a ``true'' lookahead scheme, as the information is
limited to $y_1,\ldots, y_t$. Here we focus on the utilization
of the extra information provided by future observations.

\section{Basic Lookahead Strategies}\label{sec4}

\subsection{Lookahead Weighting Algorithm}\label{sec4.1}\label{secLookaheadWeighting}
Suppose at step $t+\Delta$, we obtain a set of weight\-ed samples
$\{(\bx_{t+\Delta}^{(j)},\overline{w}{}^{(j)}_{t+\Delta}),j=1,\ldots,m\}$
properly\break weighted with respect to
$\pi_{t+\Delta}(\bx_{t+\Delta})$, using the standard concurrent
SMC. With the same weight $w_t^{(j)}\dff
\overline{w}{}^{(j)}_{t+\Delta}$, the partial chain $\bx_t^{(j)}$ is
also properly weighted with respect to the marginal distribution
$\pi_{t+\Delta}(\bx_t)$. Specifically, we have the following
algorithmic steps.\vspace*{6pt}

\noindent\rule{\columnwidth}{0.5pt}\vspace*{-4pt}
\begin{algor*}
\begin{itemize}
\item At time $t=0$, for $j=1,\ldots,m$:
\begin{itemize}
\item Draw $(x_0^{(j)}, \ldots, x_{\Delta}^{(j)})$ from
distribution $q_0(x_0) \cdot\allowbreak\prod_{s=1}^\Delta q_{s}(x_{s}\mid
\bx_{s-1})$.
\item Set
\[
w_0^{(j)}\propto\frac{\pi_{\Delta}(\bx_{\Delta}^{(j)})} {q_0(x_0)
\prod_{s=1}^\Delta q_{s}(x_{s}^{(j)}\mid\bx_{s-1}^{(j)})}.\vadjust{\goodbreak}
\]
\end{itemize}
\item At times $t=1,2,\ldots\,$, suppose we obtained $\{(\bx_{t+\Delta
-1}^{(j)},\break w_{t-1}^{(j)}),j=1,\ldots,m\}$ properly
weighted with respect to $\pi_{t+\Delta-1}(\bx_{t+\Delta-1})$.
\begin{itemize}
\item(Optional.) Resample with probability proportional to the priority scores
$\alpha_{t-1}^{(j)}=w_{t-1}^{(j)}$ to obtain a new set of weighted
samples.
%
%
\item Propagation: For $j=1,\ldots,m$:
\begin{itemize}
\item(Sampling.) Draw $x_{t+\Delta}^{(j)}$ from distribution\break
$q_{t+\Delta}(x_{t+\Delta}\mid \bx_{t+\Delta-1}^{(j)})$. Set
$\bx_{t+\Delta}^{(j)}=(\bx_{t+\Delta-1}^{(j)},\break
x_{t+\Delta}^{(j)})$.
\item(Updating weights.) Set
%
%
\[
\hspace*{-23pt}w_t^{(j)}\propto w_{t-1}^{(j)}
\frac{\pi_{t+\Delta}(\bx_{t+\Delta}^{(j)})}
{\pi_{t+\Delta-1}(\bx_{t+\Delta-1}^{(j)})
q_{t+\Delta}(x_{t+\Delta}^{(j)}\mid \bx_{t+\Delta-1}^{(j)})}.
\]
\end{itemize}
\item Inference: $E_{\pi_{t+\Delta}}(h(\bx_t))$ is estimated by
\[
\sum_{j=1}^m w_t^{(j)}h\bigl(\bx_t^{(j)}\bigr)\Big/\sum_{j=1}^m w_t^{(j)}.
\]
\hspace*{-22pt}\rule{\columnwidth}{0.5pt}%
\end{itemize}
\end{itemize}
\end{algor*}

Because the $x_t^{(j)}$ are still generated based on the
information up to step $t$, for example, $q_t(x_t\mid
\bx_{t-1})=\pi_t(x_t\mid\bx_{t-1})$, and the future information
is utilized only through weight adjustments; \citet{Chen00} called
this method the \textit{delayed-weight} method.
\citet{ClappGodsill99} called the procedure \textit{sequential
imputation with decision step}, as inference and decisions are
made separately at different time steps.

The lookahead weighting algorithm is a
simple scheme to provide a consistent estimator for\break
$E_{\pi_{t+\Delta}} (h(\bx_{t}) )$ with almost no additional
computational cost, except for some additional memory buffer. Hence,
it is often useful in real-time filtering problems
(\citecs{Chen00}; \citecs{Kantas2009}). However, when $\Delta$
is large, it is well known that such a forward algorithm is
highly inaccurate and inefficient in approximating the smoothing distribution
$\pi_{t+\Delta}(x_t)$ (e.g., \citecs{Godsill04};
\citecs{Douc2012}; \citecs{Briers2010}; \citecs{Fearnhead2010};
\citecs{Carvalho2010}).

%
\begin{figure*}

\includegraphics{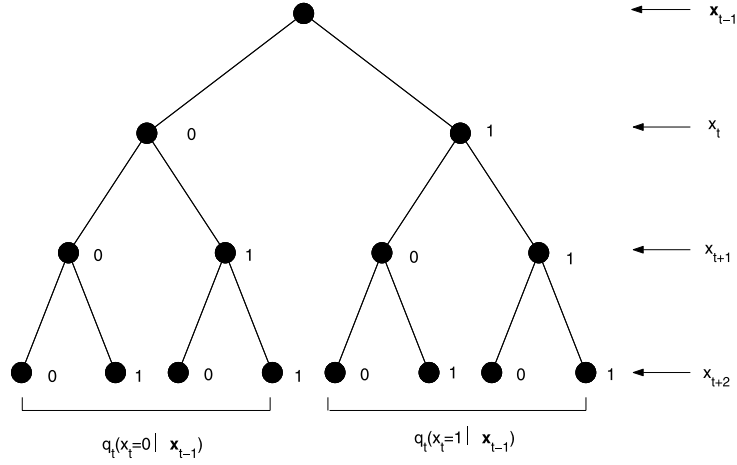}

\caption{Illustration of the
exact lookahead sampling method, in which the trail distribution
$q_t(x_t=i\mid\bx_{t-1}^{(j)})$, $i=0,1$, is proportional to the
summation of $\pi_{t+2}(x_t=i,x_{t+1},x_{t+2}\mid\bx_{t-1}^{(j)})$
for $x_{t+1},x_{t+2}=0,1$.}\label{Figdelay-sampling}
\end{figure*}

\subsection{Exact Lookahead Sampling}\label{sec4.2}
This method was proposed by
\citet{Chen00}, termed as \textit{delayed-sample} method.
Its key is to use the modified stochastic dynamic
system defined by $\pi_t^*(\bx_t)=\pi_{t+\Delta}(\bx_t)$ in
(\ref{new-SDS}) to construct the importance sampling distribution.
At step $t$, the conditional sampling distribution for $x_t^{(j)}$
is chosen to be
%
\begin{equation}
\label{exactsampling}q_t\bigl(x_t\mid\bx_{t-1}^{(j)}
\bigr)=\pi_t^*\bigl(x_t\mid\bx_{t-1}^{(j)}
\bigr)=\pi_{t+\Delta
}\bigl(x_t\mid\bx_{t-1}^{(j)}
\bigr),\hspace*{-20pt}
\end{equation}
and the weight is updated accordingly as
\begin{eqnarray*}
w_t\bigl(\bx_t^{(j)}\bigr)&=&w_{t-1}
\bigl(\bx_t^{(j)}\bigr)\frac{\pi_t^*(\bx_{t}^{(j)})} {
\pi_{t-1}^*(\bx_{t-1}^{(j)})\pi_t^*(x_t^{(j)}\mid\bx_{t-1}^{(j)})} \\
&=&w_{t-1}
\bigl(\bx_t^{(j)}\bigr)\frac{\pi_{t+\Delta}(\bx_{t-1}^{(j)})} {
\pi_{t+\Delta-1}(\bx_{t-1}^{(j)})}.
\end{eqnarray*}
Figure~\ref{Figdelay-sampling} illustrates the method with
$x_t\in{\cal A} = \{0,1\}$ and $\Delta=2$, in which the trial
distribution is
\begin{eqnarray*}
&&
q_t\bigl(x_t=i \mid\bx_{t-1}^{(j)}
\bigr)\\
&&\quad=\pi_{t+2}\bigl(x_t=i \mid\bx_{t-1}^{(j)}
\bigr) \\
&&\quad=\sum_{x_{t+1}}\sum_{x_{t+2}}
\pi_{t+2}\bigl(x_t=i,x_{t+1},x_{t+2}\mid
\bx_{t-1}^{(j)}\bigr)
\\
&&\quad\propto \sum_{x_{t+1}}\sum
_{x_{t+2}} \pi_{t+2}\bigl(\bx_{t-1}^{(j)},x_t=i,x_{t+1},x_{t+2}
\bigr)
\end{eqnarray*}
for $i=0,1$.

The exact lookahead sampling algorithm is shown as follows.\vspace*{6pt}

\noindent\rule{\columnwidth}{0.5pt}\vspace*{-4pt}
\begin{algorr*}
\begin{itemize}
\item At time $t=0$, for $j=1,\ldots,m$:
\begin{itemize}
\item Draw $x_0^{(j)}$ from distribution
$q_0(x_0)$.
\item Set $w_0^{(j)}=\pi_{\Delta}(x_0^{(j)})/q_0(x_0^{(j)})$.
\end{itemize}
\item At times $t=1,2,\ldots\,$:
\begin{itemize}
\item(Optional.) Resample
$\{\bx_{t-1}^{(j)},w_{t-1}^{(j)},j=1,\ldots,m\}$ with priority scores
$\alpha_{t-1}^{(j)}=w_{t-1}^{(j)}$.
\item Propagation: For $j=1,\ldots,m$:
\begin{itemize}
\item(Sampling.)
Draw $x_t^{(j)}$ from distribution
\begin{eqnarray*}
q_t\bigl(x_t\mid\bx_{t-1}^{(j)}\bigr)&=&
\pi_{t+\Delta}\bigl(x_t\mid\bx_{t-1}^{(j)}
\bigr)\\
&=&\frac{\pi_{t+\Delta}(
\bx_{t-1}^{(j)},x_t)}{\pi_{t+\Delta}(\bx_{t-1}^{(j)})}.
\end{eqnarray*}
\item(Updating weights.) Set
\[
w_t^{(j)}=w_{t-1}^{(j)} \frac{\pi_{t+\Delta}(\bx_{t-1}^{(j)})} {
\pi_{t+\Delta-1}(\bx_{t-1}^{(j)})}.
\]
\end{itemize}
\item Inference: $E_{\pi_{t+\Delta}}(h(\bx_t))$ is estimated by
\[
\sum_{j=1}^m w_t^{(j)}h\bigl(\bx_t^{(j)}\bigr)\Big/\sum_{j=1}^m w_t^{(j)}.
\]
\hspace*{-22pt}\rule{\columnwidth}{0.5pt}\vspace*{6pt}
\end{itemize}
\end{itemize}
\end{algorr*}

Specifically, for models with finite state space, the sampling
and weight update steps in the exact lookahead sampling method
involve evaluation of summations of the form
%
\begin{eqnarray}\label{exact111}
\pi_{t+\Delta}(\bx_t)&=& \sum_{x_{t+1},\ldots,x_{t+\Delta}}
\pi_{t+\Delta}(\bx_t,x_{t+1},\ldots,x_{t+\Delta})
\nonumber\\
&\propto& \sum_{x_{t+1},\ldots,x_{t+\Delta}} g_0(x_0)
\prod_{s=1}^{t+\Delta} g_s(x_s
\mid\bx_{s-1})\\
&&\hspace*{93.1pt}{}\cdot f_s(y_s\mid\bx_s).\nonumber
\end{eqnarray}


For continuous state space, it is more difficult to adopt this
approach, as one needs to generate samples from
%
\begin{eqnarray}
\label{exact112}\quad
&&
q_t\bigl(x_t\mid\bx_{t-1}^{(j)}
\bigr)\nonumber\\
&&\quad=\pi_{t+\Delta}\bigl(x_t\mid\bx_{t-1}^{(j)}
\bigr)\nonumber\\[-8pt]\\[-8pt]
&&\quad\propto\int\pi_{t+\Delta}\bigl(\bx_{t-1}^{(j)},x_t,x_{t+1}, \ldots,\nonumber\\
&&\quad\hspace*{50.3pt}\hspace*{58.5pt}
 x_{t+\Delta} \bigr)\,dx_{t+1} \cdots
dx_{t+\Delta}\nonumber
\end{eqnarray}
and evaluate it in order to update the weight. A~slight\-ly different
version of the algorithm was proposed in \citet{ClappGodsill99},
termed as \textit{lagged time filtering density}. Instead of
calculating the exact sampling density (\ref{exact111}) or
(\ref{exact112}),
and sample from it, they proposed to
use forward filtering backward sampling techniques of
\citet{CartKohn94} and \citet{ClappGodsill97}.

As demonstrated in \citet{Chen00} and \citet{ClappGodsill99},
the exact lookahead sampling
method can achieve a significant improvement in performance
compared to the concurrent SMC method.
\citet{Chen00} provided some heuristic justification of this
method. Here we provide a theoretical justification by showing that
the exact lookahead
sampling method generates more effective samples (or
``particles'') than any trial distribution that does not utilize
the future information.

To set up the analysis, we assume that
$\{(\bx_{t-1}^{(j)},\break w_{t-1}^{(j)})$, $j=1,\ldots,m\}$ is properly
weighted with respect to $\pi_{t-1}(\bx_{t-1})$ (not the lookahead
distribution). We compare two sampling schemes. In exact
lookahead sampling, $x_{t}^{(1,j)}$ is generated from
$\pi_{t+\Delta}(x_t\mid\bx_{t-1}^{(j)})$, and
$\bx_{t}^{(1,j)}=(\bx_{t-1}^{(j)},x_t^{(1,j)})$ is properly
weighted with respect to $\pi_{t+\Delta}(\bx_{t})$ by weight
%
\begin{equation}
\label{delay-sample-weight} w_t^{(1,j)}=w_{t-1}^{(j)}
\frac{\pi_{t+\Delta}(\bx_{t-1}^{(j)})} {
\pi_{t-1}(\bx_{t-1}^{(j)})}.
\end{equation}
Let sample $x_{t}^{(2,j)}$ be generated from a trial distribution
$q_t(x_t\mid\bx_{t-1}^{(j)})$ that uses no future information,
that is, $q_t(x_t\mid\bx_{t-1}^{(j)})$ does not depend on
$y_{t+1},\ldots,\break y_{t+\Delta}$, then
$\bx_t^{(2,j)}=(\bx_{t-1}^{(j)},x_t^{(2,j)})$ is properly weighted
with respect to $\pi_{t+\Delta}(\bx_{t})$ using the weight
%
\begin{equation}
\label{exactlookahead} w_t^{(2,j)}=w_{t-1}^{(j)}
\frac{\pi_{t+\Delta}(\bx_t^{(2,j)})} {
\pi_{t-1}(\bx_{t-1}^{(j)})q_t(x_t^{(2,j)}\mid
\bx_{t-1}^{(j)})}.
\end{equation}
%

Let the subscript $\pi_{t+\Delta}$ indicate that the corresponding operations
are to be taken conditional on
$\by_{t+\Delta}$, and let
\begin{eqnarray*}
&&
E_{\pi_{t+\Delta}} \bigl(h(\bx_t)\mid\bx_{t-1}=
\bx_{t-1}^{(j)} \bigr)\\
&&\quad=\int h\bigl(\bx_{t-1}^{(j)},x_t
\bigr)\pi_{t+\Delta}\bigl(x_t\mid\bx_{t-1}^{(j)}
\bigr)\,dx_t.
\end{eqnarray*}
We have the following proposition:
%
\begin{propo}\label{prop2}
%
\begin{equation}\label{rao2-2}
\operatorname{var}_{\pi_{t+\Delta}} \bigl( w_t^{(2,j)} \bigr)\geq
\operatorname{var}_{\pi_{t+\Delta}} \bigl(w_t^{(1,j)} \bigr)
\end{equation}
and
%
\begin{eqnarray}\qquad
\label{rao2}
&&
\operatorname{var}_{\pi_{t+\Delta}} \bigl[w_t^{(2,j)}h\bigl(
\bx_t^{(2,j)}\bigr) \bigr] \nonumber\\[-8pt]\\[-8pt]
&&\quad \geq \operatorname{var}_{\pi_{t+\Delta}}
\bigl[w_t^{(1,j)}E_{\pi_{t+\Delta}} \bigl(h(\bx_t)
\mid\bx_{t-1}=\bx_{t-1}^{(j)} \bigr) \bigr],
\nonumber\\
\label{rao2-1}
&&
\operatorname{var}_{\pi_{t+\Delta}} \bigl[w_t^{(2,j)}E_{\pi
_{t+\Delta
}}
\bigl(h(\bx_t)\mid\bx_{t-1}=\bx_{t-1}^{(j)}
\bigr) \bigr]\nonumber\\[-8pt]\\[-8pt]
&&\quad \geq \operatorname{var}_{\pi_{t+\Delta}} \bigl[w_t^{(1,j)}E_{\pi
_{t+\Delta}}
\bigl(h(\bx_t)\mid\bx_{t-1}=\bx_{t-1}^{(j)}
\bigr) \bigr].\nonumber
\end{eqnarray}
\end{propo}

The proof is presented in the \hyperref[app]{Appendix}.

Note that the right-hand sides of (\ref{rao2}) and (\ref{rao2-1})
use the Rao-Blackwellization estimator
\[
w_t^{(1,j)} E_{\pi_{t+\Delta}} \bigl(h(\bx_t)\mid
\bx_{t-1}=\bx_{t-1}^{(j)} \bigr).
\]
For finite state space, it is
often achievable since
\begin{eqnarray*}
&&
E_{\pi_{t+\Delta}} \bigl(h(\bx_t)\mid\bx_{t-1}=
\bx_{t-1}^{(j)} \bigr)\\
&&\quad =\sum_{i=1}^{|\mathcal{A}|}
h\bigl(\bx_{t-1}^{(j)},x_t=a_i\bigr)
\pi_{t+\Delta}\bigl(x_t=a_i\mid\bx_{t-1}^{(j)}
\bigr),
\end{eqnarray*}
where $\pi_{t+\Delta}(x_t=a_i\mid\bx_{t-1}^{(j)})$ have been
computed during the propagation step. Also note that (\ref{rao2})
does not provide a direct comparison between $\sum_{j=1}^m
w_{t}^{(1,j)}\cdot\allowbreak h(\bx_t^{(j)})$ and $\sum_{j=1}^m
w_{t}^{(2,j)}h(\bx_t^{(j)})$. This is because the sampling
efficiency is also related to function $h(\cdot)$. If $h(\bx_t)$
does not depend on $x_t$, then (\ref{rao2}) indeed shows that the
full lookahead sampler is always better. Otherwise, this
proposition suggests to use
\[
\frac{1}{\sum_{j=1}^m w_t^{(j)}} \sum_{j=1}^m
w_t^{(j)}E_{\pi_{t+\Delta}} \bigl(h(\bx_t)\mid
\bx_{t-1}=\bx_{t-1}^{(j)} \bigr)
\]
for estimation in the exact
lookahead sampler.


As a direct consequence of Proposition~\ref{prop2}, the following proposition
shows that exact
lookahead sampling is more efficient than lookahead weighting.\vspace*{1pt} Suppose
in lookahead weighting sample
$\bx_t^{(3,j)}=\bx_t^{(2,j)}=(\bx_{t-1}^{(j)},x_t^{(2,j)})$
is available at time $t$ and $\bx_{t+1:t+\Delta}^{(3,j)}$ is generated
from
\[
\prod_{s=t+1}^{t+\Delta}q_s
\bigl(x_s \mid\bx_{t}^{(3,j)},\bx_{t+1:s-1}
\bigr)
\]
in the next $\Delta$
steps. Let
$\bx_{t+\Delta}^{(3,j)}=(\bx_{t}^{(3,j)},\bx_{t+1:t+\Delta}^{(3,j)})$,
then the weight corresponding to the lookahead\break weighting algorithm is
\[
w_t^{(3,j)}=w_{t-1}^{(j)}\frac{\pi_{t+\Delta}(\bx_{t+\Delta}^{(3,j)})} {
\pi_{t-1}(\bx_{t-1}^{(j)})\prod_{s=t}^{t+\Delta
}q_s(x_{s}^{(3,j)}\mid
\bx_{s-1}^{(3,j)})}.
\]
We have the following proposition:
%
\begin{propo}\label{prop3}
\[
\operatorname{var}_{\pi_{t+\Delta}} \bigl( w_{t}^{(3,j)} \bigr)\geq
\operatorname{var}_{\pi_{t+\Delta}} \bigl(w_t^{(2,j)} \bigr)
\]
and for any square
integrable function $h(\bx_t)$,
\begin{eqnarray*}
&&
\operatorname{var}_{\pi_{t+\Delta}} \bigl[w_{t}^{(3,j)}h\bigl(
\bx_{t-1}^{(j)},x_t^{(2,j)}\bigr) \bigr]\\
&&\quad\geq
\operatorname{var}_{\pi_{t+\Delta}} \bigl[w_t^{(2,j)}h\bigl(
\bx_{t-1}^{(j)},x_t^{(2,j)}\bigr) \bigr].
\end{eqnarray*}
\end{propo}
The proof is presented in the \hyperref[app]{Appendix}.

In the exact lookahead sampling, the incremental weight
$U_t=\pi_{t+\Delta}(\bx_{t-1})/\pi_{t+\Delta-1}(\bx_{t-1})$
usually will
be close to 1 when $\Delta$ is large, so the variance of weights
typically decreases as $\Delta$ increases (Doucet, Briers and S{\'e}n{\'e}cal, \citeyear{Doucet2006}). The
benefit of exact lookahead sampling, however, comes at the
cost of increased analytical and computational complexities due
to the need of marginalizing out the future states
$x_{t+1},\ldots,x_{t+\Delta}$ in (\ref{new-SDS}). Often, the
computational cost grows exponentially as the lookahead step
$\Delta$ increases.


\subsection{Block Sampling}\label{sec4.3}

Doucet, Briers and S{\'e}n{\'e}cal (\citeyear{Doucet2006}) proposes a block
sampling strategy, which can be viewed as a variation of lookahead. A
slightly modified version (under our notation) is given as follows.\vspace*{6pt}

\noindent\rule{\columnwidth}{0.5pt}\vspace*{-4pt}
\begin{algorrr*}
\begin{itemize}
\item At time $t=0$, for $j=1,\ldots,m$:\vspace*{1pt}
\begin{itemize}
\item Draw $(x_0^{(j)}, \ldots, x_{\Delta}^{(j)})$ from
distribution $q_0(x_0) \cdot\allowbreak\prod_{s=1}^\Delta q_{s}(x_{s}\mid
\bx_{s-1})$.
\item Set
\[
w_0^{(j)}\propto\frac{\pi_{\Delta}(\bx_{\Delta}^{(j)})} {q_0(x_0)
\prod_{s=1}^\Delta q_{s}(x_{s}^{(j)}\mid\bx_{s-1}^{(j)})}.
\]
\end{itemize}
\item At times $t=1,2,\ldots\,$:
\begin{itemize}
\item(Optional.) Resample\vspace*{1pt} $\{\bx_{t+\Delta
-1}^{(j)},w_{t-1}^{(j)},j=1,\ldots,\allowbreak m\}$
with priority scores $\alpha_{t-1}^{(j)}=w_{t-1}^{(j)}$.
\item Propagation: For $j=1,\ldots,m$:\vspace*{1pt}
\begin{itemize}
\item(Sampling.) Draw
$\bx_{t:t+\Delta}^{*(j)}$ from
$q_t(\bx_{t:t+\Delta}^{*(j)}\mid\break
\bx_{t+\Delta-1}^{(j)})$.
\item(Updating weights.)
Set
\begin{eqnarray*}
w_t^{(j)}&=&w_{t-1}^{(j)}
\pi_{t+\Delta}\bigl(\bx_{t-1}^{(j)},\bx_{t:t+\Delta}^{*(j)}\bigr)\\
&&{}\cdot
\lambda_t\bigl(\bx_{t:t+\Delta-1}^{(j)}\mid
\bx_{t-1}^{(j)},\bx_{t:t+\Delta}^{*(j)}\bigr) \\
&&{}/ \bigl(\pi_{t+\Delta-1}\bigl(\bx_{t-1}^{(j)},
\bx_{t:t+\Delta-1}^{(j)}\bigr)\\
&&\hspace*{6.8pt}{}\cdot
q_t\bigl(\bx_{t:t+\Delta}^{*(j)}\mid\bx_{t-1}^{(j)},
\bx_{t:t+\Delta-1}^{(j)}\bigr)\bigr).
\end{eqnarray*}
\item Let
$\bx_{t+\Delta}^{(j)}=(\bx_{t-1}^{(j)},\bx_{t:t+\Delta}^{*(j)})$.\vspace*{1pt}
\end{itemize}
\item Inference: $E_{\pi_{t+\Delta}}(h(\bx_t))$ is estimated by
\[
\sum_{j=1}^m w_t^{(j)}h\bigl(\bx_t^{(j)}\bigr)\Big/\sum_{j=1}^m w_t^{(j)}.
\]
\hspace*{-22pt}\rule{\columnwidth}{0.5pt}\vspace*{6pt}
\end{itemize}
\end{itemize}
\end{algorrr*}

Here\vspace*{1pt} $\lambda_t(\bx_{t:t+\Delta-1}^{(j)}\mid
\bx_{t-1}^{(j)},\bx_{t:t+\Delta}^{*(j)})$ is called the
\textit{artificial} conditional distribution.

\citet{Doucet2006} suggested that one should choose
$q_t(\bx_{t:t+\Delta}^{*(j)}\mid\bx_{t+\Delta-1}^{(j)})=
q_t(\bx_{t:t+\Delta}^{*(j)}\mid\bx_{t-1}^{(j)})$, that is, the
trial distribution does not depend on $\bx_{t:t+\Delta-1}^{(j)}$.
Then the optimal choices of $q_t$ and $\lambda_t$ are
\[
q_t\bigl(\bx_{t:t+\Delta}^{*(j)}\mid\bx_{t-1}^{(j)},
\bx_{t:t+\Delta-1}^{(j)}\bigr) =\pi_{t+\Delta}\bigl(
\bx_{t:t+\Delta}^{*(j)}\mid\bx_{t-1}^{(j)}\bigr)
\]
and
\[
\lambda_t\bigl(\bx_{t:t+\Delta-1}^{(j)}\mid
\bx_{t-1}^{(j)},\bx_{t:t+\Delta}^{*(j)}\bigr) =
\pi_{t+\Delta-1}\bigl(\bx_{t:t+\Delta-1}^{(j)}\mid\bx_{t-1}^{(j)}
\bigr).
\]
Note that, in this case, the marginal trial distribution of
$x_t^{*(j)}$ is $\pi_{t+\Delta}(x_t^{*(j)}\mid\bx_{t-1}^{(j)})$,
and the weight is updated by
\[
w_t^{(j)}=w_{t-1}^{(j)}\frac{\pi_{t+\Delta}(\bx_{t-1}^{(j)})}{\pi
_{t+\Delta-1}(\bx_{t-1}^{(j)})}.
\]
In this case,
the blocking sampling method becomes the exact lookahead sampling.

\begin{figure*}

\includegraphics{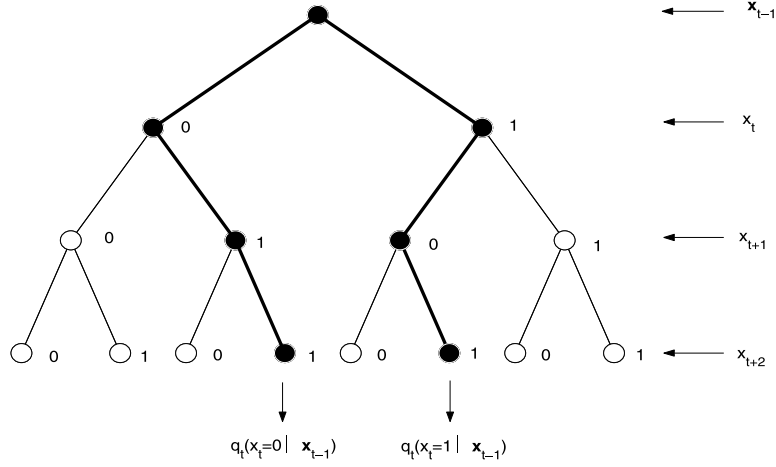}

\caption{Illustration of the
single-pilot lookahead sampling method, in which the pilot path
for $(\bx_{t-1}^{(j)},x_t=0)$ is $(x_{t+1}=1,x_{t+2}=1)$ and the
pilot path for $(\bx_{t-1}^{(j)},x_t=1)$ is
$(x_{t+1}=0,x_{t+2}=1)$.}\label{Figdelay-pilot}
\end{figure*}

In practice, we can use
\[
q_t\bigl(\bx_{t:t+\Delta}^{*(j)}\mid\bx_{t-1}^{(j)},
\bx_{t:t+\Delta-1}^{(j)}\bigr)=\widehat{\pi}_{t+\Delta}\bigl(
\bx_{t:t+\Delta
}^{*(j)}\mid\bx_{t-1}^{(j)}\bigr)
\]
and
\begin{eqnarray*}
&&
\lambda_t\bigl(\bx_{t:t+\Delta-1}^{(j)}\mid
\bx_{t-1}^{(j)},\bx_{t:t+\Delta}^{*(j)}\bigr) \\
&&\quad=
\widehat{\pi}_{t+\Delta-1}\bigl(\bx_{t:t+\Delta,t-1}^{(j)}\mid
\bx_{t-1}^{(j)}\bigr),
\end{eqnarray*}
which are low complexity approximations of the optimal $q_t$ and
$\lambda_t$.



\subsection{Pilot Lookahead Sampling}\label{sec4.4}\label{secLookaheadPilot}

Because of the desire to explore the space of future states with
controllable computational cost, \citet{Wang02SP} and
\citet{ZhangLiu02} considered the \textit{pilot} exploration method,
in which the space of future states
$\{x_{t+1},\ldots,x_{t+\Delta}\}$ is partially explored by pilot
``paths.'' The method could be viewed as a low-accuracy Monte
Carlo approximation to the exact lookahead sampling method.

The method was introduced for the case of finite state space of
$x_t\in{\cal A}=\{a_1,\ldots,a_{|{\cal A}|}\}$ in both \citet{Wang02SP}
and \citet{ZhangLiu02}. Specifically,\vspace*{1pt} suppose at time $t-1$ we have a
set of samples $\{(\bx_{t-1}^{(j)}, w_{t-1}^{(j)}),j=1,\ldots,m\}$
properly weighted with respect to $\pi_{t-1}(\bx_{t-1})$. For each
$\bx_{t-1}^{(j)}$ and each possible value $a_i$ of $x_t$, a pilot path
$\bx_{t:t+\Delta}^{(j,i)}=(x_t^{(j,i)}=a_i,x_{t+1}^{(j,i)},\ldots
,x_{t+\Delta}^{(j,i)})$
is constructed sequentially from distribution
%
\begin{equation}\label{pliot-dis}
\prod_{s=t+1}^{t+\Delta} q_s^{\mathrm{pilot}}
\bigl(x_s \mid\bx_{t-1}^{(j)},x_t=a_i,
\bx_{t+1:s-1}\bigr).
\end{equation}
Then, $x_t^{(j)}$ can be drawn from a trial distribution that utilizes the
``future information'' gathered by the pilot samples
$\bx_{t+1:t+\Delta}^{(j,i)}$, $i=1,\ldots,|{\cal A}|$.\vspace*{1pt}

Figure~\ref{Figdelay-pilot} illustrates the pilot lookahead
sampling operation, with ${\cal A} = \{0,1\}$ and $\Delta=2$, in
which the pilot path for $(\bx_{t-1}^{(j)},x_t=0)$ is
$(x_{t+1}=1,x_{t+2}=1)$ and the pilot path for
$(\bx_{t-1}^{(j)},x_t=1)$ is $(x_{t+1}=0,\break x_{t+2}=1)$; both are
shown as a dark path.

The single pilot lookahead algorithm is as follows.\vspace*{6pt}

\noindent\rule{\columnwidth}{0.5pt}\vspace*{-4pt}
\begin{algorrrr*}
\begin{itemize}
\item At time $t=0$, for $j=1,\ldots,m$:
\begin{itemize}
\item Draw $x_0^{(j)}$ from distribution
$q_0(x_0)$.
\item Set $w_0^{(j)}=\pi_0(x_0^{(j)})/q_0(x_0^{(j)})$.
\item Generate pilot path $\bx_{1:\Delta}^{(j,*)}$ from
$\prod_{s=1}^{\Delta} q_s^{\mathrm{pilot}}(x_s \mid x_0^{(j)},\bx_{1:s-1})$
and calculate
\[
\hspace*{-20pt}w_0^{\mathrm{aux}(j)}=w_0^{(j)}
\frac{\pi_{\Delta}(
x_0^{(j)},x_{1:\Delta}^{(j,*)})}{\pi_0(x_0^{(j)})\prod_{s=1}^{\Delta}
q_s^{\mathrm{pilot}}(x_s \mid x_0^{(j)},\bx_{1:s-1})}.
\]
\end{itemize}
\item At times $t=1,2,\ldots\,$:
\begin{itemize}
\item(Optional.) Resample
$\{\bx_{t-1}^{(j)},w_{t-1}^{(j)},j=1,\ldots,m\}$ with priority scores
$\alpha_{t-1}^{(j)}=w_{t-1}^{\mathrm{aux}(j)}$.
\item Propagation: For $j=1,\ldots,m$:
\begin{itemize}
\item(Generating pilots.)
For $x_t=a_i$, $i=1,\ldots, |\mathcal{A}|$, draw
$\bx_{t+1:t+\Delta}^{(j,i)}$ from (\ref{pliot-dis})
and calculate\vspace*{-4pt}
\end{itemize}
\end{itemize}
\end{itemize}
%
\begin{eqnarray}\label{cum-weight}
U_t^{(j,i)}&=&{\pi_{t+\Delta}\bigl(\bx_{t-1}^{(j)},
x_t=a_i,\bx_{t+1:t+\Delta}^{(j,i)}\bigr)
}\nonumber\\[-8pt]\\[-8pt]
&&{}/\bigl(\pi_{t-1}\bigl(\bx_{t-1}^{(j)}\bigr)Q_t^{(j,i)}\bigr),\nonumber
\end{eqnarray}
where
\begin{eqnarray*}
Q_t^{(j,i)}&=&
\prod_{s=t+1}^{t+\Delta} q_s^{\mathrm{pilot}}\bigl(x_s^{(j,i)} \mid
\bx_{t-1}^{(j)},x_t=a_i,
\bx_{t+1:s-1}^{(j,i)}\bigr).\nonumber
\end{eqnarray*}
%

\begin{itemize}
\item[]
\begin{itemize}
\item[]
\begin{itemize}
\item(Sampling.) Draw $x_t^{(j)}$ from distribution
\[
q_t\bigl(x_t=a_i\mid
\bx_{t-1}^{(j)}\bigr)=\frac{U_t^{(j,i)}}{\sum_{k=1}^{|\mathcal{A}|}
U_t^{(j,k)}}.
\]

\item(Updating weights.) We will keep two sets of weights.
Let
\[
w_{t}^{(j)}=w_{t-1}^{(j)} \frac{\pi_t(\bx_t^{(j)})}{\pi_{t-1}(\bx_{t-1}^{(j)})
q_t(x_t^{(j)}\mid\bx_{t-1}^{(j)})}
\]
and
\[
w_t^{\mathrm{aux}(j)}=w_{t-1}^{(j)}
\sum_{k=1}^{|\mathcal{A}|} U_t^{(j,k)}.
\]
\end{itemize}
\item Inference: $E_{\pi_{t+\Delta}}(h(\bx_t))$ is estimated
by\vspace*{-6pt}
\end{itemize}
\end{itemize}
%
\begin{equation}\label{estpilot}\quad
\frac{\sum_{j=1}^m w_{t-1}^{(j)}
\sum_{i=1}^{|\mathcal{A}|}U_t^{(j,i)} h (
\bx_{t-1}^{(j)},x_t=a_i )}{\sum_{j=1}^m w_{t}^{\mathrm{aux}(j)}}.
\end{equation}
\rule{\columnwidth}{0.5pt}\vspace*{6pt}
\end{algorrrr*}
%


In the algorithm we maintain two sets of weights. The
weight $w_t^{(j)}$ is being updated at each step, and the sample
$(\bx_t, w_t^{(j)})$ is
properly weighted with respect to $\pi_t(\bx_t)$, but not
$\pi_{t+\Delta}(\bx_t)$. A second set of weights,
the \textit{auxiliary weight} $w_t^{\mathrm{aux}(j)}$, is obtained
for resampling and making inference of
$E_{\pi_{t+\Delta}} (h(\bx_t) )$. We have the
following proposition:
%
\begin{propo}\label{prop4} The weighted sample
$(\bx_t^{(j)},\break w_t^{\mathrm{aux}(j)})$ obtained by the single-pilot
lookahead algorithm is properly weighted with respect %
to
$\pi_{t+\Delta}(\bx_{t})$, and estimator (\ref{estpilot})
is a consistent estimator of\break $E_{\pi_{t+\Delta}} (h(\bx_t)
)$.
\end{propo}



The proof is given in the \hyperref[app]{Appendix}.

%

The pilot scheme can be quite flexible. For example, multiple
pilots can be used for each $(\bx_{t-1}^{(j)},x_t=a_i)$. This
would be particularly useful when the size of the state space
${\cal A}$ is large. Specifically, for each
$(\bx_{t-1}^{(j)},x_t=a_i)$, multiple pilots
$\bx_{t+1:t+\Delta}^{(j,i,k)}$, $k=1,\ldots,K$, are generated from
distribution (\ref{pliot-dis}) independently and the corresponding
cumulative incremental\break weights $U_t^{(j,i,k)}$ are calculated by
\begin{eqnarray*}
U_t^{(j,i,k)}
&=&\pi_{t+\Delta}\bigl(\bx_{t-1}^{(j)},
x_t=a_i,\bx_{t+1:t+\Delta}^{(j,i,k)}\bigr)\\
&&{}/ \bigl(\pi_{t-1}\bigl(\bx_{t-1}^{(j)}\bigr)Q_t^{(j,i,k)}\bigr),
\end{eqnarray*}
where
\begin{eqnarray*}
Q_t^{(j,i,k)}&=&\prod_{s=t+1}^{t+\Delta}
q_s^{\mathrm{pilot}}\bigl(x_s^{(j,i,k)} \mid
\bx_{t-1}^{(j)},x_t=a_i,\bx_{t+1:s-1}^{(j,i,k)}\bigr).
\end{eqnarray*}
Sample
$x_{t}^{(j)}$ is then generated from distribution
%
\begin{equation}
\label{pilotsampling} q_t\bigl(x_t=a_i\mid
\bx_{t-1}^{(j)}\bigr)=\frac{\sum_{k=1}^K
U_t^{(j,i,k)}}{\sum_{i=1}^{|\mathcal{A}|}\sum_{k=1}^K
U_t^{(j,i,k)}}.
\end{equation}
The corresponding weight and auxiliary weight are updated by
\[
w_{t}^{(j)}=w_{t-1}^{(j)} \frac{\pi_t(\bx_t^{(j)})}{\pi_{t-1}(\bx_{t-1}^{(j)})
q_t(x_t^{(j)}\mid\bx_{t-1}^{(j)})}
\]
and
\[
w_t^{\mathrm{aux}(j)}= w_{t-1}^{(j)}
\sum_{i=1}^{\mathcal{A}}\frac{1}{K}\sum
_{k=1}^K U_t^{(j,i,k)},
\]
respectively.

Similar to the conclusion of Proposition~\ref{prop4}, samples
$(\bx_t^{(j)}, w_t^{\mathrm{aux}(j)})$
are properly weighted with respect to
$\pi_{t+\Delta}(\bx_{t})$.
In addition, we have the following proposition:
%
\begin{propo}\label{prop5}
Suppose sample $\bx_t^{(1,j)}$ is generated by the exact lookahead
sampling algorithm with weight $w_t^{(1,j)}$ as in
(\ref{delay-sample-weight}). Denote
$(\bx_t^{(4,j)},w_t^{(4,j)},w_t^{\mathrm{aux}(j)})$ as the weighted samples
from the $k$-pilot lookahead algorithm and $U_t^{(j,i,k)}$ are %
the
cumulative incremental weights, then
\begin{eqnarray*}
0&\leq& \operatorname{var}_{\pi_{t+\Delta}} \bigl( w_t^{\mathrm{aux}(j)} \bigr)-
\operatorname{var}_{\pi_{t+\Delta}} \bigl(w_t^{(1,j)} \bigr) \\
&\sim&
O(1/K)
\end{eqnarray*}
and
\begin{eqnarray*}
0&\leq& \operatorname{var}_{\pi_{t+\Delta}} \Biggl[ w_{t-1}^{(j)}\sum
_{i=1}^{\mathcal{A}}\frac{1}{K}\sum
_{k=1}^K U_t^{(j,i,k)} h \bigl(
\bx_{t-1}^{(j)},x_t=a_i \bigr) \Biggr]
\\
&&{} - \operatorname{var}_{\pi_{t+\Delta}} \bigl[w_t^{(1,j)}E_{\pi
_{t+\Delta}}
\bigl(h(\bx_t)\mid\bx_{t-1}=\bx_{t-1}^{(j)}
\bigr) \bigr] \\
&\sim& O(1/K).
\end{eqnarray*}
\end{propo}
The proof is in the \hyperref[app]{Appendix}.

This proposition shows that the variance of the weights under the
multiple-pilot lookahead sampling method is larger than that under
the exact lookahead sampling method, but converges to the latter
at the rate of $1/K$ as the number of pilots $K$ increases. As a
consequence, the samples generated by the multiple-pilot lookahead
sampling method are more effective than the samples generated by
the lookahead weighting method when pilot number $K$ is reasonably
large.

When the state space for $x_t$ is continuous,
it is infeasible to
explore all the possible values of $x_t$.
Finding a more efficient method to carry out
lookahead in continuous state space cases is a challenging problem
currently under investigation.

One possible approach is the following simple algorithm. For each
$j$, draw multiple samples of $x_t^{(j,i)}$, $i=1,\ldots,A$, from
$q_t(x_t\mid\bx_{t-1}^{(j)})$ and treat this set as the space of
$x_t^{(j)}$ (the possible values $x_t^{(j)}$ can take). Then we
run single or multiple pilots from each of these values and
sample $x_t^{(j)}$ according to the lookahead cumulative incremental weights,
just as in
the discrete state-space case. In the special case of $A=1$, the
sampling distribution of this lookahead method will be the same as
that in the concurrent SMC, but one would use the lookahead weight
as the resampling priority score at time $t$.

An improvement of this approach for the continuous state-space case can
be achieved if the dimension of
$x_t$ is relatively low and when the state-space model is Markovian.
That is,
\[
g_t(x_t\mid\bx_{t-1})=g_t(x_t
\mid x_{t-1})
\]
and
\[
f_t(y_t\mid
\bx_t)=f_t(y_t\mid x_t).
\]
In this case, the cumulative incremental weight $U_t^{(j,i)}$
of the pilot $(x_{t}^{(j,i)},\bx_{t+1:t+\Delta}^{(j,i)})$ can be
written as
\begin{eqnarray*}
U_t^{(j,i)}&=& \pi_{t+\Delta}\bigl(\bx_{t-1}^{(j)},x_t^{(j,i)},\bx
_{t+1:t+\Delta
}^{(j,i)}\bigr)\\
&&{}/ \bigl(\pi_{t-1}\bigl(\bx_{t-1}^{(j)}\bigr)Q_t^{(j,i)}\bigr)
\\
&\propto&
g_t\bigl(x_t^{(j,i)}\mid x_{t-1}^{(j)}\bigr)f_t\bigl(y_t\mid x_t^{(j,i)}\bigr)\\
&&{}\cdot\prod
_{s=t+1}^{t+\Delta}g_s\bigl(x_s^{(j,i)}\mid
x_{s-1}^{(j,i)}\bigr)f_s\bigl(y_s\mid x_s^{(j,i)}\bigr)/Q_t^{(j,i)}\\
\\
&\dff&V_t^{(j,i)}V_{t+1:t+\Delta}^{(j,i)},
\end{eqnarray*}
where
\begin{eqnarray*}
Q_t^{(j,i)}&=&
q_t\bigl(x_t^{(j,i)}\mid x_{t-1}^{(j)}\bigr)\prod_{s=t+1}^{t+\Delta
}q_s^{\mathrm{pilot}}\bigl(x_s^{(j,i)}\mid
x_{s-1}^{(j,i)}\bigr),\\
V_t^{(j,i)}&=&\frac{g_t(x_t^{(j,i)}\mid x_{t-1}^{(j)})f_t(y_t\mid
x_t^{(j,i)})}{q_t(x_t^{(j,i)}\mid x_{t-1}^{(j)})}
\end{eqnarray*}
and
\[
V_{t+1:t+\Delta}^{(j,i)}=\frac{
\prod_{s=t+1}^{t+\Delta}g_s(x_s^{(j,i)}\mid
x_{s-1}^{(j,i)})f_s(y_s\mid x_s^{(j,i)})} {
\prod_{s=t+1}^{t+\Delta}q_s^{\mathrm{pilot}}(x_s^{(j,i)}\mid
x_{s-1}^{(j,i)})}.
\]
Standard procedure would
choose $x_t^{(j)}$ from the generated $x_t^{(j,i)}$,
$i=1,\ldots,A$, with probability
$U_t^{(j,i)}/\allowbreak\sum_l U_t^{(j,l)}$.
However, note that
%
\begin{eqnarray}\label{EqV}
\overline{V}{}^{(j,i)}_{t+1:t+\Delta} &\dff& E\bigl(V_{t+1:t+\Delta
}^{(j,i)}
\mid\bx_{t-1}^{(j)},x_t^{(j,i)},
\by_{t+\Delta}\bigr)
\nonumber\\
&=& \int g\bigl(x_{t+1}\mid x_t^{(j,i)}\bigr)
f(y_{t+1}\mid x_{t+1}) \nonumber\\[-8pt]\\[-8pt]
&&\hspace*{8.8pt}{}\cdot\prod_{s=t+2}^{t+\Delta}
g_s(x_s\mid x_{s-1})\nonumber\\
&&\hspace*{38.5pt}{}\cdot f(y_s\mid
x_s)\,dx_{t+1}\cdots dx_{t+\Delta}
\nonumber
\end{eqnarray}
only depends on $x_t^{(j,i)}$, and
\[
V_t^{(j,i)}\overline{V}{}^{(j,i)}_{t+1:t+\Delta}\propto
\frac{\pi_{t+\Delta}(\bx_{t-1}^{(j)},x_t^{(j,i)})} {
\pi_{t-1}(\bx_{t-1}^{(j)})q_t(x_t^{(j,i)}\mid
\bx_{t-1}^{(j)})}
\]
is the lookahead cumulative incremental weight in
(\ref{exactlookahead}), which is shown to be more efficient than
$U_t^{(j,i)}=V_t^{(j,i)}V_{t+1:t+\Delta}^{(j,i)}$
as in Proposition~\ref{prop3}.

With a Markovian model, $\overline{V}{}^{(j,i)}_{t+1:t+\Delta}$ is the
function of $x_t^{(j,i)}$,
and $V_{t+1:t+\Delta}^{(j,i)}$ can be considered as a noisy version of
$\overline{V}_{t+1:t+\Delta}(x_t^{(j,i)})$. That is,
one can write
\[
V_{t+1:t+\Delta}^{(j,i)}=\overline{V}_{t+1:t+\Delta
}\bigl(x_t^{(j,i)}
\bigr)+e_t^{(j,i)},
\]
where
\[
E\bigl(e_t^{(j,i)}\mid x_t^{(j,i)}\bigr)=0.
\]
Hence, if the dimension of $x_t$ is small, one can smooth
$V_{t+1:t+\Delta}^{(j,i)}$ in the space of $x_t$ to obtain an estimate of
$\overline{V}{}^{(j,i)}_{t+1:t+\Delta}$, using all the pilot samples. The
estimate is then used
for sampling and resampling.
For example, let $\widehat{V}_{t+1:t+\Delta}^{(j,i)}$ be a nonparametric
estimate of
$\overline{V}{}^{(j,i)}_{t+1:t+\Delta}$ and let $\widehat{U}_t^{(j,i)}=V_t^{(j,i)}\widehat
{V}_{t+1:t+\Delta}^{(j,i)}$.
One can choose $x_t^{(j)}$ from $x_t^{(j,i)}$,
$i=1,\ldots,A$, with probability $\widehat{U}_t^{(j,i)}/\sum_l
\widehat{U}_t^{(j,l)}$
and weight it accordingly. Experience shows that a very accurate
smoothing method (e.g., kernel smoothing) is not necessary,
as to control computational cost. Often a piecewise
constant smoother is sufficient.

\subsection{Deterministic Piloting}\label{sec4.5}\label{secfixed-pilot}

It is also possible to use deterministic pilots in the pilot
lookahead sampling method. For example, at time $t$, the pilot
starting with $(\bx_{t-1}^{(j)},x_t=a_i)$ for each $a_i\in
\mathcal{A}$ can be a future path $\bx_{t+1:t+\Delta}^{(j,i)}$
that maximizes $\pi_{t+\Delta}(\bx_{t+1:t+\Delta}\mid
\bx_{t-1}^{(j)},x_t=a_i)$. Since such a global maximum is usually
difficult to obtain, an easily obtainable local maximum is to
sequentially,
for $s=t+1,\ldots,t+\Delta$, obtain
%
\begin{equation}\label{argmax}
x_s^{(j,i)}=\arg\max_{x_s} \pi_s
\bigl(x_s\mid\bx_{t-1}^{(j)},x_t=a_i,
\bx_{t+1:s-1}^{(j,i)}\bigr).\hspace*{-28pt}
\end{equation}


Once the pilots are drawn, the remaining steps are similar to
those in the random pilot algorithm, except that
there is usually no easy way to
obtain a proper weight with respect to $\pi_{t+\Delta}(\bx_t)$,
though a proper weight with respect to $\pi_{t}(\bx_t)$ is easily
available.
In order to make proper inference with respect to
$\pi_{t+\Delta}(\bx_t)$, one can generate an additional random
pilot path to $x_{t+\Delta}$. Specifically, we have the following
scheme.\vspace*{6pt}

\noindent\rule{\columnwidth}{0.5pt}\vspace*{-4pt}
\begin{algorrrrr*}
\begin{itemize}
\item At time $t=0$, for $j=1,\ldots,m$:
\begin{itemize}
\item Draw $x_0^{(j)}$ from distribution
$q_0(x_0)$.
\item Set $w_0^{(j)}=\pi_0(x_0^{(j)})/q_0(x_0^{(j)})$.
\item Generate deterministic pilots $\bx_{1:\Delta}^{(j,*)}$
sequentially by letting
\[
x_s^{(j,*)}=\arg\max_{x_s} \pi_s
\bigl(x_s\mid\bx_{\Delta}^{(j)},\bx_{1:s-1}^{(j,*)}
\bigr)
\]
for
$s=1,\ldots,\Delta$. Let $U_0^{(j,*)}=\pi_{\Delta}(
x_0^{(j)},\bx_{1:\Delta}^{(j,*)})/\break\pi_0(x_0^{(j)})$.
\item Set $w_0^{\mathrm{res}(j)}=w_0^{(j)}U_0^{(j,*)}$.
\end{itemize}

\item At times $t=1,2,\ldots\,$:
\begin{itemize}
\item(Optional.) Resample $\{(\bx_{t-1}^{(j)},w_{t-1}^{(j)}),j=1,\ldots
,\allowbreak m\}$
with priority scores $\alpha_{t-1}^{(j)}=w_{t-1}^{\mathrm{res}(j)}$.
\item Propagation: For $j=1,\ldots,m$:
\begin{itemize}
\item(Generating deterministic pilots.)
For $x_t=a_i$, $i=1,\ldots, |\mathcal{A}|$, obtain
$\bx_{t+1:t+\Delta}^{(j,i)}$ sequentially using (\ref{argmax})
for $s=t+1,\ldots,t+\Delta$.
\item(Sampling.) Draw $x_t^{(j)}$ from distribution
\[
q_t\bigl(x_t=a_i\mid
\bx_{t-1}^{(j)}\bigr)=U_t^{(j,i)}\Big/\sum_{i=1}^{|\mathcal{A}|}U_t^{(j,i)},
\]
where
\[
U_t^{(j,i)}= \frac{\pi_{t+\Delta}(\bx_{t-1}^{(j)},x_t=a_i,\bx
_{t+1:t+\Delta}^{(j,i)})} {
\pi_{t-1}(\bx_{t-1}^{(j)})}.
\]
\item(Updating weights.) We keep three sets of\break weights for concurrent
weighting, resampling and estimation.

(1) Concurrent weight:
\[
w_{t}^{(j)}=w_{t-1}^{(j)}
\frac{\pi_t(\bx_t^{(j)})}{\pi_{t-1}(\bx_{t-1}^{(j)})
q_t(x_t^{(j)}\mid\bx_{t-1}^{(j)})};
\]

(2) Resampling weight
\[
w_t^{\mathrm{res}(j)}=w_{t-1}^{(j)}\sum_{i=1}^{|\mathcal{A}|}U_t^{(j,i)};
\]

(3) Auxiliary weight: draw $\bx^{\mathrm{aux}(j)}_{t+1:t+\Delta}$ from
\[
\prod_{s=t+1}^{t+\Delta} q_s^{\mathrm{aux}}\bigl(x_s \mid
\bx_{t-1}^{(j)},x_t^{(j)},\bx_{t+1:s-1}\bigr)
\]
and calculate
\begin{eqnarray*}
w_t^{\mathrm{aux}(j)}&=&w_t^{(j)}{\pi_{t+\Delta}\bigl(\bx_{t-1}^{(j)},
x_t^{(j)},\bx_{t+1:t+\Delta}^{\mathrm{aux}(j)}\bigr)}\\
&&{}/\bigl(\pi_t\bigl(\bx_t^{(j)}\bigr)Q_t^{\mathrm{aux}(j)}\bigr),
\end{eqnarray*}
where
\begin{eqnarray*}
\hspace*{-20pt}Q_t^{\mathrm{aux}(j)}&=&
\prod_{s=t+1}^{t+\Delta} q_s^{\mathrm{aux}}\bigl(x_s^{\mathrm{aux}(j)} \mid
\bx_{t-1}^{(j)},
x^{(j)}_t,\bx_{t+1:s-1}^{\mathrm{aux}(j)}\bigr).
\end{eqnarray*}
\end{itemize}
\item Inference: $E_{\pi_{t+\Delta}} (h(\bx_t) )$ is estimated
by
\[
\sum_{j=1}^m w_t^{\mathrm{aux}(j)}h\bigl(\bx_t^{(j)}\bigr)\Big/
\sum_{j=1}^m w_t^{\mathrm{aux}(j)}.
\]
\hspace*{-22pt}\rule{\columnwidth}{0.5pt}\vspace*{6pt}
\end{itemize}
\end{itemize}
\end{algorrrrr*}

The above algorithm requires the generation of an additional random
pilot $\bx_{t+1:t+\Delta}^{\mathrm{aux}(j)}$ to obtain $w_t^{\mathrm{aux}(j)}$,
which is properly weighted with respect to
$\pi_{t+\Delta}(\bx_t)$. Alternatively, one can combine the
deterministic pilot scheme and the lookahead weighting method in
Section~\ref{secLookaheadWeighting} to obtain a consistent
estimate of\break $E_{\pi_{t+\Delta}}(h(\bx_t))$.

The resampling weight $w_t^{\mathrm{res}(j)}$ is served as the priority score
for resampling when needed. It retains the information from the
deterministic pilot and avoids the additional random variation from
the additional sample path required by the auxiliary weight
$w_t^{\mathrm{aux}(j)}$.

The deterministic pilots are useful because they gather future
information to guide the generation of the current state $x_t$. In
some cases, the deterministic pilots can provide a better
approximation of the distribution $\pi_{t+\Delta}(x_t\mid
\bx_{t-1}^{(j)})$ than the random pilots, especially when we can
only afford to use a single pilot for each
$(\bx_{t-1}^{(j)},x_t=a_i)$. In addition, with some proper
approximation, the deterministic pilot scheme may have lower
computational complexity. The example in Section
\ref{secexample-decoding} uses a low complexity method to generate
the deterministic pilots.

\subsection{Multilevel Pilot Lookahead Sampling}\label{sec4.6}

In case of finite state space, when the size of the state space
$\mathcal{A}$ is large, the pilot lookahead sampling method can
still be too expensive. To reduce the computational cost, we
introduce a multilevel method, which constructs a hierarchical
structure in the state space and utilizes the lookahead idea within
the structure. \citet{Guoetal04} developed a similar algorithm.

\begin{figure}

\includegraphics{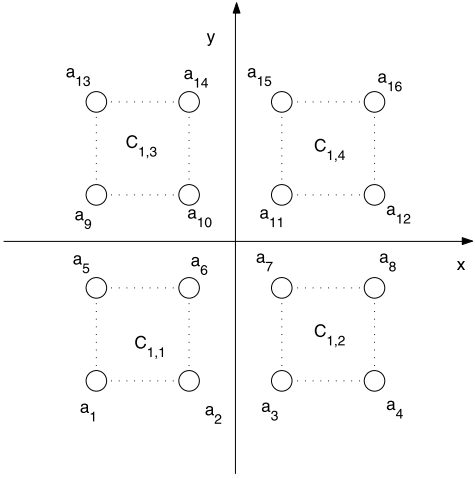}

\caption{Illustration of multilevel
structure in a 16-QAM modulation.}\label{Fig16-QAM}\vspace*{-3pt}
\end{figure}

Specifically, at time $t$, we first divide the current state space
$\mathcal{A}$ of $x_t$ into disjoint subspaces on $L+1$ different
levels, that is,
\[
{\cal A}={\cal C}_{l,1}\cup{\cal C}_{l,2}\cup\cdots\cup{\cal
C}_{l,D_l},\qquad l=0,\ldots,L.
\]
In the division, each level-$l$ subspace
${\cal C}_{l,i}$ consists of several level-$(l+1)$ sets ${\cal
C}_{l+1,j}$. On the top level \mbox{(level-$0$)}, ${\cal
C}_{0,1}={\cal A}$. On the lowest level (level-$L$), each
${\cal C}_{L,i}$ only contains a single state value $a_i\in{\cal
A}$. For example, in a 16-QAM wireless communication problem
\citep{Guoetal04}, the transmitted signal $x_{t}$ to be decoded
takes values in space ${\cal A}=\{a_i=(a_{i,1},a_{i,2})\dvtx
a_{i,1},a_{i,2}=\pm1,
\pm2\}$.
Figure~\ref{Fig16-QAM} depicts a multilevel scheme where the
state space is divided into three levels ($L=2$),
\begin{eqnarray*}
{\cal A}&=& {\cal C}_{0,1} = {\cal C}_{1,1}\cup{\cal
C}_{1,2}\cup{\cal C}_{1,3}\cup{\cal C}_{1,4} \\
&=&
\bigcup_{i=1}^{16} {\cal C}_{2,i}=
\bigcup_{i=1}^{16}\{a_i\}.
\end{eqnarray*}

At time $t$, instead of sampling $x_t^{(j)}$ directly,
we generate a length $L$ index sequence
$\{I_{t,1}^{(j)},\ldots,I_{t,L}^{(j)}\}$, in which
$I_{t,l}^{(j)}$ indicates that
$x_t^{(j)}$ belongs to level-$l$ subsets ${\cal
C}_{l,I_{t,l}^{(j)}}$. A valid index sequence\vadjust{\goodbreak}
$\{I_{t,1}^{(j)},\ldots,I_{t,L}^{(j)}\}$ needs to satisfy ${\cal
C}_{l,I_{t,l}^{(j)}}\subset{\cal C}_{l-1,I_{t,l-1}^{(j)}}$,
$l=1,\ldots,L$.
The last indicator $I_{t,L}^{(j)}$
specifies the value of $x_t^{(j)}$, as the level-$L$ subset
${\cal C}_{L,I_{t,L}^{(j)}}$ only contains one state value.

The index sequence $\{I_{t,1}^{(j)},\ldots,I_{t,L}^{(j)}\}$ is
generated sequentially, starting from the highest level, following
the trial distribution
\[
\prod_{l=1}^{L}q_{t,l}
\bigl(I_{t,l}\mid\bx_{t-1}^{(j)},I_{t,l-1}
\bigr).
\]
Here we define $I_{t,0}\equiv1$, which coincides with $x_t\in{\cal
C}_{0,1}\equiv{\cal A}$. The index\vspace*{1pt} sampling distribution
$q_{t,l}(I_{t,l}\mid\bx_{t-1}^{(j)},I_{t,l-1})$ can be constructed as
follows, using a pilot scheme.

For every $i$ such that ${\cal C}_{l,i}\subset{\cal C}_{l-1,
I_{t,l-1}}$, randomly draw a pilot path $(x_t^{(j,i)},
x_{t+1}^{(j,i)},\ldots,x_{t+\Delta}^{(j,i)})$ from the trial
distribution
%
\begin{eqnarray}\label{multi-pilot}
&&
q^{\mathrm{pilot}}_t\bigl(x_t\mid\bx_{t-1}^{(j)},I_{t,l}=i
\bigr)\nonumber\\[-8pt]\\[-8pt]
&&\quad{}\cdot\prod_{s=t+1}^{t+\Delta
}q_s^{\mathrm{pilot}}
\bigl(x_s\mid\bx_{t-1}^{(j)},\bx_{t:s-1}
\bigr),\nonumber
\end{eqnarray}
where
$q^{\mathrm{pilot}}_t(x_t\mid\bx_{t-1}^{(j)},I_{t,l}=i)$ indicates that
$x_t$ must be a member of ${\cal C}_{l,i}$, and calculate
%
\begin{eqnarray}\label{multilevel-U}
U_{t,l}^{(j,i)}&=&{ \pi_{t+\Delta}\bigl(\bx_{t-1}^{(j)},
\bx_{t:t+\Delta}^{(j,i)}\bigr) }
/ \bigl(
\pi_{t-1}\bigl(\bx_{t-1}^{(j)}\bigr)Q_{t,l}^{\mathrm{pilot}(j,i)}\bigr),\hspace*{-14pt}\nonumber\\
\end{eqnarray}
where
\begin{eqnarray*}
Q_{t,l}^{\mathrm{pilot}(j,i)}&=&
q^{\mathrm{pilot}}_t\bigl(x_t^{(j,i)}\mid
\bx_{t-1}^{(j)},I_{t,l}=i\bigr)\\
&&{}\cdot\prod_{s=t+1}^{t+\Delta
}q_s^{\mathrm{pilot}}\bigl(x_s^{(j,i)}\mid
\bx_{t-1}^{(j)},\bx_{t:s-1}^{(j,i)}\bigr).
\end{eqnarray*}
Then
sample $I_{t,l}^{(j)}$ is generated from distribution
%
\begin{eqnarray}\label{multi-level-index}
&&
q_{t,l}\bigl(I_{t,l}=i\mid\bx_{t-1}^{(j)},I_{t,l-1}^{(j)}
\bigr)\nonumber\\[-8pt]\\[-8pt]
&&\quad= \frac{U_{t,l}^{(j,i)}}{\sum_{k: {\cal C}_{l,k}\subset{\cal
C}_{l-1, I_{t,l-1}}} U_{t,l}^{(j,k)}}.\nonumber
\end{eqnarray}

Specifically, the algorithm is as follows.\vspace*{6pt}

\noindent\rule{\columnwidth}{0.5pt}\vspace*{-4pt}
\begin{algorrrrrr*}
\begin{itemize}
\item At time $t=0$, for $j=1,\ldots,m$:
\begin{itemize}
\item Draw $x_0^{(j)}$ from distribution
$q_0(x_0)$.
\item Set $w_0^{(j)}=\pi_0(x_0^{(j)})/q_0(x_0^{(j)})$.\vadjust{\goodbreak}
\item Generate pilot path $\bx_{1:\Delta}^{(j,*)}$ from
$\prod_{s=1}^{\Delta} q_s^{\mathrm{pilot}}(x_s \mid x_0^{(j)},\bx_{1:s-1})$
and calculate
\[
U_0^{(j,*)}=\frac{\pi_{\Delta}(
x_0^{(j)},x_{1:\Delta}^{(j,*)})}{\pi_0(x_0^{(j)})\prod_{s=1}^{\Delta}
q_s^{\mathrm{pilot}}(x_s \mid x_0^{(j)},\bx_{1:s-1})}.
\]
\item Set $w_0^{\mathrm{aux}(j)}=w_0^{(j)}U_0^{(j,*)}$.
\end{itemize}
\item At time $t=1,2,\ldots\,$:
\begin{itemize}
\item(Optional.) Resample $\{\bx_{t-1}^{(j)},w_{t-1}^{(j)},j=1,\ldots
,m\}$ with priority scores $\alpha_{t-1}^{(j)}=w_{t-1}^{\mathrm{aux}(j)}$.
%
\item Propagation: For $j=1,\ldots,m$:
\begin{itemize}
\item Set $I_{t,0}^{(j)}\equiv
1$. For level $l=1,2,\ldots,L$:\vspace*{1pt}
\begin{itemize}
\item[$\cdot$](Generating pilots.) For each $i$ such that ${\cal
C}_{l,i}\subset{\cal C}_{l-1,
I_{t,l-1}^{(j)}}$, generate pilot $(x_t^{(j,i)},\break
\bx_{t+1:t+\Delta}^{(j,i)})$ from\vspace*{1pt} distribution (\ref{multi-pilot})
and $U_{t,l}^{(j,i)}$ is calculated as in (\ref{multilevel-U}).
\item[$\cdot$](Sampling.) Draw $I_{t,l-1}^{(j)}$ from the trial distribution
(\ref{multi-level-index}).
\end{itemize}
\item(Updating weights.) If $x_t^{(j)}=a_{i_0}$ is chosen at last,
that is, ${\cal C}_{L,
I_{t,L}^{(j)}}=\{a_{i_0}\}$, let
\begin{eqnarray*}
\hspace*{-28pt}&\displaystyle w_t^{(j)}=w_{t-1}^{(j)}
\frac{\pi_t(\bx_t^{(j)})}{\pi_{t-1}(\bx_{t-1}^{(j)})\prod_{l=1}^L
q_{t,l}(I_{t,l}^{(j)}\mid I_{t,l-1}^{(j)},\bx_{t-1}^{(j)})},&
\\
\hspace*{-28pt}&\displaystyle w_t^{\mathrm{aux}(j)}=w_{t-1}^{(j)}
\frac{U_{t,L}^{(j,i_0)}}{\prod_{l=1}^L
q_{t,l}(I_{t,l}^{(j)}\mid I_{t,l-1}^{(j)},\bx_{t-1}^{(j)})}.&
\end{eqnarray*}
\end{itemize}

\item Inference: $E_{\pi_{t+\Delta}}(h(\bx_t))$ is estimated by
\[
\sum_{j=1}^m w_t^{\mathrm{aux}(j)}h\bigl(\bx_t^{(j)}\bigr)\Big/\sum_{j=1}^m w_t^{\mathrm{aux}(j)}.
\]
\hspace*{-22pt}\rule{\columnwidth}{0.5pt}\vspace*{6pt}
\end{itemize}
\end{itemize}
\end{algorrrrrr*}

The advantage of the multilevel method is that it
reduces the total number of probability calculations involved in
generating $x_t^{(j)}$. For example, generating $x_t^{(j)}$
directly from trial distribution $q_t(x_t\mid\bx_{t-1}^{(j)})$
requires a total of $|\mathcal{A}|$ evaluations of
$q_t(x_t=a_i\mid\bx_{t-1}^{(j)})$, $i=1,\ldots,|\mathcal{A}|$. On
the other hand, generating
$\{I_{t,1}^{(j)},\ldots,I_{t,L}^{(j)}\}$ only requires
$\sum_{l=1}^{L}n(I_{t,l-1}^{(j)})$ such evaluations, where
$n(I_{t,l-1}^{(j)})$ is the number of level-$l$ subsets contained
in level-$(l-1)$ subset\vspace*{-1pt} ${\cal C}_{l-1,I_{t,l-1}^{(j)}}$. In the
example illustrated by Figure~\ref{Fig16-QAM}, $I_{t,1}^{(j)}$
is chosen from a set of four subgroups at the first step. Given a
selected $I_{t,1}^{(j)}$, $I_{t,2}^{(j)}$ is drawn from a set of
four elements under $I_{t,1}^{(j)}$. Hence, $n(I_{t,0})=4$ and
$n(I_{t,1})=4$. In this example, a total of 8 probabilities need\vadjust{\goodbreak} to be
evaluated, reduced from 16 if $x_t^{(j)}$ were generated directly.
More generally, if $|\mathcal{A}|=4^{L}$, we can reduce the
computation to $4L$ evaluations based on such a multilevel
structure.


As discussed in Section~\ref{secfixed-pilot}, a deterministic pilot
can also be used in the multilevel method. A multilevel pilot
lookahead sampling method using deterministic pilots is applied to
the signal detection example in Section~\ref{secexample-decoding}.

\subsection{Resampling with Lookahead and Piloting}\label{sec4.7}

As discussed in \citet{LiuChen95}, \citet{LiuChen98}, although a resampling
step introduces additional Monte Carlo variations for estimating the
current state, it
enables the sampler to focus on important regions
of ``future'' spaces and can improve the effectiveness of samples in
future steps. \citet{LiuChen98} suggested that one can perform
resampling according to either a deterministic schedule or an adaptive
schedule. In the following, we consider the problem of finding the
optimal resampling priority score if resampling only takes place at time
$T,2T,3T,\ldots$ (i.e., a deterministic schedule).

Suppose we perform a standard SMC procedure. At time $t=n T$,
samples $\{(\bx_t^{(j)},w_t^{(j)}),j=1,\ldots,m\}$ properly
weighted with respect to $\pi_t(\bx_t)$ are generated, in which
$\bx_t^{(j)}$ follows the distribution $r_t(\bx_t)$, and
$w_t^{(j)}=w_t(\bx_t^{(j)})=\pi_t(\bx_t^{(j)})/r_t(\bx_t^{(j)})$.
We perform a resampling step with priority score $b(\bx_t^{(j)})$,
then the new samples $\bx_t^{*(j)},j=1,\ldots,m$, approximately
follow the distribution $\psi(\bx_t)$ that is proportional to
$r_t(\bx_t)b(\bx_t)$. In the following $T$ steps,
$x_{t+1}^{*(j)},\ldots,x_{t+T}^{*(j)}$ is generated sequentially
from distribution $q_s(x_s \mid\bx_{s-1}^{*(j)})$,
$s=t+1,\ldots,t+T$, then the corresponding weight of
$\bx_{t+T}^{*(j)}$ with respect to $\pi_{t+T}(\bx_{t+T})$ is
\begin{eqnarray*}
&&
w_{t+T}\bigl(\bx_{t+T}^{*(j)}\bigr)\\
&&\quad=
\frac{\pi_t(\bx_t^{*(j)})}{\psi_t(\bx_t^{*(j)}) } \frac{\pi_{t+T}(\bx
_{t+T}^{*(j)})}{\pi_t(\bx_t^{*(j)})\prod_{s=t+1}^{t+T}
q_s(x_s^{*(j)} \mid\bx_{s-1}^{*(j)})}
\\
&&\quad\propto \frac{\pi_t(\bx_t^{*(j)})}{r_t(\bx_t^{*(j)})
b_t(\bx_t^{*(j)})}\\
&&\qquad{}\cdot \frac{\pi_{t+T}(\bx_{t+T}^{*(j)})}{\pi_t(\bx
_t^{*(j)})\prod_{s=t+1}^{t+T}
q_s(x_s^{*(j)} \mid\bx_{s-1}^{*(j)})}.
\end{eqnarray*}
The following
proposition concerns the choice of priority score $b(\bx_t)$ that
minimizes the variance of weight $w_{t+T}(\bx_{t+T}^{*(j)})$.
%
\begin{propo}\label{prop6}
The variance of weight\break $w_{t+T}(\bx_{t+T}^{*(j)})$ is minimized when
%
\begin{equation}
\label{resampling-weight} b_t(\bx_t)\propto
w_t(\bx_t) \eta_{t,T}^{1/2}(
\bx_t),
\end{equation}
where
\begin{eqnarray*}
\eta_{t,T}(\bx_t)&=&\int\biggl[\frac{\pi_{t+T}(\bx_{t+T})}{\pi_t(\bx
_t)\prod_{s=t+1}^{t+T}
q_s(x_s \mid\bx_{s-1})}
\biggr]^2 \\
&&\hspace*{9pt}{}\cdot\prod_{s=t+1}^{t+T}
q_s(x_s \mid\bx_{s-1})\,dx_{t+1}
\cdots dx_{t+T}.
\end{eqnarray*}
\end{propo}
The proof is in the \hyperref[app]{Appendix}.

Specifically,
if we perform resampling at
every step ($T=1$), and the trial distribution is $q_s(x_s \mid
\bx_{s-1})=\pi_s(x_s \mid\bx_{s-1})$, the optimal priority score
becomes
\[
b_t(\bx_t)=w_t(\bx_t)
\frac{\pi_{t+1}(\bx_t)}{\pi_t(\bx_t)},
\]
which
is the priority score used in the sequential imputation of
\citet{Kong94} and \citet{LiuChen95}, and the auxiliary particle
filter proposed by \citet{PittShep99}.

When $T>1$, the exact value of $\eta_{t,T}(\bx_t)$ in
(\ref{resampling-weight}) is difficult to calculate. In this case,
one can use the
pilot method to find an approximation. For each\vspace*{1pt} sample
$\bx_t^{(j)}$, multiple pilots $\bx_{t+1:t+T}^{(j,i)}$,
$i=1,\ldots,K$, are generated following distribution
$\prod_{s=t+1}^{t+T} q_s(x_s \mid\bx_{s-1}^{(j,i)})$ with the
cumulative incremental weight
\[
U_t^{(j,i)}= \frac{\pi_{t+T}(\bx_t^{(j)},\bx_{t+1:t+T}^{(j,i)})} {
\pi_t(\bx_t^{(j)})\prod_{s=t+1}^{t+T}q_s(x_s^{(j,i)} \mid
\bx_t^{(j)},\bx_{t+1:s-1}^{(j,i)})}.
\]
Then $\eta(\bx_t^{(j)})$ can
be estimated by $K^{-1}\sum_{i=1}^K (U_t^{(j,i)} )^2$.

\subsection{Combined Methods}\label{sec4.8}

The lookahead schemes discussed so far can be combined to further
improve the
efficiency. For example, \citet{Wang02SP} considered a combination
of the exact lookahead sampling and the pilot lookahead sampling
methods. In this approach, the space of the immediate future states
is explored exhaustively, and the space of further future states is
explored using pilots.

\section{Adaptive Lookahead}\label{sec5}\label{secadaptive}

Many systems have structures with different local complexity. In
these systems, it may be beneficial to have different lookahead schemes
based on local information. For example, in one of the wireless
communication applications, the received signal $y_t$ can be considered
as following
\[
y_t=\xi_t x_t+ v_t,
\]
where $\{v_t\}$ is white noise with
variance $\sigma^2$, $\{x_t\}$ is the transmitted discrete symbol
sequence and $\{\xi_t\}$ is the fading channel coefficient that
varies over time. Since $\{\xi_t\}$ varies, the signal-to-noise
ratio in the system also changes. When $|\xi_t|$ is large, the
current observation $\by_t$ contains sufficient information to
decode $x_t$ accurately. In this case, lookahead is not needed.
When $|\xi_t|$ is small, the signal-to-noise ratio is low and
lookahead becomes very important to bring in future observations
to help the estimation of $\xi_t$ and $x_t$.


Lookahead strategies always
result in a better estimator provided that the Monte Carlo sample size is
sufficiently large so that $I(\Delta)$ in (\ref{MSE1}) is
negligible. 
To control computational cost, however, Monte Carlo sample size
used may not be large enough to make
$I(\Delta)$ negligible. For a fixed sample size,
$I(\Delta)$ can increase as $\Delta$ increases. Hence,
it is possible that lookahead make the performance worse with
finite Monte Carlo sample size. The
following proposition provides the condition under which one
additional lookahead step in the pilot lookahead sampling method
makes the estimator less accurate.

\begin{figure*}

\includegraphics{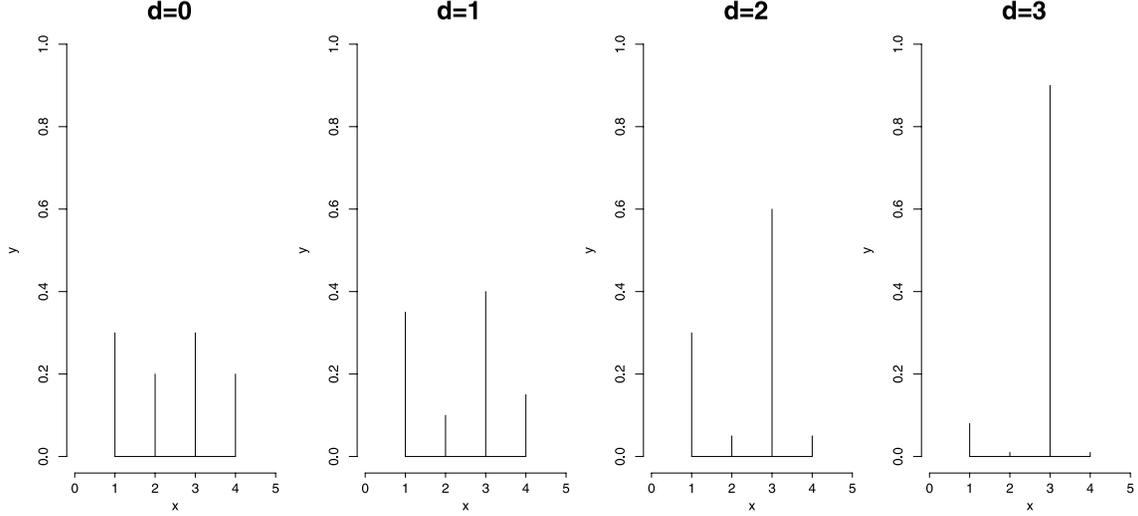}

\caption{Illustration of adaptive
lookahead criterion.}\label{Figadaptive}
\end{figure*}

Specifically, suppose in a finite state system a sample set
$\{(\bx_{t-1}^{(j)},w_{t-1}^{(j)}),j=1,\ldots,m\}$ properly
weighted with respect to $\pi_{t-1}(\bx_{t-1})$ is available at
time $t-1$. At time $t$, $\Delta$-step pilots
$\bx_{t+\Delta}^{(j,i)}=(\bx_{t-1}^{(j)},x_t=a_i,\break\bx_{t+1:t+\Delta
}^{(j,i)})$,
$j=1,\ldots,m$, $i=1,\ldots,\mathcal{A}$, are generated from
distribution (\ref{pliot-dis}) with cumulative incremental weight
\[
U_{t,\Delta}^{(j,i)}=\frac{\pi_{t+\Delta}(\bx_{t+\Delta}^{(j,i)})} {
\pi_{t-1}(\bx_{t-1}^{(j)})\prod_{s=t+1}^{t+\Delta}
q_s^{\mathrm{pilot}}(x_s^{(j,i)} \mid\bx_{s-1}^{(j,i)},\by_s)}.
\]
Then the
$\Delta$-step pilot lookahead sampling estimator of $h(\bx_t)$ is
\begin{eqnarray*}
\widehat{h}&=&\frac{1}{m}\sum_{j=1}^m
w_{t-1}^{(j)}\sum_{i=1}^{\mathcal{A}}
U_{t,\Delta}^{(j,i)}h\bigl(\bx_{t-1}^{(j)},x_t=a_i
\bigr)\\
&\rightarrow& E\bigl(h(\bx_t)\mid\by_{t+\Delta}\bigr).
\end{eqnarray*}
If we lookahead one more step and draw
$x_{t+\Delta+1}^{(j,i)}$ from trial distribution
$q_{t+\Delta+1}(x_{t+\Delta+1} \mid
\bx_{t+\Delta}^{(j,i)},\by_{t+\Delta+1})$, the proper
$(\Delta+1)$-step cumulative incremental\break weight is
\begin{eqnarray*}
U_{t,\Delta+1}^{(j,i)}&=&U_{t,\Delta}^{(j,i)}
{\pi_{t+\Delta+1}\bigl(\bx_{t+\Delta+1}^{(j,i)}\bigr)}\\
&&{}/ \bigl(\pi_{t+\Delta}\bigl(\bx_{t+\Delta}^{(j,i)}\bigr)\\
&&\hspace*{6.7pt}{}\cdot q_{t+\Delta
+1}\bigl(x_{t+\Delta+1}^{(j,i)}
\mid\bx_{t+\Delta}^{(j,i)},\by_{t+\Delta+1}\bigr)\bigr).
\end{eqnarray*}
Then
the $(\Delta+1)$-step pilot lookahead sampling estimator of $h(\bx_t)$
is
\begin{eqnarray*}
\widehat{h}^*&=&\frac{1}{m}\sum_{j=1}^m
w_{t-1}^{(j)}\sum_{i=1}^{\mathcal{A}}
U_{t,\Delta+1}^{(j,i)}h\bigl(\bx_{t-1}^{(j)},x_t=a_i
\bigr)\\
&\rightarrow& E\bigl(h(\bx_t)\mid\by_{t+\Delta+1}\bigr).
\end{eqnarray*}

\begin{propo}\label{prop7}
Let
$\bx_{t+\Delta}^{(j,i=1:\mathcal{A})}=\{\bx_{t+\Delta
}^{(j,i)},i=\break1,\ldots,\mathcal{A}\}$
and suppose $\bx_{t-1}^{(j)}$, $j=1,\ldots,m$, are i.i.d. given
$\by_{t+\Delta}$. When
%
\begin{eqnarray}
\label{C-1}
&&
\frac{1}{m} E \Biggl[\operatorname{var} \Biggl(w_{t-1}^{(j)}
\sum_{i=1}^{\mathcal{A}} U_{t,\Delta+1}^{(j,i)}\nonumber\\
&&{}\hspace*{68pt}\cdot h\bigl(\bx_{t-1}^{(j,i)}\bigr)\mid\bx_{t+\Delta}^{(j,i=1:\mathcal{A})},
\by_{t+\Delta} \Biggr) \Bigm| \by_{t+\Delta} \Biggr]
\\
&&\quad\geq\biggl(1+\frac{1}{m} \biggr) \operatorname{var} \bigl[E \bigl( h(
\bx_{t})\mid\by_{t+\Delta+1} \bigr) \mid \by_{t+\Delta} \bigr],\nonumber
\end{eqnarray}
we have
\[
E \bigl[ \bigl(\widehat{h}^*-h(\bx_t) \bigr)^2 \mid
\by_{t+\Delta
} \bigr]\geq E \bigl[ \bigl(\widehat{h}-h(\bx_t)
\bigr)^2 \mid \by_{t+\Delta} \bigr].
\]
\end{propo}
The proof is in the \hyperref[app]{Appendix}.

Condition (\ref{C-1}) may be difficult to check in practice.
However, when $p(\bx_t \mid\by_{t+\Delta})= p(\bx_t \mid
\by_{t+\Delta+1})$, that is, $y_{t+\Delta+1}$ is independent of
the current state $\bx_t$ given $\by_{t+\Delta}$, the condition
always holds since\break $\operatorname{var} [E (h(\bx_t)\mid
\by_{t+\Delta+1} ) \mid \by_{t+\Delta} ]=0$.



Proposition~\ref{prop7} suggests that, with a fixed number of samples, the
performance of the SMC estimator can be optimized by choosing a
proper lookahead step. Here we use a heuristic criteria, depicted in
Figure~\ref{Figadaptive}. Suppose that the
state space of $x_t$ takes four possible values and the
distribution $\pi_{t+d}(x_t)$ for different lookahead $d=0,1,2,3$
is as shown in Figure~\ref{Figadaptive}, then we can conclude that the
information available at $t$ (i.e., $\by_t$) is not
sufficiently strong for making inference on $x_t$, and the samples
we generate for $x_t$ at this time ($d=0$) may not be useful as
the system propagates. However, as $d$ increases, the distribution
becomes less diffused, showing the accumulation of information
about $x_t$ from the future $y_{t+d}$. It also shows that further
lookahead beyond $d=3$ is probably not necessary.
The details
of this adaptive criteria are as follows:
\begin{itemize}
\item In a finite state space model, consider lookahead steps
$\Delta=0,1,2,\ldots\,$. Stop if $\Delta\geq N$
or the estimated posterior distribution satisfies
%
\begin{eqnarray}
\label{adpt-lookahead-1}
&&\max_i \bigl\{\widehat{\pi}_{t+\Delta}(x_t=a_i)
\bigr\}\nonumber\\[-8pt]\\[-8pt]
&&\quad= \max_i \biggl\{ \frac{\sum_{j} w_{t-1}^{(j)}U_{t,\Delta}^{(j,i)}
}{\sum_{l,j}w_{t-1}^{(j)}U_{t,\Delta}^{(j,l)}} \biggr
\}>p_0,\nonumber
\end{eqnarray}
where $N$ is the maximum number of lookahead steps we will perform,
$0<p_0<1$ is a threshold close to~1, and $U_t^{(j,i)}$ are the
cumulative incremental weights defined in (\ref{cum-weight}).
\item In a continuous state space model, try lookahead steps
$\Delta=0,1,2,\ldots\,$. Stop if $\Delta\geq N$
or the estimated variance $\operatorname{var}_{\pi_{t+\Delta}}(x_t)$ satisfies
%
\begin{eqnarray}
\label{adpt-lookahead-2} \widehat{\mathrm{var}}_{\pi_{t+\Delta}}(x_t)&=&
\frac{\sum_{i,j}w_{t-1}^{(j)}U_{t,\Delta}^{(j,i)} (x_t^{(j,i)} )^2} {
\sum_{i,j}w_{t-1}^{(j)}U_{t,\Delta}^{(j,i)}}\nonumber\\
&&{}- \biggl(\frac{\sum
_{i,j}w_{t-1}^{(j)}U_{t,\Delta}^{(j,i)}x_t^{(j,i)}} {
\sum_{i,j}w_{t-1}^{(j)}U_{t,\Delta}^{(j,i)}} \biggr)^2\\
&<&
\sigma_0^2,\nonumber
\end{eqnarray}
where $\sigma_0^2$ is a given threshold,
$x_t^{(j,i)}$ are samples of current state generated from each
$\bx_{t-1}^{(j)}$ under the pilot scheme and $U_{t,\Delta}^{(j,i)}$ are
the corresponding cumulative
incremental weights.
\end{itemize}
Some examples of using adaptive lookahead in finite state space
models and continuous state space models are presented in Section
\ref{secexample}.

\section{Applications}\label{sec6}\label{secexample}

In this section we demonstrate the property of lookahead and make performance
comparisons. In all cases, $\delta$,
$\Delta$ and $\Delta'$ are used to denote the numbers of
lookahead steps in lookahead weighting, exact lookahead sampling
and pilot lookahead sampling, respectively.

\subsection{Signal Detection over Flat-Fading Channel}\label{sec6.1}\label
{secexample-decoding}

In a digital wireless communication problem
(Chen and Liu, \citeyear{ChenLiu00}; \citecs{Wang02SP}), the received signal sequence
$\{y_t\}$ is modeled as
\[
y_t=\xi_t x_t+ v_t,
\]
where $\{x_t\}$ is the transmitted complex digital symbol
sequence, $\{v_t\}$ is the white complex Gaussian noise with variance
$\sigma^2$ and independent real and complex components,
and $\{\xi_t\}$ is the transmitted channel, which can be modeled as
an ARMA process
\begin{eqnarray*}
&&
\xi_t+ \phi_1\xi_{t-1}+\cdots+
\phi_r \xi_{t-r} \\
&&\quad=\theta_0 u_t +
\theta_1 u_{t-1} +\cdots+ \theta_r
u_{t-r},
\end{eqnarray*}
where $\{u_t\}$ is a unit white complex Gaussian noise. In this
example, we assume $\{\xi_t\}$ follows the\break $\operatorname{ARMA}(3,3)$ process
\citep{Guoetal04}
\begin{eqnarray*}
&&\xi_t - 2.37409\xi_{t-1}+1.92936 \xi_{t-2} -
0.53208 \xi_{t-3}
\\
&&\quad= 10^{-2} (0.89409 u_t + 2.68227 u_{t-1} \\
&&\hspace*{24.5pt}\qquad{}+
2.68227 u_{t-2} + 0.89409 u_{t-3} ).
\end{eqnarray*}
This system can be turned into a conditional dynamic linear model
(CDLM) as follows:
\begin{eqnarray*}
\bz_{t} & = & \bF\bz_{t-1}+ \bg u_t,
\\
y_t & = & \xi_t x_t +v_t =
\bh^H \bz_t x_t + v_t,
\end{eqnarray*}
where
\begin{eqnarray*}
\bF&=&\pmatrix{ %
-\phi_1 & -\phi_{2} &
\cdots& -\phi_{r} & 0
\cr
1 & 0 & \cdots& 0 & 0
\cr
0 & 1 & \cdots& 0 &
0
\cr
\vdots& \vdots& \ddots& \vdots& \vdots
\cr
0 & 0 & \cdots& 1 & 0},\quad \bg=
\pmatrix{ 1
\cr
0
\cr
\vdots
\cr
0},
\\
\bh&=& [\theta_0 \theta_{1} \cdots\theta_{r}]^H.
\end{eqnarray*}

Here we consider a high-constellation system with a 256-QAM
modulation, thus the symbol space is ${\mathcal
A}=\{a_i=(a_{i,1},a_{i,2})\dvtx  a_{i,1},a_{i,2}=\pm1,\pm
3,\ldots,\pm15 \}$,\break where $a_{i,1}$ and $a_{i,2}$ are the real
and imaginary parts of symbol $a_i$, respectively. We decode
$\{x_t\}$ from received $\{y_t\}$ under the framework of the mixture
Kal\-man filter of \citet{ChenLiu00} and the ``optimal-resampling''
scheme of \citet{FearnheadClifford03}.

Because the symbol space is large ($|{\mathcal A}|=256$), we use a
combination of the multilevel pilot lookahead sampling method and
the lookahead weighting method. The multilevel structure used is
similar to that of 16-QAM presented in Figure~\ref{Fig16-QAM}.
The symbol space is divided into subspaces of five different
levels ($L=4$). Hence, at time $t$, we generate
$(I_{t,1}^{(j)},I_{t,2}^{(j)},I_{t,3}^{(j)},I_{t,4}^{(j)})$ to
obtain $x_t^{(j)}$ for given $\bx_{t-1}^{(j)}$ sequentially.

To construct\vspace*{1pt} the conditional trial distribution\break
$q_{t,l}(I_{t,l}\mid\bx_{t-1}^{(j)},I_{t,l-1}^{(j)})$, we
generate a deterministic pilot $(x_t^{(j,I_{t,l})}, \ldots,
x_{t+\Delta'}^{(j,I_{t,l})})$ for every possible $I_{t,l}$ given
$(\bx_{t-1}^{(j)},I_{t,1}^{(j)},\ldots,I_{t,l-1}^{(j)})$
generated. The steps to generate the deterministic pilot are as
follows:
\begin{itemize}
\item Predict channel $\xi_t$ by $\widehat{\xi}_{t}^{(j)}=E(\xi_{t}
\mid
\bx_{t-1}^{(j)},Y_{t-1})$. Let $x_t^{(j,I_{t,l})}$ be the symbol
$a_i\in{\cal C}_{l,I_{t,l}}$ closest to
$y_{t}/\widehat{\xi}_{t}^{(j)}$.
\item For $s=t+1,\ldots,t+\Delta'$,
repeat the following:
\begin{itemize}
\item Predict channel $\xi_s$ by $\widehat{\xi
}_{s}^{(j,I_{t,l})}=E(\xi_{s} \mid
\bx_{t-1}^{(j)},
x_t^{(j,I_{t,l})},\break\ldots,x_{s-1}^{(j,I_{t,l})},\by_{s-1})$.
\item Choose symbol $a_i\in\mathcal{A}$ closest to
$y_{s}/\widehat{\xi}_{s}^{(j,I_{t,l})}$ as $x_s^{(j,I_{t,l})}$.
\end{itemize}
\end{itemize}
Letting $ U_t^{(j,I_{t,l})}=
\pi_{t+\Delta'}(\bx_{t-1}^{(j)},x_t^{(j,I_{t,l})},\bx_{t+1:t+\Delta
'}^{(j,I_{t,l})})/\break
\pi_{t-1}(\bx_{t-1}^{(j)})$, the trial distribution is
\[
q_{t,l}\bigl(I_{t,l}\mid\bx_{t-1}^{(j)},I_{t,l-1}^{(j)}
\bigr)=\frac{U_t^{(j,I_{t,l})}}{\sum_{k:
{\cal C}_{l,k}\subset{\cal C}_{l-1, I_{t,l-1}^{(j)}}}
U_t^{(j,k)}}.
\]

For comparison, SMC without using the multilevel structure and
lookahead pilot is also considered. More computational details of
this problem can be found in \citet{Wang02SP}.

In the simulation, the length of transmitted symbol sequences is
500. To avoid phase ambiguities, differential decoding is used.
Specifically, suppose the information symbol sequence is
$\{d_t\}$. The actual transmitted symbol sequence $\{x_t\}$ is
constructed as follows: given the 256 QAM transmitted symbol
$x_{t-1}$ and information symbol $d_t$, we first map them to four
QPSK symbols
$(r_{x_{t-1},1},r_{x_{t-1},2},r_{x_{t-1},3},r_{x_{t-1},4})$ and
$(r_{d_t,1},r_{d_t,2},r_{d_t,3},r_{d_t,4})$, respectively. Let
$r_{x_{t},i}=r_{d_{t},i}r_{x_{t-1},i}$, $i=1,2,3,4$, and we map
these four QPSK symbols
$(r_{x_{t},1},r_{x_{t},2},r_{x_{t},3},r_{x_{t},4})$ back to
256-QAM as the transmitted symbol $x_t$. The differential receiver
calculates
$r_{\widehat{d}_{t},i}=r_{\widehat{x}_{t},i}r_{\widehat{x}_{t-1},i}^*$,
where $(\widehat{x}_{t-1}, \widehat{x}_{t})$ are estimated
$(x_{t-1}, x_t)$ at the receiver, then decodes the information
symbol $d_t$ as the 256-QAM symbol corresponding to
$(r_{\widehat{d}_t,1},r_{\widehat{d}_t,2}$,
$r_{\widehat{d}_t,3},r_{\widehat{d}_t,4})$. To improve the
decoding accuracy of this high-constellation system, we also
insert $10\%$ symbols that are known to the receiver into the
transmitted symbol sequences periodically. The experiment is
repeated 100 times. A~total of 50,000 symbols (400,000 bit information)
are decoded.

\begin{figure}

\includegraphics{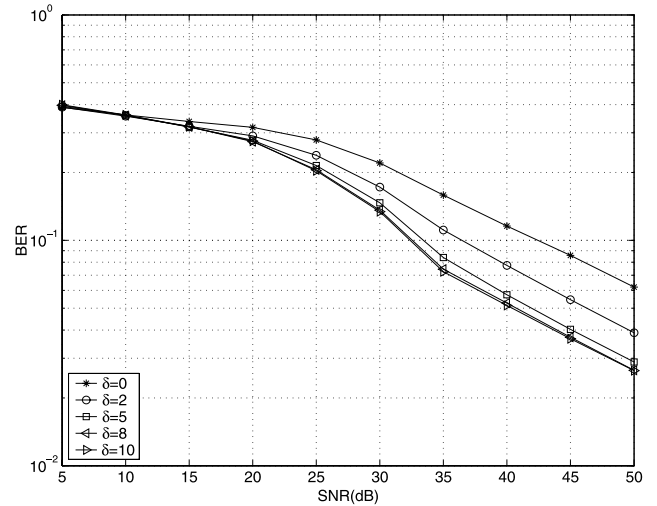}

\caption{BER
performance of the lookahead weighting method with $\Delta=0$,
$\Delta'=0$, $m=200$ and different $\delta$ in a 256-QAM
system.}\label{Figexample-1-1}
\end{figure}

Figure~\ref{Figexample-1-1} reports the bit-error-ratio (BER)
performance of a different lookahead step $\delta$ of the lookahead
weighting method with standard concurrent SMC sampling
($\Delta=0,\Delta'=0$). $m=200$ samples are used. It is seen that
the BER performance does not improve further after $\delta\geq8$
lookahead steps. We use $\delta=10$ in the following comparison.

\begin{table*}
\caption{Average $\mathrm{RMSE}_1$ for SMC with different lookahead
methods. The same numbers of samples ($m=3000$)\break are used in
different methods. We use a single pilot lookahead ($K=1$) unless
stated otherwise.\break Average lookahead steps in the adaptive
lookahead method are reports in the~parentheses}\label{table1}
\begin{tabular*}{\tablewidth}{@{\extracolsep{4in minus 4in}}lccccccc@{}}
\hline
& \multicolumn{6}{c}{$\bolds{\Delta'+\delta}$} &\\[-4pt]
& \multicolumn{6}{c}{\rule{282pt}{1pt}}\\
$\bolds{\mathrm{RMSE}_1}$ & \textbf{0} & \textbf{1} & \textbf{2} & \textbf{3} &
\textbf{5} & \textbf{7} & \textbf{Time (sec.)}\\
\hline
%
%
SMC ($\Delta'=0$) & 3.128 & 1.011 & 0.828 & 0.817 & 0.818 & 0.819 &
0.113 \\
[4pt]
SMC ($\Delta'=1,A=10,K=16$) & -- & 1.009 & 0.824 & 0.813 & 0.812 & 0.813
& 5.952 \\
[4pt]
SMC ($\Delta'=1,A=3$) & -- & 1.011 & 0.831 & 0.826 & 0.831 & 0.839 &
0.319 \\
SMC ($\Delta'=2,A=3$) & -- & -- & 0.838 & 0.844 & 0.860 & 0.876 & 0.405
\\
SMC ($\Delta'=3,A=3$) & -- & -- & -- & 0.846 & 0.885 & 0.913 & 0.504 \\
[4pt]
SMC-S ($\Delta'=1,A=1$) & -- & 1.009 & 0.825 & 0.815 & 0.814 & 0.815 &
0.170 \\
SMC-S ($\Delta'=2,A=1$) & -- & -- & 0.825 & 0.815 & 0.815 & 0.815 &
0.197 \\
SMC-S ($\Delta'=3,A=1$) & -- & -- & -- & 0.816 & 0.816 & 0.816 & 0.224
\\
SMC-S ($\operatorname{adpt}\Delta'(0.244),A=1$) & 0.995 & 0.834 & 0.815 & 0.814 &
0.816 & 0.817 & 0.147 \\
[4pt]
SMC-S ($\Delta'=1,A=3$) & -- & 1.009 & 0.824 & 0.813 & 0.813 & 0.813 &
0.421 \\
SMC-S ($\Delta'=2,A=3$) & -- & -- & 0.824 & 0.813 & 0.813 & 0.813 &
0.498 \\
SMC-S ($\Delta'=3,A=3$) & -- & -- & -- & 0.814 & 0.814 & 0.813 & 0.576
\\
\hline
\end{tabular*}
\end{table*}

\begin{figure}[b]

\includegraphics{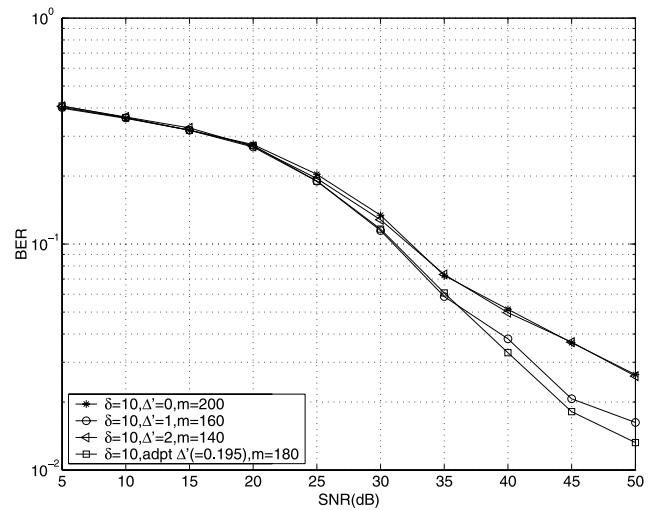}

\caption{BER performance of the multilevel pilot lookahead sampling
method with $\Delta=0$, $\delta=10$ but different $\Delta'$ and number
of samples $m$ in the 256-QAM system. The number of samples are chosen
so that each of the methods takes approximately the same CPU
time.}\label{Figexample-1-2}
\end{figure}

BER performance of pilot lookahead sampling methods with different
lookahead steps $\Delta'$ is shown in Figure
\ref{Figexample-1-2}. The number of Monte Carlo samples is
adjusted so that each method takes approximately the same CPU
time. From the result, it is seen that the multilevel pilot
lookahead sampling method with $\Delta'=1$ has smaller BER
%
than SMC without using lookahead pilots. But when we use $\Delta'=
2$, the performance is worse. One of the reasons is that we use
the predicted channel to construct the pilot, which could be very
different %
from the true channel and severely mislead the sampling,
especially when the number of lookahead steps is large. We also
implement the adaptive method. Here we use adaptive stop criteria
(\ref{adpt-lookahead-1}) with $p_0=0.90$. The resulting average
number of lookahead steps is 0.195. Due to the saving in the smaller
number of lookahead steps, larger Monte Carlo sample size is used
with the same computational time. Its BER performance is slightly
better than using the fixing pilot lookahead step $\Delta'=1$.

\subsection{Nonlinear Filtering}\label{sec6.2}

Consider the following nonlinear state space model
\citep{Gordonetal93}:
\begin{eqnarray*}
&&\mbox{state equation: } \\
&&\quad x_t=0.5x_{t-1}+25x_{t-1}/
\bigl(1+x_{t-1}^2\bigr)\\
&&\qquad{}+8\cos\bigl(1.2(t-1)
\bigr)+u_t,
\\
&&\mbox{observation equation:}\quad
y_t=x_t^2/20+v_t,
\end{eqnarray*}
where $u_{t}\sim N(0,\sigma^2 )$, $v_t \sim
N(0,\eta^2 )$ are Gaussian\break white noise. In the simulation, we let
$\sigma=1$ and $\eta=1$, and the length of observations is
$T=100$. We compare the performance of different lookahead
strategies.


In this nonlinear system, $\pi_t(x_t\mid\bx_{t-1})$ cannot be easily
sampled from. Here we use the
simple trial distribution
\[
q_t(x_t\mid\bx_{t-1})= \pi_{t-1}(x_t\mid
\bx_{t-1})=g_t(x_t\mid x_{t-1})
\]
and
\[
q_s^{\mathrm{pilot}}(x_s\mid\bx_{s-1})=g_s(x_s
\mid x_{s-1}).
\]
We use SMC to denote the pilot lookahead sampling method for the continuous
state space case. The implementation with smoothing step presented in
Section~\ref{secLookaheadPilot} is denoted as
SMC-S. A simple piecewise constant function with interval width 0.5 is
used for smoothing. Resampling is applied at every step.

\begin{table*}
\caption{Average $\mathrm{RMSE}_2$ for SMC with different lookahead
methods. The same numbers of samples ($m=3000$)\break are used in
different methods}\label{table2}
\begin{tabular*}{\tablewidth}{@{\extracolsep{4in minus 4in}}lccccccc@{}}
\hline
& \multicolumn{6}{c}{$\bolds{\Delta'+\delta}$} &\\[-4pt]
& \multicolumn{6}{c}{\rule{282pt}{1pt}}\\
$\bolds{\mathrm{RMSE}_2}$ & \textbf{0} & \textbf{1} & \textbf{2} & \textbf{3} &
\textbf{5} & \textbf{7} & \textbf{Time (sec.)}\\
\hline
%
%
SMC ($\Delta'=0$) & 0.137 & 0.055 & 0.057 & 0.066 & 0.078 & 0.090 &
0.113 \\
[4pt]
SMC ($\Delta'=1,A=10,K=16$) & -- & 0.023 & 0.027 & 0.032 & 0.038 & 0.043
& 5.952 \\
[4pt]
SMC ($\Delta'=1,A=3$) & -- & 0.070 & 0.105 & 0.138 & 0.174 & 0.203 &
0.319 \\
SMC ($\Delta'=2,A=3$) & -- & -- & 0.156 & 0.220 & 0.278 & 0.326 & 0.405
\\
SMC ($\Delta'=3,A=3$) & -- & -- & -- & 0.240 & 0.356 & 0.417 & 0.504 \\
[4pt]
SMC-S ($\Delta'=1,A=1$) & -- & 0.043 & 0.048 & 0.053 & 0.062 & 0.072 &
0.170 \\
SMC-S ($\Delta'=2,A=1$) & -- & -- & 0.051 & 0.063 & 0.066 & 0.075 &
0.197 \\
SMC-S ($\Delta'=3,A=1$) & -- & -- & -- & 0.073 & 0.081 & 0.090 & 0.224
\\
[4pt]
SMC-S ($\Delta'=1,A=3$) & -- & 0.029 & 0.032 & 0.036 & 0.041 & 0.048 &
0.421 \\
SMC-S ($\Delta'=2,A=3$) & -- & -- & 0.031 & 0.039 & 0.042 & 0.047 &
0.498 \\
SMC-S ($\Delta'=3,A=3$) & -- & -- & -- & 0.045 & 0.050 & 0.055 & 0.576
\\
\hline
\end{tabular*}
\end{table*}

\begin{table*}[b]
\caption{Average $\mathrm{RMSE}_1$ for SMC with different lookahead
methods. The numbers of samples are chosen so that each method
used approximately the same CPU time. Average lookahead steps in the adaptive
lookahead method are reports in the parentheses}\label{table3}
\begin{tabular*}{\tablewidth}{@{\extracolsep{4in minus 4in}}lccccccc@{}}
\hline
& \multicolumn{6}{c}{$\bolds{\Delta'+\delta}$} &\\[-4pt]
& \multicolumn{6}{c}{\rule{252pt}{1pt}}\\
$\bolds{\mathrm{RMSE}_1}$ & \textbf{0} & \textbf{1} & \textbf{2} & \textbf{3} &
\textbf{5} & \textbf{7} & \textbf{Time (sec.)}\\
\hline
%
%
SMC ($m=3000,\Delta'=0$) & 3.128 &1.011& 0.828& 0.817& 0.818& 0.819 &
0.113\\
[4pt]
SMC ($m=60,\Delta'=1,A=10,K=16$) & --&1.079& 0.911& 0.906& 0.912& 0.920
& 0.125\\
[4pt]
SMC-S ($m=2000,\Delta'=1,A=1$) & -- & 1.010 & 0.826 & 0.817 & 0.817 &
0.818 & 0.117 \\
SMC-S ($m=1700,\Delta'=2,A=1$) & -- & -- & 0.827 & 0.818 & 0.817 &
0.819 & 0.116 \\
SMC-S ($m=1500,\Delta'=3,A=1$) & -- & -- & -- & 0.820 & 0.822 & 0.823 &
0.118 \\
SMC-S ($m=2400$, $\operatorname{adpt}\Delta'(0.245),A=1$) & 0.994 & 0.835 & 0.816
&\textbf{0.815} & 0.817 & 0.818 & 0.104 \\
[4pt]
SMC-S ($m=800,\Delta'=1,A=3$) & -- & 1.015 & 0.832 & 0.821 & 0.821 &
0.822 & 0.108 \\
SMC-S ($m=700,\Delta'=2,A=3$) & -- & -- & 0.827 & 0.817 & 0.816 & 0.817
& 0.111 \\
SMC-S ($m=600,\Delta'=3,A=3$) & -- & -- & -- & 0.819 & 0.819 & 0.820 &
0.119 \\
\hline
\end{tabular*}
\end{table*}

We repeat the experiment 1000 times. The good\-ness-of-fit
measures used are
\[
\mathrm{RMSE}_1 = \Biggl[\frac
{1}{T}\sum
_{t=1}^T ( \widehat{x}_t-x_t
)^2 \Biggr]^{1/2}
\]
and
\[
\mathrm{RMSE}_2 =
\Biggl[\frac{1}{T}\sum_{t=1}^T
\bigl( \widehat{x}_t-\widetilde{E}_{\pi_{t+\delta+\Delta'}}(x_t)
\bigr)^2 \Biggr]^{1/2},
\]
where $\mathrm{RMSE}_2$ is a measurement of estimation variance,
$I(\delta+\Delta')$ in (\ref{MSE1}). Here
$\widetilde{E}_{\pi_{\delta+\Delta'}}(x_t)$ is obtained by SMC
($\Delta'=0$) with a large number of samples ($m=200\mbox{,}000$) and the
lookahead weighting method with lookahead steps
$\delta^*=\delta+\Delta'$.
Tables~\ref{table1} and~\ref{table2} report average $\mathrm{RMSE}_1$
and $\mathrm{RMSE}_2$ and the associated CPU time of using different
sampling methods and $m=3000$ samples. It can be seen that the delayed
methods can greatly reduce $\mathrm{RMSE}_1$ for small $\delta+\Delta'$, but no
further improvement can be found when $\delta+\Delta'\geq3$. SMC with
$\Delta'=1,A=10,K=16$ is an approximation of the exact lookahead
sampling method with $\Delta=1$. It has the smallest $\mathrm{RMSE}_1$ at the
cost of extensive computation, which confirms Proposition~\ref{prop3}.
The performance of SMC with a single pilot ($K=1$) is poor because
the future state space cannot be efficiently explored by the small
number of pilots. With the smoothing step, SMC-S can achieve better
performance than the simple lookahead weighting method (SMC,
$\Delta'=0$). SMC-S with $A=3$ has better performance than SMC-S with
$A=1$, because when using $A=1$, the pilot only affects resampling and
estimation, but not the sampling procedure. However, SMC-S with $A=3$
also takes a longer CPU time.

\begin{table*}
\caption{Average $\mathrm{RMSE}_2$ for SMC with different lookahead
methods. The numbers of samples are chosen so that\break each method
used approximately the same CPU time}\label{table4}
\begin{tabular*}{\tablewidth}{@{\extracolsep{4in minus 4in}}lccccccc@{}}
\hline
& \multicolumn{6}{c}{$\bolds{\Delta'+\delta}$} &\\[-4pt]
& \multicolumn{6}{c}{\rule{260pt}{1pt}}\\
$\bolds{\mathrm{RMSE}_2}$ & \textbf{0} & \textbf{1} & \textbf{2} & \textbf{3} &
\textbf{5} & \textbf{7} & \textbf{Time (sec.)}\\
\hline
%
%
SMC ($m=3000,\Delta'=0$) &0.137& 0.055& 0.057& 0.066& 0.078 & 0.090 &
0.113\\
[4pt]
SMC ($m=60,\Delta'=1,A=10,K=16$) & --&0.228& 0.254& 0.277& 0.306& 0.334
& 0.125\\
[4pt]
SMC-S ($m=2000,\Delta'=1,A=1$) & -- & 0.054 & 0.058 & 0.064 & 0.075 &
0.087 & 0.117 \\
SMC-S ($m=1700,\Delta'=2,A=1$) & -- & -- & 0.066 & 0.083 & 0.085 &
0.098 & 0.116 \\
SMC-S ($m=1500,\Delta'=3,A=1$) & -- & -- & -- & 0.103 & 0.114 & 0.126 &
0.118 \\
[4pt]
SMC-S ($m=800, \Delta'=1,A=3$) & -- & 0.062 & 0.067 & 0.074 & 0.084 &
0.096 & 0.108 \\
SMC-S ($m=700, \Delta'=2,A=3$) & -- & -- & 0.061 & 0.078 & 0.082 & 0.094
& 0.111 \\
SMC-S ($m=600, \Delta'=3,A=3$) & -- & -- & -- & 0.097 & 0.109 & 0.121 &
0.119 \\
\hline
\end{tabular*}
\end{table*}

We also use the adaptive stop criteria
(\ref{adpt-lookahead-2}) (adpt) to choose the lookahead steps
adaptively. In the criteria, we let $\sigma_0^2=4$. The
adaptive method has similar performance to the fixed-step pilot lookahead
sampling method, but much fewer
average lookahead steps (average lookahead steps are only 0.244) and
less CPU time.


For a fair comparison, Tables~\ref{table3} and~\ref{table4} report
average $\mathrm{RMSE}_1$ and $\mathrm{RMSE}_2$ of different methods with
different numbers of samples, which are chosen so that each method used
approximately the same CPU time. In this table, SMC with $A=1$ and the
adaptive lookahead scheme has the smallest $\mathrm{RMSE}_1$, which demonstrates
the effectiveness of the adaptive lookahead strategy. It also shows
that SMC-1 with $\Delta'=1,A=10,K=16$ has a large $\mathrm{RMSE}_1$, because of
its high computational cost per sample.



\subsection{Target Tracking in Clutter}\label{sec6.3}\label{secexample-tracking}

Consider the problem of tracking a single target in clutter
\citep{Avit95}. In this example, the target moves with random\vadjust{\goodbreak}
acceleration in one dimension. The state equation can be written
as
\[
\pmatrix{ x_{t,1}
\cr
x_{t,2} } %
=
\pmatrix{ 1 & 1
\cr
0 &1 } %
\pmatrix{ x_{t-1,1}
\cr
x_{t-1,2} } %
+ %
\pmatrix{ 1/2
\cr
1 }
u_t,
\]
where $x_{t,1}$ and $x_{t,2}$ denote the one-dimensional location
and velocity of the target, respectively; $u_t\sim N(0,\sigma^2)$
is the random acceleration.

At each time $t$, the target can be observed with probability
$p_d$ independently. If the target is observed, the observation is
\[
z_t=x_{t,1}+v_t,
\]
where $v_t\sim N(0,r^2)$.

In additional to the true observation, there are false signals.
Observation of false signals follows a spatially homogeneous Poisson
process with rate $\lambda$. Suppose the observation window is
wide and centers around the predicted location of the target. Let
$D$ be the range of the observation window. The actual
observation $y_t$ includes $n_t$ detected signals, among which at
most one is the true observation. Therefore, $n_t$ follows a
Bernoulli($p_d$)${}+{}$Poisson($\lambda D$) distribution.

Define an indicator variable $I_t$ as follows:
\[
I_t= %
\cases{ 0,& if the target is not detected at time
$t$,
\cr
k, & if the $k$th signal in $y_t$\cr
& is the true observation,}
\]
then we have
\begin{eqnarray*}
&&
p(y_t, I_t\mid x_t )\\
&&\quad\propto%
\cases{ (1-p_d)\lambda, \quad \mbox{if $I_t=0$},
\vspace*{1pt}\cr
p_d\bigl(2\pi r^2\bigr)^{-1/2}\operatorname{exp}\bigl
\{-(y_{t,k}-x_t)^2/2r^2\bigr\},\vspace*{1pt}\cr
\hspace*{61.2pt}\mbox{if $I_t=k>0$}.} %
\end{eqnarray*}

In this system, given $\bI_t=(I_1,\ldots,I_t)$, it becomes a linear
Gaussian state space model. In such a
system, the mixture Kalman filter (MKF) can be applied. The
mixture Kalman filter only generates samples of the indicators
$\bI_t^{(j)}$ and considers the state space as discrete. Conditional on
$\bI_t^{(j)}$ and $\by_t$, the state
variable $x_t$ is normally distributed. The mean and the variance of
$p(x_{t-\delta}\mid\bI_t^{(j)},\by_t)$ can be exactly calculated
through the Kalman filter. To perform lookahead strategies in MKF,
suppose we can obtain samples
$\{(\bI_{t+\Delta}^{(j)},w_t^{j}),j=1,\ldots,m\}$ properly
weighted with respect to
$\pi_{t+\Delta}(\bI_{t+\Delta})=p(\bI_{t+\Delta}\mid
\by_{t+\Delta})$, then
\[
\frac{\sum_{j=1}^m
w_t^{(j)}E_{\pi_{t+\Delta}}(x_{t-\delta}\mid
\bI_{t+\Delta}=\bI_{t+\Delta}^{(j)})}{\sum_{j=1}^m w_t^{(j)}}
\]
is a consistent estimator of $E_{\pi_{t+\Delta}}(x_{t-\delta})$,
$\delta=0,1,\ldots\,$. More details of MKF and MKF with lookahead can be
found in \citet{ChenLiu00} and \citet{Wang02SP}.

\begin{table*}
\caption{$\mathrm{MAE}_1$ and $\mathrm{MAE}_2$ for different lookahead methods. The
same numbers of samples ($m=200$) are used.\break The CPU time used in
each experiment is 0.341 seconds for the lookahead weighting
method (MKF, $\Delta=0$);\break 3.554 seconds for the exact lookahead
sampling method (MKF, $\Delta=3$); 0.783 seconds for the pilot
lookahead sampling~method (MKF, $\Delta'=3$), and 0.788 seconds for
MKF-S ($\Delta'=3$)}\label{table5}
\begin{tabular*}{\tablewidth}{@{\extracolsep{4in minus 4in}}lcccccccccc@{}}
\hline
& &\multicolumn{9}{c@{}}{$\bolds{\Delta+\delta/\Delta'+\delta}$} \\[-4pt]
& & \multicolumn{9}{l@{}}{\rule{380pt}{1pt}}\\
& & \textbf{0} & \textbf{1}
& \textbf{2} & \textbf{3} & \textbf{5} & \textbf{8} & \textbf{10} &
\textbf{13} & \textbf{15}\\
\hline
%
%
$\mathrm{MAE}_1$ & MKF ($\Delta=0$) & 1.0300 & 0.7890 & 0.6560 & 0.5830 & 0.5180 & 0.4750
& 0.4590 & 0.4470 & 0.4450 \\
& MKF ($\Delta=3$) & -- & -- & -- & 0.5780 & 0.5150 & 0.4710 &
0.4540 & 0.4410 & 0.4370 \\
& MKF ($\Delta'=3$) & -- & -- & -- & 0.5780 & 0.5150 & 0.4730 & 0.4560 &
0.4440 & 0.4420 \\
& MKF-S ($\Delta'=3$) & -- & -- & -- & 0.5730 & 0.5120 & 0.4690 & 0.4530
& 0.4410 & 0.4370 \\
[6pt]
$\mathrm{MAE}_2$ & MKF ($\Delta=0$) & 0.0932 & 0.0760 & 0.0618 & 0.0525 & 0.0460 & 0.0453
& 0.0467 & 0.0500 & 0.0520 \\
& MKF ($\Delta=3$) & -- & -- & -- & 0.0463 & 0.0357 & 0.0307 &
0.0298 & 0.0298 & 0.0305 \\
& MKF ($\Delta'=3$) & -- & -- & -- & 0.0575 & 0.0503 & 0.0472 & 0.0480 &
0.0503 & 0.0525 \\
& MKF-S ($\Delta'=3$) & -- & -- & -- & 0.0490 & 0.0398 & 0.0360 & 0.0353
& 0.0365 & 0.0375 \\
\hline
\end{tabular*}\vspace*{-4pt}
\end{table*}

In this example, we can also use the smoothing step presented
in Section~\ref{secLookaheadPilot} to improve the performance of
the pilot lookahead sampling method. It can be shown that $\overline
{V}{}^{(j,i)}_{t+1:t+\Delta}=
E(V_{t+1:t+\Delta}^{(j,i)}\mid\bI_{t-1}^{(j)},I_t=i,\by_{t+\Delta})$
in (\ref{EqV}) only depends on
the mean $\mu_t^{(j,i)}$ and the variance $\Sigma_t^{(j,i)}$
of the normal distribution $p(x_t\mid\bI_{t-1}^{(j)},I_t=i,\by_t)$.
For simplicity, we approximately assume $\overline{V}{}^{(j,i)}_{t+1:t+\Delta
}$ only depends on
$\mu_t^{(j,i)}=(\mu_{t,1}^{(j,i)},\mu_{t,2}^{(j,i)})$, that is,
\[
V_{t+1:t+\Delta}^{(j,i)}\approx\overline{V}_{t+1:t+\Delta}\bigl(
\mu_t^{(j,i)}\bigr)+e_t^{(j,i)}.
\]
We then\vspace*{1pt} use the smoothed $\overline{V}{}^{(j,i)}_{t+1:t+\Delta}$
to reduce the variation introduced by random pilots. We denoted this
method by MKF-S.
We used the piecewise cons\-tant smoother to estimate $\overline
V{}^{(j,i)}_{t+1:t+\Delta}$. In the smoother, the space
$ [\min\{\mu_{t,1}^{(j,i)}\},\max\{\mu_{t,1}^{(j,i)}\}
]\times
[\min\{\mu_{t,2}^{(j,i)}\},\break\max\{\mu_{t,2}^{(j,i)}\} ]$ is
divided into
$10\times10$ equal parts.

In this example, we let $\sigma^2=0.1$, $r^2=1.0$, $p_d=0.8$,
$\lambda=0.1$, and $D=100 r$. The length of the observation
period is $T=100$. We repeat the experiment 500 times. The
resampling step is applied when the effective sample size is less
than $0.1 m$.

Following \citet{Avit95}, we use the median absolute error (MAE) as
the performance measurement. Define
\[
\mathrm{MAE}_1=\operatorname{median} \bigl\{|\widehat{x}_{t,1}-x_{t,1}|\bigr\}
\]
and
\[
\mathrm{MAE}_2=\operatorname{median} \bigl\{\bigl|
\widehat{x}_{t,1}-\widetilde{E}_{\pi
_{t+\Delta+\delta}}(x_{t,1})\bigr|
\bigr\},
\]
where $\widehat{x}_{t,1}$ is the consistent estimation of\break
$E_{\pi_{t+\Delta+\delta}}(x_{t,1})$ using different lookahead
methods, and $\widetilde{E}_{\pi_{t+\Delta+\delta}}(x_{t,1})$ is
obtained by the lookahead weighting method using a large number of
samples ($m=20\mbox{,}000$).

We first compare the performance of different look\-ahead methods using
the same number of samples ($m=200$). Table~\ref{table5} reports
$\mathrm{MAE}_1$ and $\mathrm{MAE}_2$ for the lookahead weighting
method (MKF, $\Delta=0$), the exact lookahead sampling method (MKF,
$\Delta=3$) and the single pilot lookahead sampling method (MKF,
$\Delta'=3$ and MKF-S, $\Delta'=3$). From the result, $\mathrm{MAE}_1$
decreases as the number of lookahead steps increases, which shows the
effectiveness of the lookahead strategies. The exact lookahead sampling
method (MKF, $\Delta=3$) has the smallest $\mathrm{MAE}_2$,\vadjust{\goodbreak} which
confirms Pro\-positions~\ref{prop3} and~\ref{prop5}, although
its computational cost is the highest. We can also
see that MKF-S ($\Delta'=3$) performs better than MKF ($\Delta'=3$).

Then we compare the performance of different\break methods
under similar computational cost. The number of samples is
adjusted so that each method takes approximately the same CPU
time. Table~\ref{table6} reports the quantiles of
absolute estimation errors $|\widehat{x}_{t,1}-x_{t,1}|$ for
different lookahead methods with lookahead steps
\mbox{$\Delta+\delta=15$} (or $\Delta'+\delta=15$). The performance does
not improve further when $\Delta'+\delta\geq15$. Under the same CPU time,
the lookahead sampling method has the largest
absolute estimation error because of its high computational cost. The
%
\begin{table*}
\caption{Quantiles of absolute estimation errors $|\widehat
{x}_{t,1}-x_{t,1}|$ for different
lookahead methods. The numbers of samples\break are chosen so that each
method used approximately the same CPU time. In the adaptive pilot
lookahead\break sampling method (MKF-S, $\operatorname{adpt}\Delta'$),\allowbreak the average
number of lookahead steps is 1.572}\label{table6}
\begin{tabular*}{\tablewidth}{@{\extracolsep{4in minus 4in}}lcccccc@{}}
\hline
& \multicolumn{6}{c@{}}{\textbf{Quantiles} $\bolds{(\Delta+\delta=15)}$} \\[-4pt]
& \multicolumn{6}{l@{}}{\rule{366pt}{1pt}}\\
& $\bolds{0.05}$ & $\bolds{0.25}$& $\bolds{0.50}$& $\bolds{0.75}$&
$\bolds{0.95}$& \textbf{Time (sec.)}\\
\hline
%
MKF ($m=450,\Delta=0$) & 0.0400 & 0.2040 & 0.4420 & 0.7910 & 1.7885 &
0.791 \\
MKF ($m=50,\Delta=3$) & 0.0400 & 0.2080 & 0.4490 & 0.8120 & 2.1610 &
0.851 \\
MKF-S ($m=200,\Delta'=3$) & 0.0390 & 0.2030 & 0.4370 & 0.7790 & 1.6590
& 0.788 \\
MKF-S ($m=280$, $\operatorname{adpt}\Delta'$) & 0.0390 & 0.2020 & 0.4340 &
0.7700 & 1.6295 & 0.802 \\
\hline
\end{tabular*}
\end{table*}
pilot lookahead
sampling method (MKF-S, $\Delta'=3$) has better performance than the
simple lookahead weighting meth\-od (MKF, \mbox{$\Delta=0$}).
We then use the stop criteria (\ref{adpt-lookahead-2}) to choose
lookahead steps in the pilot lookahead sampling method adaptively
(MKF-S, $\operatorname{adpt}\Delta'$).\break When we set $\sigma_0^2=1.5 r^2$ in
criteria (\ref{adpt-lookahead-2}), the average number of lookahead
steps is $1.572$. The result
shows that the adaptive pilot lookahead sampling method
performs the best under the same CPU time.

\begin{appendix}\label{app}
\section*{Appendix}

\begin{pf*}{Proof of Proposition~\ref{prop1}}
For any $\Delta_2> \Delta_1\geq0$, we have
\begin{eqnarray*}
\hspace*{-4pt}&&
E_{\pi_t} \bigl[E \bigl(h(\bx_t)\mid\by_{t+\Delta_1}
\bigr)-h(\bx_t) \bigr]^2
\\
\hspace*{-4pt}&&\quad= E_{\pi_t} \bigl[E \bigl(h(\bx_t)\mid\by_{t+\Delta_1}
\bigr)-E \bigl(h(\bx_t)\mid\by_{t+\Delta_2} \bigr)
\bigr]^2\\
\hspace*{-4pt}&&\qquad{} + E_{\pi_t} \bigl[E \bigl(h(\bx_t)\mid
\by_{t+\Delta_2} \bigr) -h(\bx_t) \bigr]^2
\\
\hspace*{-4pt}&&\qquad{}+2E_{\pi_t} \bigl\{ \bigl[E \bigl(h(\bx_t)\mid
\by_{t+\Delta_1} \bigr)-E \bigl(h(\bx_t)\mid\by_{t+\Delta_2}
\bigr) \bigr] \\
\hspace*{-4pt}&&\hspace*{104.5pt}{}\cdot\bigl[E \bigl(h(\bx_t)\mid\by_{t+\Delta_2} \bigr)
-h(\bx_t) \bigr] \bigr\}.
\end{eqnarray*}
Because
\begin{eqnarray*}
\hspace*{-4pt}&&E_{\pi_t} \bigl\{ \bigl[E \bigl(h(\bx_t)\mid
\by_{t+\Delta_1} \bigr)-E \bigl(h(\bx_t)\mid\by_{t+\Delta_2}
\bigr) \bigr]\\[-1pt]
\hspace*{-4pt}&&\hspace*{64pt}{}\cdot \bigl[E \bigl(h(\bx_t)\mid\by_{t+\Delta_2} \bigr)
-h(\bx_t) \bigr] \bigr\}
\\[-1pt]
\hspace*{-4pt}&&\quad= E \bigl\{E \bigl\{ \bigl[E \bigl(h(\bx_t)\mid\by_{t+\Delta_1}
\bigr)-E \bigl(h(\bx_t)\mid\by_{t+\Delta_2} \bigr) \bigr]\\
\hspace*{-4pt}&&\hspace*{32pt}\qquad{}\cdot \bigl[E
\bigl(h(\bx_t)\mid\by_{t+\Delta_2} \bigr) -h(\bx_t)
\bigr] \mid \by_{t+\Delta_2} \bigr\} \mid \by_t \bigr\}\\[-1pt]
\hspace*{-4pt}&&\quad=0,
\end{eqnarray*}
the conclusion holds.
\end{pf*}
\begin{pf*}{Proof of Proposition~\ref{prop2}}
It is easily seen that
\begin{eqnarray*}
&&
E_{\pi_{t+\Delta}} \bigl[w_t^{(2,j)}h\bigl(
\bx_t^{(2,j)}\bigr)\mid\bx_{t-1}^{(j)}
\bigr]\\[-1.2pt]
&&\quad= E_{\pi_{t+\Delta}} \biggl[w_{t-1}^{(j)}
\frac{\pi_{t+\Delta}(\bx_t^{(2,j)})} {
\pi_{t-1}(\bx_{t-1}^{(j)})q_t(x_t^{(2,j)}\mid\bx_{t-1}^{(j)})} \\[-1.2pt]
&&\hspace*{127pt}{}\cdot h\bigl
(\bx_t^{(2,j)}\bigr)\bigm|
\bx_{t-1}^{(j)} \biggr]
\\[-2pt]
&&\quad= w_{t-1}^{(j)}\frac{\pi_{t+\Delta}(\bx_{t-1}^{(j)})} {
\pi_{t-1}(\bx_{t-1}^{(j)})} \\[-1.2pt]
&&\qquad{}\cdot E_{\pi_{t+\Delta}} \biggl[h
\bigl(\bx_t^{(2,j)}\bigr) \frac{\pi_{t+\Delta}(x_t^{(2,j)}\mid\bx
_{t-1}^{(j)})} {
q_t(x_t^{(2,j)}\mid\bx_{t-1}^{(j)})} \Bigm|
\bx_{t-1}^{(j)} \biggr]
\\[-1.2pt]
&&\quad= w_t^{(1,j)}E_{\pi_{t+\Delta}} \bigl(h(\bx_t)
\mid\bx_{t-1}=\bx_{t-1}^{(j)} \bigr).
\end{eqnarray*}
Then (\ref{rao2}) is a
direct result of Rao-Blackwellization. By replacing $h(\bx_t)$
with $E_{\pi_{t+\Delta}} (h(\bx_t)\mid\bx_{t-1} )$ and\break
$h(\bx_t)=1$ in (\ref{rao2}), we obtain (\ref{rao2-1}) and
(\ref{rao2-2}), respectively.
\end{pf*}
\begin{pf*}{Proof of Proposition~\ref{prop3}} Since
\begin{eqnarray*}
&&E_{\pi_{t+\Delta}} \bigl\{w_t^{(3,j)}h\bigl(
\bx_{t-1}^{(j)},x_t^{(2,j)}\bigr)\mid
\bx_{t-1}^{(j)},x_t^{(2,j)} \bigr\}
\\
&&\quad= E_{\pi_{t+\Delta}} \biggl\{h\bigl(\bx_{t-1}^{(j)},x_t^{(2,j)}
\bigr) \\
&&\qquad{}\cdot\biggl[w_{t-1}^{(j)}\frac{\pi_{t+\Delta}(\bx_{t+\Delta
}^{(3,j)})} {
\pi_{t-1}(\bx_{t-1}^{(j)})\prod_{s=t}^{t+\Delta
}q_s(x_{s}^{(3,j)}\mid
\bx_{s-1}^{(3,j)})} \biggr] \Bigm|
\\
&&\qquad\hspace*{152pt}\bx_{t-1}^{(j)},x_t^{(2,j)} \biggr\}
\\
&&\quad= h\bigl(\bx_{t-1}^{(j)},x_t^{(2,j)}\bigr)
w_{t-1}^{(j)}\frac{\pi_{t+\Delta}(\bx_{t-1}^{(j)},x_t^{(2,j)})} {
\pi_{t-1}(\bx_{t-1}^{(j)})q_t(x_t^{(2,j)}\mid\bx_{t-1}^{(j)})}\\
&&\qquad{}\cdot E_{\pi
_{t+\Delta}} \biggl\{
\frac{\pi_{t+\Delta}(\bx_{t+1:t+\Delta}^{(3,j)}\mid
\bx_{t-1}^{(j)},x_t^{(2,j)})} {
\prod_{s=t+1}^{t+\Delta}q_s(x_{s}^{(3,j)}\mid
\bx_{s-1}^{(3,j)})} \Bigm|\\
&&\qquad\hspace*{129pt} \bx_{t-1}^{(j)},x_t^{(2,j)}
\biggr\}
\\
&&\quad= w_{t}^{(2,j)} h\bigl(\bx_{t-1}^{(j)},x_t^{(2,j)}
\bigr),
\end{eqnarray*}
we have
\begin{eqnarray*}
&&
\operatorname{var}_{\pi_{t+\Delta}} \bigl[w_{t}^{(3,j)}h\bigl(
\bx_{t-1}^{(j)},x_t^{(2,j)}\bigr) \bigr]\\
&&\quad\geq
\operatorname{var}_{\pi_{t+\Delta}} \bigl[w_t^{(2,j)}h\bigl(
\bx_{t-1}^{(j)},x_t^{(2,j)}\bigr) \bigr]
\end{eqnarray*}
according to the Rao-Blackwellization theorem.
\end{pf*}
\begin{pf*}{Proof of Proposition~\ref{prop4}} To prove
$(\bx_t^{(j)},\break w_t^{\mathrm{aux}(j)})$ is properly weighted with respect to
distribution $\pi_{t+\Delta}(\bx_{t})$, we only need to prove
\[
E_{\pi_{t+\Delta}} \bigl[w_t^{\mathrm{aux}(j)}h\bigl(
\bx_t^{(j)}\bigr) \bigr]=E_{\pi
_{t+\Delta}} \bigl[h(
\bx_t) \bigr].
\]

According to the sampling distribution of the lookahead pilot
$\bx_{t+\Delta}^{(j,i)}$ and calculation of the corresponding
cumulative incremental weight $U_t^{(j,i)}$,
\begin{eqnarray*}
&&
E_{\pi_{t+\Delta}} \bigl[w_t^{\mathrm{aux}(j)}h\bigl(
\bx_t^{(j)}\bigr)\mid\bx_{t-1}^{(j)}
\bigr]\\
&&\quad= E_{\pi_{t+\Delta}} \Biggl[w_{t-1}^{(j)}\sum
_{i=1}^{\mathcal{A}} U_t^{(j,i)} h\bigl(
\bx_{t-1}^{(j)},x_t=a_i\bigr)\mid
\bx_{t-1}^{(j)} \Biggr]
\\
&&\quad= w_{t-1}^{(j)}\sum_{i=1}^{\mathcal{A}}
\frac{
\pi_{t+\Delta}(\bx_{t-1}^{(j)},x_t=a_i)h(\bx_{t-1}^{(j)},x_t=a_i)}{\pi
_{t-1}(\bx_{t-1}^{(j)})}
\\
&&\quad= w_{t-1}^{(j)} \frac{\pi_{t+\Delta}(\bx_{t-1}^{(j)})}{\pi_{t-1}(\bx
_{t-1}^{(j)})} E_{\pi_{t+\Delta}} \bigl[h(
\bx_t)\mid\bx_{t-1}^{(j)} \bigr].
\end{eqnarray*}
Because sample $(\bx_{t-1}^{(j)},w_{t-1}^{(j)})$ is properly
weighted with respect to $\pi_{t-1}(\bx_{t-1})$,
\begin{eqnarray*}
&&
E_{\pi_{t+\Delta}} \biggl[w_{t-1}^{(j)}
\frac{\pi_{t+\Delta}(\bx_{t-1}^{(j)})}{\pi_{t-1}(\bx_{t-1}^{(j)})}
E_{\pi_{t+\Delta}} \bigl(h(\bx_t)\mid
\bx_{t-1}^{(j)} \bigr) \biggr]\\
&&\quad=E_{\pi_{t-1}} \biggl[
\frac{\pi_{t+\Delta}(\bx_{t-1})}{\pi_{t-1}(\bx_{t-1})} E_{\pi_{t+\Delta
}} \bigl(h(\bx_t)\mid
\bx_{t-1} \bigr) \biggr]
\\
&&\quad=E_{\pi_{t+\Delta}} \bigl[h(\bx_t) \bigr];
\end{eqnarray*}
the proposition follows.
\end{pf*}
\begin{pf*}{Proof of Proposition~\ref{prop5}} Because
$w_t^{\mathrm{aux}(j)}$ and $w_t^{(1,j)}$ both are importance weights, we
have\break
$E_{\pi_{t+\Delta}} (w_t^{\mathrm{aux}(j)} )=E_{\pi_{t+\Delta
}}
(w_t^{(1,j)} )=1$.
Hence,
\begin{eqnarray*}
&&
\operatorname{var}_{\pi_{t+\Delta}} \bigl( w_t^{\mathrm{aux}(j)} \bigr)-
\operatorname{var}_{\pi_{t+\Delta}} \bigl(w_t^{(1,j)} \bigr) \\
&&\quad=
E_{\pi_{t+\Delta}} \bigl( w_t^{\mathrm{aux}(j)} \bigr)^2-
E_{\pi_{t+\Delta}} \bigl(w_t^{(1,j)} \bigr)^2
\\
&&\quad= E_{\pi_{t+\Delta}} \bigl\{E_{\pi_{t+\Delta}} \bigl[ \bigl(
w_t^{\mathrm{aux}(j)} \bigr)^2 \mid \bx_{t-1}^{(j)}
\bigr]\\
&&\qquad\hspace*{30.2pt}{}-E_{\pi_{t+\Delta}} \bigl[ \bigl( w_t^{(1,j)} \bigr)^2 \mid
\bx_{t-1}^{(j)} \bigr] \bigr\}.
\end{eqnarray*}
Now we consider the difference between\break
$E_{\pi_{t+\Delta}} [ ( w_t^{\mathrm{aux}(j)} )^2 \mid
\bx_{t-1}^{(j)} ]$ and $E_{\pi_{t+\Delta}} [ (
w_t^{(1,j)} )^2 \mid \bx_{t-1}^{(j)} ]$. Let
\[
\varepsilon^{(j,i,k)}=U_t^{(j,i,k)}-\frac{\pi_{t+\Delta}(\bx
_{t-1}^{(j)},x_t=a_i)}{\pi_{t-1}(\bx_{t-1}^{(j)})}.
\]
Because
\[
E_{\pi_{t+\Delta}} \bigl[ U_t^{(j,i,k)} \mid
\bx_{t-1}^{(j)}
\bigr]=\frac{\pi_{t+\Delta}(\bx_{t-1}^{(j)},x_t=a_i)}{\pi_{t-1}(\bx_{t-1}^{(j)})},
\]
we have $E_{\pi_{t+\Delta}} [ \varepsilon^{(j,i,k)}
\mid \bx_{t-1}^{(j)} ]=0$. In addition,
$\varepsilon^{(j,i,k)}$, $i=1,\ldots,\mathcal{A}$,
$k=1,\ldots,K$ are independent condition\-al on $\bx_{t-1}^{(j)}$, and
for fixed $i$, $\varepsilon^{(j,i,k)}$, $k=1,\ldots,K$, follow the
same distribution:
\begin{eqnarray*}
&&
E_{\pi_{t+\Delta}} \bigl[ \bigl( w_t^{\mathrm{aux}(j)}
\bigr)^2 \mid \bx_{t-1}^{(j)} \bigr]
\\
&&\quad= \bigl(w_{t-1}^{(j)} \bigr)^2E_{\pi_{t+\Delta}}
\Biggl[ \Biggl( \sum_{i=1}^{\mathcal{A}}
\frac{\pi_{t+\Delta}(\bx_{t-1}^{(j)},x_t=a_i)}{\pi_{t-1}(\bx
_{t-1}^{(j)})}\\
&&\qquad\hspace*{82pt}{}+\sum_{i=1}^{\mathcal{A}}
\frac{1}{K}\sum_{k=1}^K
\varepsilon^{(j,i,k)} \Biggr)^2 \Bigm| \bx_{t-1}^{(j)}
\Biggr]
\\
&&\quad= \bigl(w_{t-1}^{(j)} \bigr)^2E_{\pi_{t+\Delta}}
\biggl[ \biggl( \frac{\pi_{t+\Delta}(\bx_{t-1}^{(j)})}{\pi_{t-1}(\bx
_{t-1}^{(j)})} \biggr)^2 \Bigm| \bx_{t-1}^{(j)}
\biggr] \\
&&\qquad\hspace*{0pt}{}+ \bigl(w_{t-1}^{(j)} \bigr)^2E_{\pi_{t+\Delta}}
\Biggl[ \Biggl( \sum_{i=1}^{\mathcal{A}}
\frac{1}{K}\sum_{k=1}^K
\varepsilon^{(j,i,k)} \Biggr)^2 \Bigm| \bx_{t-1}^{(j)}
\Biggr]
\\
&&\quad= E_{\pi_{t+\Delta}} \bigl[ \bigl( w_t^{(1,j)}
\bigr)^2 \mid \bx_{t-1}^{(j)} \bigr] \\
&&\qquad{}+
\frac{1}{K^2} \bigl(w_{t-1}^{(j)} \bigr)^2 \sum
_{i=1}^{\mathcal{A}} \sum
_{k=1}^K E_{\pi_{t+\Delta}} \bigl[ \bigl(
\varepsilon^{(j,i,k)} \bigr)^2 \mid \bx_{t-1}^{(j)}
\bigr]
\\
&&\quad= E_{\pi_{t+\Delta}} \bigl[ \bigl( w_t^{(1,j)}
\bigr)^2 \mid \bx_{t-1}^{(j)} \bigr]\\
&&\qquad{} +
\frac{1}{K} \bigl(w_{t-1}^{(j)} \bigr)^2 \sum
_{i=1}^{\mathcal{A}} E_{\pi_{t+\Delta}} \bigl[ \bigl(
\varepsilon^{(i,j,k=1)} \bigr)^2 \mid \bx_{t-1}^{(j)}
\bigr].
\end{eqnarray*}
Then we have
\begin{eqnarray*}
&&
E_{\pi_{t+\Delta}} \bigl[ \bigl( w_t^{\mathrm{aux}(j)}
\bigr)^2 \mid \bx_{t-1}^{(j)} \bigr]-E_{\pi_{t+\Delta}}
\bigl[ \bigl( w_t^{(1,j)} \bigr)^2 \mid
\bx_{t-1}^{(j)} \bigr]\\
&&\quad=\frac{1}{K} \bigl(w_{t-1}^{(j)}
\bigr)^2 \sum_{i=1}^{\mathcal{A}}
E_{\pi_{t+\Delta}} \bigl[ \bigl(\varepsilon^{(i,j,k=1)} \bigr)^2 \mid
\bx_{t-1}^{(j)} \bigr],
\end{eqnarray*}
hence,
\begin{eqnarray*}
0&\leq& \operatorname{var}_{\pi_{t+\Delta}} \bigl( w_t^{\mathrm{aux}(j)} \bigr)-
\operatorname{var}_{\pi_{t+\Delta}} \bigl(w_t^{(1,j)} \bigr) \\
&\sim& O(1/K).
\end{eqnarray*}
With a similar method, we can prove
\begin{eqnarray*}
0&\leq& \operatorname{var}_{\pi_{t+\Delta}} \Biggl[ w_{t-1}^{(j)}\sum
_{i=1}^{\mathcal{A}}\frac{1}{K}\sum
_{k=1}^K U_t^{(j,i,k)} h \bigl(
\bx_{t-1}^{(j)},x_t=a_i \bigr) \Biggr]
\\
&&{} - \operatorname{var}_{\pi_{t+\Delta}} \bigl[w_t^{(1,j)}E_{\pi
_{t+\Delta}}
\bigl(h\bigl(x_{t-1}^{(j)},x_t\bigr)\mid
x_{t-1}^{(j)} \bigr) \bigr] \\
&\sim& O(1/K).
\end{eqnarray*}
\upqed\end{pf*}
\begin{pf*}{Proof of Proposition~\ref{prop6}} Let
$\psi_t(\bx_t)\propto r_t(\bx_t)\cdot\allowbreak b_t(\bx_t)$ be the distribution of
samples after resampling, Because
$E_{\pi_{t+T}} (w_{t+T}^{*(j)} ) \equiv1$, we only
consider minimizing $E_{\pi_{t+T}} (w_{t+T}^{*(j)} )^2$.
We have
\begin{eqnarray*}
&&E_{\pi_{t+T}} \bigl(w_{t+T}^{*(j)} \bigr)^2
\\
&&\quad=\int\frac{\pi_t^2(\bx_t)}{\psi_t^2(\bx_t)} \biggl[\frac{\pi_{t+T}(\bx
_{t+T})}{\pi_t(\bx_t)\prod_{s=t+1}^{t+T}
q_s(x_s \mid
\bx_{s-1})} \biggr]^2
\psi_t(\bx_t)\\
&&\hspace*{8.8pt}\qquad{}\cdot\prod_{s=t+1}^{t+T}
q_s(x_s \mid\bx_{s-1}) \,d\bx_t
\,dx_{t+1} \cdots dx_{t+T}
\\
&&\quad= \int\frac{\pi_t^2(\bx_t)}{\psi_t(\bx_t)} \eta(\bx_t) \,d\bx_t.
\end{eqnarray*}
According\vspace*{1pt} to Jensen's inequality, to minimize\break
$E_{\pi_{t+T}} (w_{t+T}^{*(j)} )^2$, $\psi_t(\bx_t)$
needs to be\vspace*{1pt} proportional to
$\pi_t(\bx_t)\eta^{1/2}(\bx_t)$.\
\end{pf*}
\begin{pf*}{Proof of Proposition~\ref{prop7}} For estimator
$\widehat{h}^*$, because $\widehat{h}^*$ and $h(\bx_{t})$ are
independent conditional on $\by_{t+\Delta+1}$, we have
\begin{eqnarray*}
\hspace*{-4pt}&&E \bigl[ \bigl(\widehat{h}^*-h(\bx_{t}) \bigr)^2\mid
\by_{t+\Delta
} \bigr]
\\
\hspace*{-4pt}&&\quad=E \bigl\{E \bigl[ \bigl[\widehat{h}^*-E\bigl(h(\bx_{t})\mid
\by_{t+\Delta+1}\bigr) \\
\hspace*{-4pt}&&\hspace*{0pt}\qquad{}+ E\bigl(h(\bx_{t})\mid\by_{t+\Delta+1}
\bigr)- h(\bx_{t}) \bigr]^2\mid\by_{t+\Delta+1} \bigr]
\mid \by_{t+\Delta
} \bigr\}
\\
\hspace*{-4pt}&&\quad=E \bigl[\operatorname{var} \bigl(\widehat{h}^*\mid\by_{t+\Delta+1} \bigr) \mid
\by_{t+\Delta} \bigr]\\
\hspace*{-4pt}&&\hspace*{0pt}\qquad{} + E \bigl[ \bigl[E\bigl(h(\bx_{t})\mid
\by_{t+\Delta+1}\bigr)- h(\bx_{t}) \bigr]^2\mid
\by_{t+\Delta} \bigr]
\end{eqnarray*}
and for estimator $\widehat{h}$,
\begin{eqnarray*}
&&E \bigl[ \bigl(\widehat{h}-h(\bx_{t}) \bigr)^2 \mid
\by_{t+\Delta} \bigr]\\
&&\quad=\operatorname{var} [\widehat{h}\mid\by_{t+\Delta} ]\\
&&\qquad{}+ E \bigl[
\bigl[E\bigl(h(\bx_{t})\mid\by_{t+\Delta}\bigr)- h(
\bx_{t}) \bigr]^2\mid\by_{t+\Delta} \bigr].
\end{eqnarray*}
Similar to the proof of Proposition~\ref{prop4}, we have
\begin{eqnarray*}
&&E \bigl[ \bigl[E\bigl(h(\bx_{t})\mid\by_{t+\Delta}\bigr)- h(
\bx_{t}) \bigr]^2\mid\by_{t+\Delta} \bigr]\\[1.2pt]
&&\qquad{}-E \bigl[
\bigl[E\bigl(h(\bx_{t})\mid\by_{t+\Delta+1}\bigr)- h(
\bx_{t}) \bigr]^2\mid\by_{t+\Delta} \bigr]
\\[1.2pt]
&&\quad= \operatorname{var} \bigl[E \bigl( h(\bx_{t})\mid\by_{t+\Delta+1} \bigr) \mid
\by_{t+\Delta} \bigr].
\end{eqnarray*}
Hence,
%
\renewcommand{\theequation}{\arabic{equation}}
\begin{eqnarray}
\label{P7-1}
&&
E \bigl[ \bigl(\widehat{h}^*-h(\bx_{t})
\bigr)^2\mid\by_{t+\Delta
} \bigr]-E \bigl[ \bigl(\widehat{h}-h(
\bx_{t}) \bigr)^2 \mid\by_{t+\Delta} \bigr]
\nonumber\\[1.2pt]
&&\quad=E \bigl[\operatorname{var} \bigl(\widehat{h}^*\mid\by_{t+\Delta+1} \bigr) \mid
\by_{t+\Delta} \bigr]
-\operatorname{var} [\widehat{h}\mid\by_{t+\Delta} ]\nonumber\\[1.2pt]
&&\qquad{}-\operatorname{var} \bigl[E
\bigl( h(\bx_{t})\mid\by_{t+\Delta+1} \bigr) \mid  \by_{t+\Delta}
\bigr]
\nonumber\\[1.2pt]
&&\quad=\frac{1}{m} E \Biggl[\operatorname{var} \Biggl(w_{t-1}^{(j)}\sum
_{i=1}^{\mathcal{A}} U_{t,\Delta+1}^{(j,i)}\nonumber\\[1.2pt]
&&\qquad\hspace*{80pt}{}\cdot h
\bigl(\bx_{t-1}^{(j,i)}\bigr) \mid  \by_{t+\Delta+1} \Biggr) \Bigm|
\by_{t+\Delta} \Biggr]
\nonumber
\\[1.2pt]
&&\qquad{}-\frac{1}{m} \operatorname{var} \Biggl[w_{t-1}^{(j)}\sum
_{i=1}^{\mathcal{A}} U_{t,\Delta}^{(j,i)}h\bigl(
\bx_{t-1}^{(j,i)}\bigr) \mid  \by_{t+\Delta}
\Biggr]\\[1.2pt]
&&\qquad{}-\operatorname{var} \bigl[E
\bigl( h(\bx_{t})\mid\by_{t+\Delta+1} \bigr) \mid
\by_{t+\Delta} \bigr]
\nonumber\\[1.2pt]
&&\quad=\frac{1}{m} \Biggl\{\operatorname{var} \Biggl[w_{t-1}^{(j)}\sum
_{i=1}^{\mathcal{A}} U_{t,\Delta+1}^{(j,i)}h
\bigl(\bx_{t-1}^{(j,i)}\bigr) \mid  \by_{t+\Delta} \Biggr]\nonumber\\[1.2pt]
&&\hspace*{15pt}\qquad{}-\operatorname{var}
\bigl[E \bigl( h(\bx_{t})\mid\by_{t+\Delta+1} \bigr) \mid
\by_{t+\Delta} \bigr]
\nonumber
\\[1.2pt]
&&\hspace*{15.1pt}\qquad{}-\operatorname{var} \Biggl[w_{t-1}^{(j)}\sum_{i=1}^{\mathcal{A}}
U_{t,\Delta}^{(j,i)}h\bigl(\bx_{t-1}^{(j,i)}\bigr)
\mid  \by_{t+\Delta} \Biggr] \Biggr\}\nonumber\\[1.2pt]
&&\qquad{}-\operatorname{var} \bigl[E \bigl( h(\bx_{t})
\mid\by_{t+\Delta+1} \bigr) \mid  \by_{t+\Delta} \bigr]
\nonumber\\[1.2pt]
&&\quad=\frac{1}{m} \Biggl\{\operatorname{var} \Biggl[w_{t-1}^{(j)}\sum
_{i=1}^{\mathcal{A}} U_{t,\Delta+1}^{(j,i)}h
\bigl(\bx_{t-1}^{(j,i)}\bigr) \mid  \by_{t+\Delta} \Biggr]\nonumber\\[1.2pt]
&&\hspace*{15.1pt}\qquad{}-\operatorname{var}
\Biggl[w_{t-1}^{(j)}\sum_{i=1}^{\mathcal{A}}
U_{t,\Delta}^{(j,i)}h\bigl(\bx_{t-1}^{(j,i)}\bigr)
\mid  \by_{t+\Delta} \Biggr] \Biggr\}
\nonumber\\[1.2pt]
&&\qquad{}- \biggl(1+\frac{1}{m} \biggr) \operatorname{var} \bigl[E \bigl( h(\bx_{t})
\mid\by_{t+\Delta+1} \bigr) \mid  \by_{t+\Delta} \bigr].\nonumber
\end{eqnarray}
In the pilot lookahead sampling method,
because
\begin{eqnarray*}
U_{t,\Delta+1}^{(j,i)}&=&U_{t,\Delta}^{(j,i)}
{p\bigl(\bx_{t+\Delta
+1}^{(j,i)}\mid\by_{t+\Delta+1}\bigr)}\\[1.2pt]
&&{}\Big/ \biggl(
p\bigl(\bx_{t+\Delta}^{(j,i)}\mid
\by_{t+\Delta}\bigr)\\[1.2pt]
&&\hspace*{13pt}{}\cdot q_{t+\Delta+1}\bigl(x_{t+\Delta+1}^{(j,i)}\mid
\bx_{t+\Delta}^{(j,i)},\by_{t+\Delta+1}\bigr)\biggr),%
\end{eqnarray*}
we have
\begin{eqnarray*}
&&E \bigl(U_{t,\Delta+1}^{(j,i)}h\bigl(\bx_{t}^{(j,i)}
\bigr)\mid\bx_{t+\Delta}^{(j,i)},\by_{t+\Delta} \bigr)
\\
&&\quad= U_{t,\Delta}^{(j,i)}h\bigl(\bx_{t}^{(j,i)}
\bigr)\\
&&\qquad{}\cdot\int\bigl(p\bigl(\bx_{t+\Delta+1}^{(j,i)}\mid\by_{t+\Delta+1}\bigr)\\
&&\qquad\hspace*{22pt}{}/
\bigl(p\bigl(\bx_{t+\Delta}^{(j,i)}\mid
\by_{t+\Delta}\bigr)\\
&&\hspace*{52.5pt}{}\cdot q_{t+\Delta+1}\bigl(x_{t+\Delta+1}^{(j,i)}\mid
x_{t+\Delta}^{(j,i)},\by_{t+\Delta+1}\bigr)\bigr)\bigr)
\\
&&\hspace*{14.5pt}\qquad{} \cdot q_{t+\Delta+1}\bigl(x_{t+\Delta+1}^{(j,i)}
\mid x_{t+\Delta}^{(j,i)},\by_{t+\Delta+1}\bigr) \\
&&\qquad\hspace*{14.5pt}{}\cdot p(y_{t+\Delta+1}
\mid\by_{t+\Delta}) \,dx_{t+\Delta+1}^{(j,i)}\,dy_{t+\Delta+1}
\\
&&\quad= U_{t,\Delta}^{(j,i)}h\bigl(\bx_{t}^{(j,i)}
\bigr),
\end{eqnarray*}
according to the Rao-Blackwellization theorem,
%
\begin{eqnarray}
\label{P7-2}
&&
\operatorname{var} \Biggl[w_{t-1}^{(j)}\sum
_{i=1}^{\mathcal{A}} U_{t,\Delta+1}^{(j,i)}h\bigl(
\bx_{t-1}^{(j,i)}\bigr) \mid  \by_{t+\Delta} \Biggr] \nonumber\\
&&\qquad{}- \operatorname{var}
\Biggl[w_{t-1}^{(j)}\sum_{i=1}^{\mathcal{A}}
U_{t,\Delta}^{(j,i)}h\bigl(\bx_{t-1}^{(j,i)}\bigr)
\mid  \by_{t+\Delta} \Biggr]
\nonumber\\[-8pt]\\[-8pt]
&&\quad= E \Biggl[\operatorname{var} \Biggl(w_{t-1}^{(j)}\sum
_{i=1}^{\mathcal{A}} U_{t,\Delta+1}^{(j,i)}\nonumber\\
&&\hspace*{60pt}{}\cdot h\bigl(
\bx_{t-1}^{(j,i)}\bigr)\mid\bx_{t+\Delta}^{(j,i=1:\mathcal{A})},
\by_{t+\Delta} \Biggr) \Bigm|  \by_{t+\Delta} \Biggr].\nonumber
\end{eqnarray}
Combining (\ref{P7-1}) and (\ref{P7-2}), the conclusion holds.
\end{pf*}
\end{appendix}

\section*{Acknowledgments}

Ming Lin's research is supported by the National Nature Science
Foundation of China Grant 11101341 and by the Fundamental Research
Funds for the Central Universities 2010221093. Rong Chen's research is
sponsored in part by NSF Grants DMS-08-00183, DMS-09-05076 and
DMS-09-15139. Jun Liu's research was supported in part by NSF Grants
DMS-07-06989 and DMS-10-07762.


%


\begin{thebibliography}{47}

\bibitem[\protect\citeauthoryear{Andrieu, Doucet and
Holenstein}{2010}]{Andrieu2010}
%
\begin{barticle}[mr]
\bauthor{\bsnm{Andrieu},~\bfnm{Christophe}\binits{C.}},
\bauthor{\bsnm{Doucet},~\bfnm{Arnaud}\binits{A.}} \AND
\bauthor{\bsnm{Holenstein},~\bfnm{Roman}\binits{R.}}
(\byear{2010}).
\btitle{Particle {M}arkov chain {M}onte {C}arlo methods}.
\bjournal{J. R. Stat. Soc. Ser. B Stat. Methodol.}
\bvolume{72}
\bpages{269--342}.
\bid{doi={10.1111/j.1467-9868.2009.00736.x}, issn={1369-7412}, mr={2758115}}
\bptok{imsref}%
\end{barticle}
%
\endbibitem

\bibitem[\protect\citeauthoryear{Avitzour}{1995}]{Avit95}
%
\begin{barticle}[author]
\bauthor{\bsnm{Avitzour},~\bfnm{D.}\binits{D.}}
(\byear{1995}).
\btitle{Stochastic simulation Bayesian approach to multitarget tracking}.
\bjournal{IEE Proceedings on Radar, Sonar and Navigation}
\bvolume{142}
\bpages{41--44}.
\bptok{imsref}%
\end{barticle}
%
\endbibitem

\bibitem[\protect\citeauthoryear{Briers, Doucet and
Maskell}{2010}]{Briers2010}
%
\begin{barticle}[mr]
\bauthor{\bsnm{Briers},~\bfnm{Mark}\binits{M.}},
\bauthor{\bsnm{Doucet},~\bfnm{Arnaud}\binits{A.}} \AND
\bauthor{\bsnm{Maskell},~\bfnm{Simon}\binits{S.}}
(\byear{2010}).
\btitle{Smoothing algorithms for state-space models}.
\bjournal{Ann. Inst. Statist. Math.}
\bvolume{62}
\bpages{61--89}.
\bid{doi={10.1007/s10463-009-0236-2}, issn={0020-3157}, mr={2577439}}
\bptok{imsref}%
\end{barticle}
%
\endbibitem

\bibitem[\protect\citeauthoryear{Carpenter, Clifford and
Fearnhead}{1999}]{Carpenter99}
%
\begin{barticle}[author]
\bauthor{\bsnm{Carpenter},~\bfnm{J.}\binits{J.}},
\bauthor{\bsnm{Clifford},~\bfnm{P.}\binits{P.}} \AND
\bauthor{\bsnm{Fearnhead},~\bfnm{P.}\binits{P.}}
(\byear{1999}).
\btitle{An improved particle for non-linear problems}.
\bjournal{IEE Proceedings on Radar, Sonar and Navigation}
\bvolume{146}
\bpages{2--7}.
\bptok{imsref}%
\end{barticle}
%
\endbibitem

\bibitem[\protect\citeauthoryear{Carter and Kohn}{1994}]{CartKohn94}
%
\begin{barticle}[mr]
\bauthor{\bsnm{Carter},~\bfnm{C.~K.}\binits{C.~K.}} \AND
\bauthor{\bsnm{Kohn},~\bfnm{R.}\binits{R.}}
(\byear{1994}).
\btitle{On {G}ibbs sampling for state space models}.
\bjournal{Biometrika}
\bvolume{81}
\bpages{541--553}.
\bid{doi={10.1093/biomet/81.3.541}, issn={0006-3444}, mr={1311096}}
\bptok{imsref}%
\end{barticle}
%
\endbibitem

\bibitem[\protect\citeauthoryear{Carvalho et~al.}{2010}]{Carvalho2010}
%
\begin{barticle}[mr]
\bauthor{\bsnm{Carvalho},~\bfnm{Carlos~M.}\binits{C.~M.}},
\bauthor{\bsnm{Johannes},~\bfnm{Michael~S.}\binits{M.~S.}},
\bauthor{\bsnm{Lopes},~\bfnm{Hedibert~F.}\binits{H.~F.}} \AND
\bauthor{\bsnm{Polson},~\bfnm{Nicholas~G.}\binits{N.~G.}}
(\byear{2010}).
\btitle{Particle learning and smoothing}.
\bjournal{Statist. Sci.}
\bvolume{25}
\bpages{88--106}.
\bid{doi={10.1214/10-STS325}, issn={0883-4237}, mr={2741816}}
\bptok{imsref}%
\end{barticle}
%
\endbibitem

\bibitem[\protect\citeauthoryear{Chen and Liu}{2000}]{ChenLiu00}
%
\begin{barticle}[mr]
\bauthor{\bsnm{Chen},~\bfnm{Rong}\binits{R.}} \AND
\bauthor{\bsnm{Liu},~\bfnm{Jun~S.}\binits{J.~S.}}
(\byear{2000}).
\btitle{Mixture {K}alman filters}.
\bjournal{J. R. Stat. Soc. Ser. B Stat. Methodol.}
\bvolume{62}
\bpages{493--508}.
\bid{doi={10.1111/1467-9868.00246}, issn={1369-7412}, mr={1772411}}
\bptok{imsref}%
\end{barticle}
%
\endbibitem

\bibitem[\protect\citeauthoryear{Chen, Wang and Liu}{2000}]{Chen00}
%
\begin{barticle}[mr]
\bauthor{\bsnm{Chen},~\bfnm{Rong}\binits{R.}},
\bauthor{\bsnm{Wang},~\bfnm{Xiaodong}\binits{X.}} \AND
\bauthor{\bsnm{Liu},~\bfnm{Jun~S.}\binits{J.~S.}}
(\byear{2000}).
\btitle{Adaptive joint detection and decoding in flat-fading channels via
mixture {K}alman filtering}.
\bjournal{IEEE Trans. Inform. Theory}
\bvolume{46}
\bpages{2079--2094}.
\bid{doi={10.1109/18.868479}, issn={0018-9448}, mr={1781368}}
\bptok{imsref}%
\end{barticle}
%
\endbibitem

\bibitem[\protect\citeauthoryear{Chopin}{2002}]{Chopin2002}
%
\begin{barticle}[mr]
\bauthor{\bsnm{Chopin},~\bfnm{Nicolas}\binits{N.}}
(\byear{2002}).
\btitle{A sequential particle filter method for static models}.
\bjournal{Biometrika}
\bvolume{89}
\bpages{539--551}.
\bid{doi={10.1093/biomet/89.3.539}, issn={0006-3444}, mr={1929161}}
\bptok{imsref}%
\end{barticle}
%
\endbibitem

\bibitem[\protect\citeauthoryear{Chopin}{2004}]{Chopin04}
%
\begin{barticle}[mr]
\bauthor{\bsnm{Chopin},~\bfnm{Nicolas}\binits{N.}}
(\byear{2004}).
\btitle{Central limit theorem for sequential {M}onte {C}arlo methods
and its
application to {B}ayesian inference}.
\bjournal{Ann. Statist.}
\bvolume{32}
\bpages{2385--2411}.
\bid{doi={10.1214/009053604000000698}, issn={0090-5364}, mr={2153989}}
\bptok{imsref}%
\end{barticle}
%
\endbibitem

\bibitem[\protect\citeauthoryear{Clapp and Godsill}{1997}]{ClappGodsill97}
%
\begin{bmisc}[author]
\bauthor{\bsnm{Clapp},~\bfnm{T.~C.}\binits{T.~C.}} \AND
\bauthor{\bsnm{Godsill},~\bfnm{S.~J.}\binits{S.~J.}}
(\byear{1997}).
\bhowpublished{Bayesian blind deconvolution for mobile communications.
In \textit{Proceedings of IEE Colloqium on Adaptive Signal Processing
for Mobile
Communication Systems}
\textbf{9} 1--6. IET}.
\bptok{imsref}%
\end{bmisc}
%
\endbibitem

\bibitem[\protect\citeauthoryear{Clapp and Godsill}{1999}]{ClappGodsill99}
%
\begin{bincollection}[author]
\bauthor{\bsnm{Clapp},~\bfnm{T.~C.}\binits{T.~C.}} \AND
\bauthor{\bsnm{Godsill},~\bfnm{S.~J.}\binits{S.~J.}}
(\byear{1999}).
\btitle{Fixed-lag smoothing using sequential importance sampling}.
In \bbooktitle{Bayesian Statistics 6}
(\beditor{\bfnm{J.~M.}\binits{J.~M.}~\bsnm{Bernardo}},
\beditor{\bfnm{J.~O.}\binits{J.~O.}~\bsnm{Berger}},
\beditor{\bfnm{A.~P.}\binits{A.~P.}~\bsnm{Dawid}} \AND
\beditor{\bfnm{A.~F.~M.}\binits{A.~F.~M.}~\bsnm{Smith}}, eds.)
\bpages{743--752}.
\bpublisher{Oxford Univ. Press}, \baddress{Oxford}.
\bptok{imsref}%
\end{bincollection}
%
\endbibitem

\bibitem[\protect\citeauthoryear{Crisan and Lyons}{2002}]{CrisanLyons2002}
%
\begin{barticle}[mr]
\bauthor{\bsnm{Crisan},~\bfnm{Dan}\binits{D.}} \AND
\bauthor{\bsnm{Lyons},~\bfnm{Terry}\binits{T.}}
(\byear{2002}).
\btitle{Minimal entropy approximations and optimal algorithms}.
\bjournal{Monte Carlo Methods Appl.}
\bvolume{8}
\bpages{343--355}.
\bid{doi={10.1515/mcma.2002.8.4.343}, issn={0929-9629}, mr={1943203}}
\bptok{imsref}%
\end{barticle}
%
\endbibitem

\bibitem[\protect\citeauthoryear{Del~Moral}{2004}]{DelMoral04}
%
\begin{bbook}[mr]
\bauthor{\bsnm{Del~Moral},~\bfnm{Pierre}\binits{P.}}
(\byear{2004}).
\btitle{Feynman--{K}ac Formulae:
Genealogical and Interacting Particle Systems with Applications}.
\bpublisher{Springer}, \baddress{New York}.
\bid{doi={10.1007/978-1-4684-9393-1}, mr={2044973}}
\bptok{imsref}%
\end{bbook}
%
\endbibitem

\bibitem[\protect\citeauthoryear{Douc et~al.}{2009}]{Douc2012}
%
\begin{bmisc}[author]
\bauthor{\bsnm{Douc},~\bfnm{R.}\binits{R.}},
\bauthor{\bsnm{Garivier},~\bfnm{E.}\binits{E.}},
\bauthor{\bsnm{Moulines},~\bfnm{E.}\binits{E.}} \AND
\bauthor{\bsnm{Olsson},~\bfnm{J.}\binits{J.}}
(\byear{2009}).
\bhowpublished{On the forward filtering backward smoothing particle
approximations of
the smoothing distribution in general state space models.
Working paper, Institut T\'{e}l\'{e}com, Paris.}
\bptok{imsref}%
\end{bmisc}
%
\endbibitem

\bibitem[\protect\citeauthoryear{Doucet, Briers and S{\'e}n{\'e}cal}{2006}]{Doucet2006}
%
\begin{barticle}[mr]
\bauthor{\bsnm{Doucet},~\bfnm{Arnaud}\binits{A.}},
\bauthor{\bsnm{Briers},~\bfnm{Mark}\binits{M.}} \AND
\bauthor{\bsnm{S{\'e}n{\'e}cal},~\bfnm{St{\'e}phane}\binits{S.}}
(\byear{2006}).
\btitle{Efficient block sampling strategies for sequential {M}onte {C}arlo
methods}.
\bjournal{J. Comput. Graph. Statist.}
\bvolume{15}
\bpages{693--711}.
\bid{doi={10.1198/106186006X142744}, issn={1061-8600}, mr={2280152}}
\bptok{imsref}%
\end{barticle}
%
\endbibitem

\bibitem[\protect\citeauthoryear{Doucet, de~Freitas and
Gordon}{2001}]{Douc01}
%
\begin{bbook}[mr]
\beditor{\bsnm{Doucet},~\bfnm{Arnaud}\binits{A.}}, \beditor
{\bparticle{de}
\bsnm{Freitas},~\bfnm{Nando}\binits{N.}} \AND
\beditor{\bsnm{Gordon},~\bfnm{Neil}\binits{N.}}, eds.
(\byear{2001}).
\btitle{Sequential {M}onte {C}arlo Methods in Practice}.
\bpublisher{Springer}, \baddress{New York}.
\bid{mr={1847783}}
\bptok{imsref}%
\end{bbook}
%
\endbibitem

\bibitem[\protect\citeauthoryear{Fearnhead}{2002}]{Fearnhead2002}
%
\begin{barticle}[mr]
\bauthor{\bsnm{Fearnhead},~\bfnm{Paul}\binits{P.}}
(\byear{2002}).
\btitle{Markov chain {M}onte {C}arlo, sufficient statistics, and particle
filters}.
\bjournal{J. Comput. Graph. Statist.}
\bvolume{11}
\bpages{848--862}.
\bid{doi={10.1198/106186002321018821}, issn={1061-8600}, mr={1951601}}
\bptok{imsref}%
\end{barticle}
%
\endbibitem

\bibitem[\protect\citeauthoryear{Fearnhead and
Clifford}{2003}]{FearnheadClifford03}
%
\begin{barticle}[mr]
\bauthor{\bsnm{Fearnhead},~\bfnm{Paul}\binits{P.}} \AND
\bauthor{\bsnm{Clifford},~\bfnm{Peter}\binits{P.}}
(\byear{2003}).
\btitle{On-line inference for hidden {M}arkov models via particle filters}.
\bjournal{J. R. Stat. Soc. Ser. B Stat. Methodol.}
\bvolume{65}
\bpages{887--899}.
\bid{doi={10.1111/1467-9868.00421}, issn={1369-7412}, mr={2017876}}
\bptok{imsref}%
\end{barticle}
%
\endbibitem

\bibitem[\protect\citeauthoryear{Fearnhead, Wyncoll and
Tawn}{2010}]{Fearnhead2010}
%
\begin{barticle}[mr]
\bauthor{\bsnm{Fearnhead},~\bfnm{Paul}\binits{P.}},
\bauthor{\bsnm{Wyncoll},~\bfnm{David}\binits{D.}} \AND
\bauthor{\bsnm{Tawn},~\bfnm{Jonathan}\binits{J.}}
(\byear{2010}).
\btitle{A sequential smoothing algorithm with linear computational cost}.
\bjournal{Biometrika}
\bvolume{97}
\bpages{447--464}.
\bid{doi={10.1093/biomet/asq013}, issn={0006-3444}, mr={2650750}}
\bptok{imsref}%
\end{barticle}
%
\endbibitem

\bibitem[\protect\citeauthoryear{Fong et~al.}{2002}]{Fong02}
%
\begin{barticle}[author]
\bauthor{\bsnm{Fong},~\bfnm{W.}\binits{W.}},
\bauthor{\bsnm{Godsill},~\bfnm{S.~J.}\binits{S.~J.}},
\bauthor{\bsnm{Doucet},~\bfnm{A.}\binits{A.}} \AND
\bauthor{\bsnm{West},~\bfnm{M}\binits{M.}}
(\byear{2002}).
\btitle{Monte {C}arlo smoothing with application to speekch enhancement}.
\bjournal{IEEE Trans. Signal Process.}
\bvolume{50}
\bpages{438--449}.
\bptok{imsref}%
\end{barticle}
%
\endbibitem

\bibitem[\protect\citeauthoryear{Gilks and
Berzuini}{2001}]{GilksBerzuini2001}
%
\begin{barticle}[mr]
\bauthor{\bsnm{Gilks},~\bfnm{Walter~R.}\binits{W.~R.}} \AND
\bauthor{\bsnm{Berzuini},~\bfnm{Carlo}\binits{C.}}
(\byear{2001}).
\btitle{Following a moving target---{M}onte {C}arlo inference for dynamic
{B}ayesian models}.
\bjournal{J. R. Stat. Soc. Ser. B Stat. Methodol.}
\bvolume{63}
\bpages{127--146}.
\bid{doi={10.1111/1467-9868.00280}, issn={1369-7412}, mr={1811995}}
\bptok{imsref}%
\end{barticle}
%
\endbibitem

\bibitem[\protect\citeauthoryear{Godsill, Doucet and West}{2004}]{Godsill04}
%
\begin{barticle}[author]
\bauthor{\bsnm{Godsill},~\bfnm{S.~J.}\binits{S.~J.}},
\bauthor{\bsnm{Doucet},~\bfnm{A.}\binits{A.}} \AND
\bauthor{\bsnm{West},~\bfnm{M}\binits{M.}}
(\byear{2004}).
\btitle{Monte {C}arlo smoothing for non-linear time series}.
\bjournal{J. Amer. Statist. Assoc.}
\bvolume{50}
\bpages{438--449}.
\bptok{imsref}%
\end{barticle}
%
\endbibitem

\bibitem[\protect\citeauthoryear{Godsill and
Vermaak}{2004}]{GodsillVermaak2004}
%
\begin{barticle}[author]
\bauthor{\bsnm{Godsill},~\bfnm{S.~J.}\binits{S.~J.}} \AND
\bauthor{\bsnm{Vermaak},~\bfnm{J.}\binits{J.}}
(\byear{2004}).
\btitle{Models and algorithms for tracking using trans-dimensional sequential
Monte Carlo}.
\bjournal{Proc. IEEE ICASSP}
\bvolume{3}
\bpages{976--979}.
\bptok{imsref}%
\end{barticle}
%
\endbibitem

\bibitem[\protect\citeauthoryear{Gordon, Salmond and Smith}{1993}]{Gordonetal93}
%
\begin{barticle}[author]
\bauthor{\bsnm{Gordon},~\bfnm{N.~J.}\binits{N.~J.}},
\bauthor{\bsnm{Salmond},~\bfnm{D.~J.}\binits{D.~J.}} \AND
\bauthor{\bsnm{Smith},~\bfnm{A.~F.~M.}\binits{A.~F.~M.}}
(\byear{1993}).
\btitle{Novel approach to nonlinear/non-Gaussian Bayesian state
estimation}.
\bjournal{IEE Proceedings on Radar and Signal Processing}
\bvolume{140}
\bpages{107--113}.
\bptok{imsref}%
\end{barticle}
%
\endbibitem

\bibitem[\protect\citeauthoryear{Guo, Wang and Chen}{2004}]{Guoetal04}
%
\begin{barticle}[mr]
\bauthor{\bsnm{Guo},~\bfnm{Dong}\binits{D.}},
\bauthor{\bsnm{Wang},~\bfnm{Xiaodong}\binits{X.}} \AND
\bauthor{\bsnm{Chen},~\bfnm{Rong}\binits{R.}}
(\byear{2004}).
\btitle{Multilevel mixture {K}alman filter}.
\bjournal{EURASIP J. Appl. Signal Process.}
\bvolume{15}
\bpages{2255--2266}.
\bid{doi={10.1155/S1110865704403229}, issn={1110-8657}, mr={2116229}}
\bptok{imsref}%
\end{barticle}
%
\endbibitem

\bibitem[\protect\citeauthoryear{H{\"{u}}rzeler and
K{\"{u}}nsch}{1995}]{HurzKuns95}
%
\begin{bmisc}[author]
\bauthor{\bsnm{H{\"{u}}rzeler},~\bfnm{M.}\binits{M.}} \AND
\bauthor{\bsnm{K{\"{u}}nsch},~\bfnm{H.~R.}\binits{H.~R.}}
(\byear{1995}).
\bhowpublished{Monte {C}arlo approximations for general state space models.
Research Report 73, ETH, Z\"{u}rich.}
\bptok{imsref}%
\end{bmisc}
%
\endbibitem

\bibitem[\protect\citeauthoryear{Ikoma et~al.}{2001}]{Ikomaetal01}
%
\begin{bmisc}[author]
\bauthor{\bsnm{Ikoma},~\bfnm{N.}\binits{N.}},
\bauthor{\bsnm{Ichimura},~\bfnm{N.}\binits{N.}},
\bauthor{\bsnm{Higuchi},~\bfnm{T.}\binits{T.}} \AND
\bauthor{\bsnm{Maeda},~\bfnm{H.}\binits{H.}}
(\byear{2001}).
\bhowpublished{Maneuvering target tracking by using particle filter.
In \textit{Joint} 9\textit{th IFSA World Congress and} 20\textit{th NAFIPS International
Conference}
\textbf{4} 2223--2228. IEEE}.
\bptok{imsref}%
\end{bmisc}
%
\endbibitem

\bibitem[\protect\citeauthoryear{Kantas et~al.}{2009}]{Kantas2009}
%
\begin{bmisc}[author]
\bauthor{\bsnm{Kantas},~\bfnm{N.}\binits{N.}},
\bauthor{\bsnm{Doucet},~\bfnm{A.}\binits{A.}},
\bauthor{\bsnm{Singh},~\bfnm{S.~S.}\binits{S.~S.}} \AND
\bauthor{\bsnm{Maciejowski},~\bfnm{J.~M.}\binits{J.~M.}}
(\byear{2009}).
\bhowpublished{An overview of sequential Monte Carlo methods for parameter estimation
in general state-space models.
In 15\textit{th IFAC Symposium on System Identification}.}
\bptok{imsref}%
\end{bmisc}
%
\endbibitem

\bibitem[\protect\citeauthoryear{Kim, Shephard and Chib}{1998}]{Kim1998}
%
\begin{barticle}[author]
\bauthor{\bsnm{Kim},~\bfnm{S.}\binits{S.}},
\bauthor{\bsnm{Shephard},~\bfnm{N.}\binits{N.}} \AND
\bauthor{\bsnm{Chib},~\bfnm{S.}\binits{S.}}
(\byear{1998}).
\btitle{Stochastic volatility: Likelihood inference and comparison
with ARCH
models}.
\bjournal{Review of Economic Studies}
\bvolume{65}
\bpages{361--393}.
\bptok{imsref}%
\end{barticle}
%
\endbibitem

\bibitem[\protect\citeauthoryear{Kitagawa}{1996}]{Kita96}
%
\begin{barticle}[mr]
\bauthor{\bsnm{Kitagawa},~\bfnm{Genshiro}\binits{G.}}
(\byear{1996}).
\btitle{Monte {C}arlo filter and smoother for non-{G}aussian nonlinear state
space models}.
\bjournal{J. Comput. Graph. Statist.}
\bvolume{5}
\bpages{1--25}.
\bid{doi={10.2307/1390750}, issn={1061-8600}, mr={1380850}}
\bptok{imsref}%
\end{barticle}
%
\endbibitem

\bibitem[\protect\citeauthoryear{Kong, Liu and Wong}{1994}]{Kong94}
%
\begin{barticle}[author]
\bauthor{\bsnm{Kong},~\bfnm{A.}\binits{A.}},
\bauthor{\bsnm{Liu},~\bfnm{J.~S.}\binits{J.~S.}} \AND
\bauthor{\bsnm{Wong},~\bfnm{W.~H.}\binits{W.~H.}}
(\byear{1994}).
\btitle{Sequential imputations and Bayesian missing data problems}.
\bjournal{J. Amer. Statist. Assoc.}
\bvolume{89}
\bpages{278--288}.
\bptok{imsref}%
\end{barticle}
%
\endbibitem

\bibitem[\protect\citeauthoryear{Kotecha and Djuri{\'c}}{2003}]{KoteDjur03}
%
\begin{barticle}[mr]
\bauthor{\bsnm{Kotecha},~\bfnm{Jayesh~H.}\binits{J.~H.}} \AND
\bauthor{\bsnm{Djuri{\'c}},~\bfnm{Petar~M.}\binits{P.~M.}}
(\byear{2003}).
\btitle{Gaussian sum particle filtering}.
\bjournal{IEEE Trans. Signal Process.}
\bvolume{51}
\bpages{2602--2612}.
\bid{doi={10.1109/TSP.2003.816754}, issn={1053-587X}, mr={2003067}}
\bptok{imsref}%
\end{barticle}
%
\endbibitem

\bibitem[\protect\citeauthoryear{Liang, Chen and Zhang}{2002}]{Liang02}
%
\begin{barticle}[author]
\bauthor{\bsnm{Liang},~\bfnm{J}\binits{J.}},
\bauthor{\bsnm{Chen},~\bfnm{R.}\binits{R.}} \AND
\bauthor{\bsnm{Zhang},~\bfnm{J.}\binits{J.}}
(\byear{2002}).
\btitle{Statistical geometry of packing defects of lattice chain
polymer from
enumeration and sequential Monte Carlo method}.
\bjournal{J. Chem. Phys.}
\bvolume{117}
\bpages{3511--3521}.
\bptok{imsref}%
\end{barticle}
%
\endbibitem

\bibitem[\protect\citeauthoryear{Lin et~al.}{2005}]{Linetal05}
%
\begin{barticle}[mr]
\bauthor{\bsnm{Lin},~\bfnm{Ming~T.}\binits{M.~T.}},
\bauthor{\bsnm{Zhang},~\bfnm{Junni~L.}\binits{J.~L.}},
\bauthor{\bsnm{Cheng},~\bfnm{Qiansheng}\binits{Q.}} \AND
\bauthor{\bsnm{Chen},~\bfnm{Rong}\binits{R.}}
(\byear{2005}).
\btitle{Independent particle filters}.
\bjournal{J. Amer. Statist. Assoc.}
\bvolume{100}
\bpages{1412--1421}.
\bid{doi={10.1198/016214505000000349}, issn={0162-1459}, mr={2236451}}
\bptok{imsref}%
\end{barticle}
%
\endbibitem

\bibitem[\protect\citeauthoryear{Liu}{2001}]{Liu01}
%
\begin{bbook}[mr]
\bauthor{\bsnm{Liu},~\bfnm{Jun~S.}\binits{J.~S.}}
(\byear{2001}).
\btitle{Monte {C}arlo Strategies in Scientific Computing}.
\bpublisher{Springer}, \baddress{New York}.
\bid{mr={1842342}}
\bptok{imsref}%
\end{bbook}
%
\endbibitem

\bibitem[\protect\citeauthoryear{Liu and Chen}{1995}]{LiuChen95}
%
\begin{barticle}[author]
\bauthor{\bsnm{Liu},~\bfnm{J.~S.}\binits{J.~S.}} \AND
\bauthor{\bsnm{Chen},~\bfnm{R.}\binits{R.}}
(\byear{1995}).
\btitle{Blind deconvolution via sequential imputations}.
\bjournal{J. Amer. Statist. Assoc.}
\bvolume{90}
\bpages{567--576}.
\bptok{imsref}%
\end{barticle}
%
\endbibitem

\bibitem[\protect\citeauthoryear{Liu and Chen}{1998}]{LiuChen98}
%
\begin{barticle}[mr]
\bauthor{\bsnm{Liu},~\bfnm{Jun~S.}\binits{J.~S.}} \AND
\bauthor{\bsnm{Chen},~\bfnm{Rong}\binits{R.}}
(\byear{1998}).
\btitle{Sequential {M}onte {C}arlo methods for dynamic systems}.
\bjournal{J. Amer. Statist. Assoc.}
\bvolume{93}
\bpages{1032--1044}.
\bid{doi={10.2307/2669847}, issn={0162-1459}, mr={1649198}}
\bptok{imsref}%
\end{barticle}
%
\endbibitem

\bibitem[\protect\citeauthoryear{Liu, Chen and Wong}{1998}]{LiuChenWong98}
%
\begin{barticle}[mr]
\bauthor{\bsnm{Liu},~\bfnm{Jun~S.}\binits{J.~S.}},
\bauthor{\bsnm{Chen},~\bfnm{Rong}\binits{R.}} \AND
\bauthor{\bsnm{Wong},~\bfnm{Wing~Hung}\binits{W.~H.}}
(\byear{1998}).
\btitle{Rejection control and sequential importance sampling}.
\bjournal{J. Amer. Statist. Assoc.}
\bvolume{93}
\bpages{1022--1031}.
\bid{doi={10.2307/2669846}, issn={0162-1459}, mr={1649197}}
\bptok{imsref}%
\end{barticle}
%
\endbibitem

\bibitem[\protect\citeauthoryear{Marshall}{1956}]{Mars56}
%
\begin{binproceedings}[author]
\bauthor{\bsnm{Marshall},~\bfnm{A.~W.}\binits{A.~W.}}
(\byear{1956}).
\btitle{The use of multi-stage sampling schemes in Monte Carlo computations}.
In \bbooktitle{Symposium on Monte Carlo Methods}
(\beditor{\bfnm{M.~A.}\binits{M.~A.}~\bsnm{Meyer}}, ed.)
\bpages{123--140}.
\bpublisher{Wiley}, \baddress{New York}.
\bptok{imsref}%
\end{binproceedings}
%
\endbibitem

\bibitem[\protect\citeauthoryear{Pitt}{2002}]{Pitt2002}
%
\begin{bmisc}[author]
\bauthor{\bsnm{Pitt},~\bfnm{M.~K.}\binits{M.~K.}}
(\byear{2002}).
\bhowpublished{Smooth particle filters for likelihood and maximisation.
Technical report, Univ. Warwick}.
\bptok{imsref}%
\end{bmisc}
%
\endbibitem

\bibitem[\protect\citeauthoryear{Pitt and Shephard}{1999}]{PittShep99}
%
\begin{barticle}[mr]
\bauthor{\bsnm{Pitt},~\bfnm{Michael~K.}\binits{M.~K.}} \AND
\bauthor{\bsnm{Shephard},~\bfnm{Neil}\binits{N.}}
(\byear{1999}).
\btitle{Filtering via simulation: Auxiliary particle filters}.
\bjournal{J. Amer. Statist. Assoc.}
\bvolume{94}
\bpages{590--599}.
\bid{doi={10.2307/2670179}, issn={0162-1459}, mr={1702328}}
\bptok{imsref}%
\end{barticle}
%
\endbibitem

\bibitem[\protect\citeauthoryear{Rosenbluth and
Rosenbluth}{1955}]{RoseRose55}
%
\begin{barticle}[author]
\bauthor{\bsnm{Rosenbluth},~\bfnm{M.~N.}\binits{M.~N.}} \AND
\bauthor{\bsnm{Rosenbluth},~\bfnm{A.~W.}\binits{A.~W.}}
(\byear{1955}).
\btitle{Monte Carlo calculation of the average extension of molecular chains}.
\bjournal{J. Chem. Phys.}
\bvolume{23}
\bpages{356--359}.
\bptok{imsref}%
\end{barticle}
%
\endbibitem

\bibitem[\protect\citeauthoryear{van~der Merwe et~al.}{2002}]{Merweetal02}
%
\begin{binproceedings}[author]
\bauthor{\bparticle{van~der} \bsnm{Merwe},~\bfnm{R.}\binits{R.}},
\bauthor{\bsnm{Doucet},~\bfnm{A.}\binits{A.}}, \bauthor{\bparticle{de}
\bsnm{Freitas},~\bfnm{N.}\binits{N.}} \AND
\bauthor{\bsnm{Wan},~\bfnm{E.}\binits{E.}}
(\byear{2002}).
\btitle{The unscented particle filter}.
In \bbooktitle{Advances in Neural Information Processing Systems (NIPS13)}
(\beditor{T. K. Leen}, \beditor{T. G. Dietterich} and \beditor{V.
Tresp}, eds.).
\bpublisher{MIT Press}, \baddress{Cambridge, MA}.
\bptok{imsref}%
\end{binproceedings}
%
\endbibitem

\bibitem[\protect\citeauthoryear{Wang, Chen and Guo}{2002}]{Wang02SP}
%
\begin{barticle}[author]
\bauthor{\bsnm{Wang},~\bfnm{X.}\binits{X.}},
\bauthor{\bsnm{Chen},~\bfnm{R.}\binits{R.}} \AND
\bauthor{\bsnm{Guo},~\bfnm{D.}\binits{D.}}
(\byear{2002}).
\btitle{Delayed pilot sampling for mixture Kalman filter with
application in fading channels}.
\bjournal{IEEE Trans. Signal Process.}
\bvolume{50}
\bpages{241--264}.
\bptok{imsref}%
\end{barticle}
%
\endbibitem

\bibitem[\protect\citeauthoryear{Zhang and Liu}{2002}]{ZhangLiu02}
%
\begin{barticle}[author]
\bauthor{\bsnm{Zhang},~\bfnm{J.~L.}\binits{J.~L.}} \AND
\bauthor{\bsnm{Liu},~\bfnm{J.~S.}\binits{J.~S.}}
(\byear{2002}).
\btitle{A new sequential importance sampling method and its
application to the
two-dimensional hydrophobic-hydrophilic model}.
\bjournal{J. Chem. Phys.}
\bvolume{117}
\bpages{3492--3498}.
\bptok{imsref}%
\end{barticle}
%
\endbibitem

\end{thebibliography}
\end{document}